\documentclass[a4paper,11pt]{article}
\usepackage[utf8]{inputenc}
\usepackage[english]{babel}
\usepackage[left=3cm,right=3cm,top=3cm,bottom=3.5cm]{geometry}

\usepackage{amsmath}
\usepackage{amsfonts}
\usepackage{amssymb}

\usepackage{soul}

\usepackage{color}
\definecolor{nicered}{rgb}{0.7,0.1,0.1}
\definecolor{nicegreen}{rgb}{0.1,0.5,.1}
\definecolor{niceblue}{rgb}{0.1,0.2,0.8}

\usepackage{hyperref}
\hypersetup{colorlinks,citecolor=nicegreen,linkcolor=nicered}
\hypersetup{colorlinks=true}

\usepackage{graphicx}

\usepackage{slashed}

\usepackage[makeroom]{cancel}

\usepackage[table]{xcolor}

\usepackage{fancyhdr}

\usepackage{authblk}

\usepackage{stackrel}

\usepackage{comment}

\usepackage{float}
\usepackage{mathtools}
\usepackage{bm}
\usepackage{mathrsfs}
\usepackage{esvect}
\usepackage{feynmf}
\usepackage{simplewick}
\usepackage{anyfontsize}
\usepackage{setspace}
\usepackage{sectsty}
\usepackage{isotope}
\usepackage{afterpage}
\usepackage{xcolor}
\usepackage{multirow}

\numberwithin{equation}{section}

\usepackage{simpler-wick}

\usepackage{rotating}

\title{Symmetry-restoring finite counterterms of SMEFT four-fermion operator insertions at one loop}

\author[a]{Sergio~Ferrando~Solera}
\author[b]{Sebastian~J\"ager}
\author[c]{Luiz~Vale~Silva}

\affil[a]{\it Departament de F\'{i}sica Te\`{o}rica, Instituto de F\'{i}sica Corpuscular,

Universitat de Val\`encia -- Consejo Superior de Investigaciones Cient\'{i}ficas,

Parc Cient\'{i}fic, Catedr\'{a}tico Jos\'{e} Beltr\'{a}n 2, E-46980 Paterna, Valencia, Spain}

\affil[b]{\it University of Sussex, Department of Physics and Astronomy, Falmer, Brighton BN1 9QH, UK}

\affil[c]{\it Departamento de Matem\'{a}ticas, F\'{i}sica y Ciencias Tecnol\'{o}gicas,

Universidad Cardenal Herrera-CEU, CEU Universities,

46115 Alfara del Patriarca, Val\`{e}ncia, Spain}

\begin{document}

{\let\newpage\relax\maketitle}

\maketitle

\noindent
\textbf{Abstract.} Some effects induced by SMEFT operators at one loop have attracted a lot of attention in recent years,
in particular, the renormalization of divergences by physical operators in single insertions of dimension-6 operators. Important non-logarithmically enhanced
contributions must also be calculated.
We discuss dimensional regularization in the Breitenlohner-Maison-'t Hooft-Veltman scheme.
The goal here consists of determining in this scheme quantum effects in chiral theories at one loop.
Namely, the determination of finite counterterms at one loop that reestablish the Slavnov-Taylor identities,
which follow from gauge symmetries.
These counterterms are necessary due to the presence of evanescent symmetry-breaking terms in the classical Lagrangian needed to regularize fermion propagators.
We consider a
technique that allows an easier calculation of such finite effects, relying on the identification of $(D-4)/(D-4)$ terms of one-loop amplitudes with an external ghost leg.
We focus on dimension-6 four-fermion operators, identifying all finite counterterms in the Breitenlohner-Maison-'t Hooft-Veltman scheme at one loop,
and as expected find no obstructions to the Slavnov-Taylor identities that cannot be cured by finite counterterms.
This represents one step towards moving to higher order calculations.

\section{Introduction}

%%%%%%%%%%%%%%% Why moving beyond %%%%%%%%%%%%%%%%%%%
The current Standard Model (SM) of particle physics leaves many problems opened,
including the matter/anti-matter asymmetry puzzle and dark matter.
Also, the hierarchy among some of its free parameters
points at a more fundamental dynamics able to account for the emergence of this structure.
One has attempted various directions to extend the SM,
which so far lack of direct evidence.

We will exploit here the Standard Model Effective Field Theory (SMEFT). This consists of the full set of higher-dimensional operators that respect the gauge symmetries of the SM, and built exclusively from the SM degrees of freedom.
The associated scale of New Physics (NP) is assumed to remain sufficiently above the electroweak (EW) scale,
and the Wilson Coefficients (WCs) of the SMEFT operators are assumed to be perturbative, i.e., small enough so that one is able to truncate the series of higher-dimensional operators to achieve a good precision when performing predictions from the SMEFT.
These operators are organized according to inverse positive powers of the NP scale, namely, operators of dimension-$(4+n)$ are suppressed by $n$ powers of this NP scale.
For most purposes, one can consider operators of dimension up to six.
The basis of dimension-6 operators
consists of 2499 operators when enforcing total baryon and lepton numbers global symmetries.
When writing down the minimal basis of operators of a certain dimension, one applies Equations of Motion (EoMs), algebraic identities and integration by parts.
Different bases are possible, the most common currently used one of dimension-6 operators is the so-called Warsaw basis \cite{Grzadkowski:2010es}.

%%%%%%%%%%%%%%% Direct searches in colliders %%%%%%%%%%%%%%%%%%%

At the tree level, many phenomenological bounds exist.
According to the category of operators, their WCs are more or less constrained: e.g., the WCs of flavour-conserving operators are typically less constrained than the WCs of flavour-changing ones.
%
%%%%%%%%%%%%%%% One-loop renormalization %%%%%%%%%%%%%%%%%%%
%
The full one-loop renormalization of divergences by physical operators is available \cite{Jenkins:2013zja,Jenkins:2013wua,Alonso:2013hga,Alonso:2014zka}, unlocking further possibilities for phenomenology.
This is so because, for instance, operators of different categories mix under renormalization. Such a mixing may then lead to more constraining phenomenological bounds on the WCs, since operators of different categories, such as distinct flavour structures, are constrained differently.

%%%%%%%%%%%%%%% One-loop ren.: finite counterterms %%%%%%%%%%%%%%%%%%%

In practice, one often wishes that the NP scale is not prohibitively too large to prevent the search for direct effects of the new heavy degrees of freedom (or that the related WCs are large enough for the same reason).
Therefore, a leading-logarithmic analysis is likely to be incomplete: indeed, in numerical studies logarithmic effects
may be of the same order of finite contributions, that are sub-leading in the logarithmic counting.
At the very least, the quest for precision motivates the study of sub-leading effects.
%For different reasons discussed next,
%these finite contributions are also necessary.
Furthermore, one-loop
finite terms may become crucial in phenomenological applications,
in particular where some one-loop mixing among physical operators caused by the renormalization of divergences vanishes.
Therefore, such finite effects alone can also lead to important phenomenological bounds on the WCs.
We note that other one-loop finite matching effects, which are not the focus of the present work, have been computed in the context of the low-energy EFT starting from SMEFT
in a scheme mostly consisting of naive dimensional regularization
\cite{Dekens:2019ept}.

%%%%%%%%%%%%%%% Clifford algebra %%%%%%%%%%%%%%%%%%%
%%%%%%%%%%%%%%% Spurious anomalies %%%%%%%%%%%%%%%%%%%

In this paper, we discuss dimensional regularization \cite{tHooft:1972tcz,Bollini:1972ui,Cicuta:1972jf,Ashmore:1972uj,Collins:1984xc}.
When computing one-loop diagrams of certain operators of SMEFT one deals with the algebra of gamma matrices.
This raises delicate issues, as illustrated by the well-known case of anomalous chiral symmetries, which is a physical effect with observable consequences.
To compute finite terms, one has to provide a consistent treatment of the Clifford algebra of gamma matrices in dimensional regularization, which has been achieved firstly by Breitenlohner-Maison \cite{Breitenlohner:1977hr,Breitenlohner:1975hg,Breitenlohner:1976te}.
There are effects analogous in a sense to true anomalies that at first sight could seem to imply that a larger set of symmetries becomes anomalous, e.g., local symmetries. If present, such true anomalies would be manifested by the irremediable failure to satisfy
Ward-Takahashi or
Slavnov-Taylor identities relating Green's functions, thus invalidating the physical theory.
Indeed, local gauge symmetries are needed for dealing with massless vector bosons.
More generally,
these identities express the properties of a theory that respects symmetries, which if broken at the quantum level cannot be exploited to perform predictions from the theory.

%%%%%%%%%%%%%%% Curing spurious anomalies %%%%%%%%%%%%%%%%%%%

The so-called ``spurious anomalies'' (also called ``trivial anomalies'') can, however, be cured by the consideration of appropriate local counterterms, which restore the symmetries at the quantum level, see e.g. Ref.~\cite{Barnich:2000zw}.
In practice, one follows the following steps: if certain symmetries are not anomalous (as in the case of the SM local symmetries, see e.g. Ref.~\cite{Weinberg:1996kr}), one computes the spurious anomalous terms, and renormalizes them away via available finite counterterms; the fact that these symmetries are free of anomalies guarantees that such local counterterms exist (were they not available, the symmetry would be necessarily anomalous).
Non-anomalous global symmetries in the low-energy EFT are studied in great details in Ref.~\cite{Naterop:2023dek}, using different methods compared to the ones depicted below, that we use to study local symmetries; see also Refs.~\cite{Naterop:2024cfx,Naterop:2025cwg}.

%%%%%%%%%%%%%%% Cohomology %%%%%%%%%%%%%%%%%%%

Ultimately, building the action and renormalizing it is an exercise of classifying operators and counterterms of distinct ghost numbers and mass dimensions according to the Becchi-Rouet-Stora-Tyutin (BRST) transformation.
At the classical level, the action can be expressed as a gauge-invariant part of ghost number $0$ (built from matter and gauge fields, for which a BRST transformation can be deduced from an infinitesimal gauge transformation) plus the result of the BRST operator acting on a polynomial of ghost number $-1$ that plays the role of the gauge-fixing functional.
Since the BRST operator is nilpotent (of index two), the physical content of the theory is therefore set by the kernel of the BRST operator modulo BRST-exact operators (this notion can be extended to operators that are total derivatives when discussing the Lagrangian density, see e.g. Ref.~\cite{Barnich:2000zw}), or in other words its cohomology; see Sec.~\ref{sec:Usual_Renormalization}.
Possible counterterms taking care of infinities in a vector-like theory can be listed systematically by building up all possible symmetric
operators of ghost number $0$, including those built from Batalin-Vilkovisky antifields \cite{Dixon:1974ss,Kluberg-Stern:1974iel,Joglekar:1975nu,Joglekar:1976eb,Joglekar:1976pe,Collins:1984xc,Barnich:2000zw}.
Operators that vanish following the use of EoMs are formally treated when such antifields are considered \cite{Barnich:2000zw}.
After the introduction of the BMHV scheme in Sec.~\ref{sec:BMHV}
and the identification of a symmetry-breaking term in dimensional regularization already at the classical level in Sec.~\ref{sec:BMHV_breaking},
we turn to local, finite counterterms, which are also in the space of ghost number $0$ (derived first in the formalism described below as local functionals of ghost number $1$, that admit being recast in BRST-exact form), but that are not gauge-invariant.
This will be further discussed in Sec.~\ref{sec:Restoring_Symmetry}, where we introduce the framework of algebraic renormalization, which reestablishes, in the absence of true anomalies,
the Slavnov-Taylor identities.
The renormalization of infinities of non-power counting renormalizable theories are discussed in e.g. Refs.~\cite{Collins:1984xc,Weinberg:1996kr}, while cohomology results for EFTs are also stated in Ref.~\cite{Barnich:2000zw}.

%%%%%%%%%%%%%%% Anomalies, consistent conditions %%%%%%%%%%%%%%%%%%%

We note that possible anomalies can also be expressed in cohomological language following the Wess-Zumino consistency conditions, this time in the space of ghost number $1$. Whether anomalies are present or not, these conditions must be respected. Therefore, the consistency conditions offer a sanity check that we employed at intermediate steps of our calculations, i.e., before committing to find finite counterterms. We note that since the SMEFT is built on the basis of the same field content and local symmetries of the SM the latter local symmetries will remain anomaly free, see Refs.~\cite{Minn:1986ba,Feruglio:2020kfq} and references therein, while Ref.~\cite{Bonnefoy:2020tyv} discusses the same result in the case of SMEFT operators of the category $ \psi^2 H^2 D $. See also the recent discussion of Ref.~\cite{Cohen:2023gap}.
Non-renormalizable interactions are also discussed in Ref.~\cite{Anselmi:2015sqa}.

%%%%%%%%%%%%%%% Strategy %%%%%%%%%%%%%%%%%%%

There is a practical algorithm available for the computation of such counterterms which does not require the determination of finite contributions from the original set of diagrams that break the symmetries of the theory, but rather the infinities of a related set of Feynman diagrams (as it will be made explicit later in the text, the former set of diagrams has no external ghost line, while the latter does, with a modified vertex being employed).
This is possible because such finite counterterms correspond in dimensional regularization to poles in the regulator (from the Feynman integral) that are compensated by the same power of the regulator in the numerator (from the Dirac algebra, when effects of the symmetry-breaking term are considered); employing the usual notation, the spurious anomalies at one loop come from $\varepsilon / \varepsilon$ terms, where $\varepsilon = 4 - D$,
$D$ being the dimension of the Lagrangian density in dimensional regularization.
This simplifies enormously the technical work, and allows a more straightforward automation, which is welcomed for the simultaneous determination of the leading (the ones that renormalize infinities) and sub-leading (the ones that restore Slavnov-Taylor identities) counterterms.
The usage of the approach due to Bonneau \cite{Bonneau:1979jx,Bonneau:1980zp} will be discussed in Sec.~\ref{sec:Restoring_Symmetry_Bonneau}; to the best of our knowledge, it is the first time algebraic renormalization and the simplifying approach by Bonneau are discussed in the context of SMEFT.
Although the $\varepsilon / \varepsilon$ amplitudes of ghost number $1$ are uniquely defined, the finite counterterms of ghost number $0$ are determined up to a BRST-invariant ambiguity. Other scheme choices will be spelled out in the following sections.

%%%%%%%%%%%%%%% Starting point for pheno/for this whole project %%%%%%%%%%%%%%%%%%%

We focus on a class of such finite effects inspired by preliminary two-loop calculations \cite{Silva:2020vyx,ValeSilva:2022nzm}.
Our interest therein was the determination of the two-loop anomalous dimension describing the mixing of certain four-fermion operators into dipole operators, in cases where the mixing was absent at one loop (we also discussed dimension-6 Yukawa operators in naive dimensional regularization in Ref.~\cite{Jager:2019wkc}).
In that preliminary computation, we further focused on those cases that could avoid an Yukawa suppression, and thus limited the analysis to a subset of the four-fermion operators of the Warsaw basis.
We considered the BMHV scheme for dealing with $\gamma_5$, thus requiring the introduction of finite counterterms at one loop, likely contributing to the validation of some cross-checks based on relations among Green's functions at two-loops.
We will here identify all finite counterterms necessary to restore the Slavnov-Taylor identities at one loop in presence of any four-fermion operator of the Warsaw basis (or any other linearly related basis, while evanescent differences among operators do not produce any effect at the considered one-loop order), reproduced and extended to include right-handed neutrinos in Sec.~\ref{sec:operators}.

%%%%%%%%%%%%%%% Literature %%%%%%%%%%%%%%%%%%%

Such finite counterterms are not necessary at the leading-logarithmic order, which should explain why such effects have been scarcely discussed in the literature of non-power counting renormalizable NP operators in the past, Refs.~\cite{Durieux:2018ggn,Boughezal:2019xpp,Degrande:2020evl,Feruglio:2020kfq} being noticeable exceptions nonetheless.
Very recently, one witnessed a growing interest for their systematic discussion, as shown by the aforementioned Ref.~\cite{Naterop:2023dek} in the case of the low-energy EFT, and Ref.~\cite{Fuentes-Martin:2025meq} in the case of SMEFT, which we will later discuss more closely, where a different technique with respect to the one depicted above is employed.
The enforcement of chiral symmetries at the quantum level has been discussed for renormalizable interactions in Refs.~\cite{Ferrari:1994ct,Trueman:1995ca,Martin:1999cc,Grassi:1999tp,Belusca-Maito:2020ala,Cornella:2022hkc,OlgosoRuiz:2024dzq};
notably, the methods discussed in the present paper are detailed and applied beyond the one-loop order in the framework of algebraic renormalization in a series of papers \cite{Belusca-Maito:2021lnk,Belusca-Maito:2022wem,Stockinger:2023ndm,Kuhler:2025znv,vonManteuffel:2025swv}.
It is clear from e.g. Ref.~\cite{Belusca-Maito:2020ala} that the discussion of anomalies and finite counterterms allows a simultaneous calculation (although if anomalies of local symmetries are present, they make the theory not viable).
It should not be surprising that the two aspects are related \cite{Belusca-Maito:2023wah}: true anomalies are in principle possible, so a (consistent) regularization and renormalization scheme cannot manifestly respect chiral symmetry invariance; conversely, this fact can lead to the presence of symmetry breaking effects in anomaly-free theories that can be cured by finite counterterms.
Finally, we mention that 4-dimensional schemes may also require a special treatment of $ \gamma_5 $, see e.g. Refs.~\cite{Bruque:2018bmy,Cherchiglia:2021uce,Rosado:2024pyy,Rosado:2024gqp} for the recent literature.

%%%%%%%%%%%%%%% Comparison to one-loop SMEFT %%%%%%%%%%%%%%%%%%%

In the absence of true anomalies, the workload is similar to the one needed in order to renormalize divergences in single insertions at one loop of dimension-6 operators of SMEFT \cite{Jenkins:2013zja,Jenkins:2013wua,Alonso:2013hga,Alonso:2014zka}. Namely: (i) following the method by Bonneau \cite{Bonneau:1979jx,Bonneau:1980zp}, compute all divergences of single insertions of dimension-6 operators that also carry single insertions of the symmetry-breaking operator (in its ``inverse-hat'' form; the final results are displayed in Sec.~\ref{sec:results_CTs} and App.~\ref{app:example}); (ii) build a complete basis of dimension-6 operators that do not vanish under the BRST transformation (a dedicated discussion follows in App.~\ref{sec:list_CTs}); (iii) find the linear combination of the latter non-invariant operators that acts as a counterterm to compensate for the results of the first step (a detailed discussion in one particular example, the most challenging one, follows in App.~\ref{app:example}).
Fortunately, in (i) a series of more difficult topologies does not contribute (see App.~\ref{sec:properties_Greens_functions}).

%%%%%%%%%%%%%%% Index %%%%%%%%%%%%%%%%%%%

After discussing basic concepts in Sec.~\ref{sec:Formalism}, we discuss the computation of the symmetry breaking terms, and present solutions for their finite renormalization in Sec.~\ref{sec:results_CTs}.
Our conclusions are found in Sec.~\ref{sec:conclusions}. A series of appendices discuss more technical aspects.

\section{Formalism}\label{sec:Formalism}

\subsection{Renormalization of gauge theories}\label{sec:Usual_Renormalization}

We first introduce the central physical objects of our work and briefly discuss their meaning; a longer discussion can be found e.g. in Refs.~\cite{Weinberg:1996kr,Itzykson:1980rh,Zinn-Justin:2002ecy}.
At the classical level a gauge theory is described by a Lagrangian \(\mathcal{L}_{\mathrm{Tree}}=\mathcal{L}_{\mathrm{Tree}}\left(A_{\mu}^{\alpha},\psi_{l}^r\right)\) that contains information about the interactions of the so-called gauge \(A_{\mu}^{\alpha}\left(x\right)\) and matter fields \(\psi_l^r\left(x\right)\). This Lagrangian must be invariant under the infinitesimal gauge transformation (no sum over $r$ is understood on the right-hand side of the following equation)

\begin{eqnarray}
\label{Gauge_Transformation}
    \begin{array}{l}
        \delta\psi_l^r\left(x\right)=i\,\varepsilon^{\alpha}\left(x\right)\left(t_{\alpha}^r\right)_l^m\psi^r_m\left(x\right),  \\ \\
        \delta A_{\mu}^{\beta}\left(x\right)=\partial_{\mu}\varepsilon^{\beta}\left(x\right)+C^{\beta}_{\gamma\alpha}\varepsilon^{\alpha}\left(x\right)A^{\gamma}_{\mu}\left(x\right),
    \end{array}
\end{eqnarray}

\noindent where \(\varepsilon^{\alpha}\left(x\right)\) is an infinitesimal function, \(t_{\alpha}^r\) with \(\alpha=1,\dots,n\) are generators of
a Lie group
and \(C^{\gamma}_{\alpha\beta}\) the associated structure constants defined by the relation \(\left[t_{\alpha},t_{\beta}\right]=i\,C^{\gamma}_{\alpha\beta}t_{\gamma}\).
This means that the number of gauge fields \(n\)
is equal to the dimension of the Lie group, also referred to as the gauge group. On the other hand, we are using the superscript \(r\) to indicate that we can have different
matter fields (in the following taken to be irreducible), which transform under different representations of the gauge group.
To shorten expressions, we will be omitting gauge couplings everywhere in this section.
It is important to remark that it is only possible to construct gauge invariant kinetic terms for the gauge fields for certain Lie groups:
the ones that can be written as a direct product of compact simple Lie groups (which are classified in the Cartan catalog) and
$U(1)$ factors.
For those groups it is always possible to find a basis of generators for which the structure constants are totally antisymmetric. We will use this prescription in this text, writing $ C^{\beta}_{\gamma\alpha} = C_{\beta\gamma\alpha} $.

To quantize this theory in a general gauge a new set of fields is introduced: the so-called ghosts \(c_{\alpha}\left(x\right)\) and antighosts \(c_{\alpha}^{\ast}\left(x\right)\), as well as the Nakanishi-Lautrup fields \(h_{\alpha}\left(x\right)\).
The ghosts and antighosts are independent anticommuting variables which have an associated charge
called ghost number, whose value is \(1\) for ghosts and \(-1\) for antighosts. The Nakanishi-Lautrup fields are just an auxiliary set of commuting variables. The infinitesimal gauge transformation is substituted by
the so-called BRST transformation

\begin{eqnarray}
    \label{BRST_Transformation}
    \begin{array}{l}
        \delta_{\theta}\psi_{l}^r\left(x\right)=i\,\theta c_{\alpha}\left(x\right)\left(t^r_{\alpha}\right)^m_l\psi^r_m\left(x\right),\\ \\
        \delta_{\theta}A^{\mu}_{\alpha}\left(x\right)=\theta\left[\partial^{\mu}c_{\alpha}\left(x\right)+C_{\alpha\beta\gamma}A^{\mu}_{\beta}\left(x\right) c_{\gamma}\left(x\right)\right], \\ \\
        \delta_{\theta} c_{\alpha}^{\ast}\left(x\right)=-\theta h_{\alpha}\left(x\right), \\ \\
        \delta_{\theta} c_{\alpha}\left(x\right)=-\dfrac{1}{2}\theta C_{\alpha\beta\gamma} c_{\beta}\left(x\right) c_{\gamma}\left(x\right), \\ \\
        \delta_{\theta} h_{\alpha}\left(x\right)=0,
    \end{array}
\end{eqnarray}
where \(\theta\) is an anticommuting infinitesimal constant.
As we can see, for the matter and gauge fields the BRST transformation is just the infinitesimal gauge transformation, but with the correspondence \(\varepsilon_{\alpha}\left(x\right)=\theta c_{\alpha}\left(x\right)\). It is also usual to define the operator \(s\) as \(\delta_{\theta}F\coloneqq\theta\,\left(sF\right)\). The main properties of the BRST transformation are that it is non-linear (in contrast with the gauge transformation) when acting on matter, gauge and ghost fields, it is nilpotent \(ssF=0\) and it increases the ghost number by one unit \(\mathrm{gh}\left(sF\right)=1+\mathrm{gh}\left(F\right)\).

In proving renormalizability,
gauge invariance is then substituted by BRST invariance at the quantum level
for every choice of the gauge fixing functional. As a consequence, assume that the action that describes the theory which has a gauge symmetry at the classical level is any functional of ghost number $0$ that is invariant under the associated BRST transformation. It is easy to show that the most general form of such a functional is (we will later briefly comment on so-called antifields)

\begin{equation}
    \label{Action_BRST}
    I\left[A_{\mu}^{\alpha},\psi_l^r,c_{\alpha},c^{\ast}_{\alpha},h_{\alpha}\right]=I_0\left[A_{\mu}^{\alpha},\psi_l^r\right]+s\Psi\left[A_{\mu}^{\alpha},\psi_l^r,c_{\alpha},c^{\ast}_{\alpha},h_{\alpha}\right],
\end{equation}

\noindent where \(I_0\) is a gauge invariant action and \(\Psi\) a functional of ghost number \(-1\). Furthermore, in order to describe a theory at the quantum level it is convenient to use the path integral formulation. For this purpose, we introduce the vacuum-vacuum amplitude in the presence of external currents \(Z\left[J\right]\). Denoting all of the fields (including ghosts and Nakanishi-Lautrup fields) of the theory as \(\chi^n\left(x\right)\), we can write that amplitude as 

\begin{equation}
    \label{Path_Integral}
    Z\left[J\right]=\int\mathcal{D}\chi\exp\left\{iI\left[\chi\right]+i\int{\rm d}^4x\chi^n\left(x\right)J_{n}\left(x\right)\right\}.
\end{equation}
All of the correlation functions of the theory are obtained from functional derivatives of \(Z\left[J\right]\) with respect to the currents. It is important to remark that \(J_n\left(x\right)\) has the same bosonic or fermionic nature (commuting or anticommuting) as the associated \(\chi^n\left(x\right)\);
under the BRST transformation,
$ s J_n = 0 $.

Now, define the functional \(W\left[J\right]\) by the condition \(Z\left[J\right]\coloneqq\exp\left(iW\left[J\right]\right)\). It is easy to show that \(W\left[J\right]\) is the contribution to the vacuum-vacuum amplitude in presence of classical currents \(J_n\left(x\right)\) from connected Feynman diagrams. Then, define the classical fields \(X^n_J\left(x\right)\) as

\begin{equation}
    \label{Classical_Field}
    X^n_J\left(x\right)\coloneqq-\frac{i}{Z\left[J\right]}\frac{\delta_R}{\delta J_n\left(x\right)}Z\left[J\right]=\frac{\delta_R}{\delta J_n\left(x\right)}W\left[J\right],
\end{equation}

\noindent where the subscript \(R\) represents differentiation from the right-hand side. Assuming the previous equation can be inverted, we can define the current \(\left(J_{X}\right)_n\left(x\right)\) as the one for which \(X_J^n\left(x\right)\) takes the particular value \(X^n\left(x\right)\). Finally, define the quantum effective action as the Legendre transformation of \(W\left[J\right]\):

\begin{equation}
    \label{Quantum_Effective_Action}
    \Gamma\left[X\right]\coloneqq-\int{\rm d}^4xX^n\left(x\right)\left(J_X\right)_n\left(x\right)+W\left[J_X\right],
\end{equation}

\noindent which satisfies the condition

\begin{equation}
    \label{Derivative_QEA}
    \frac{\delta_L\Gamma\left[X\right]}{\delta X^n\left(x\right)}=-\left(J_X\right)_n\left(x\right),
\end{equation}

\noindent where the subscript \(L\) represents differentiation from the left-hand side.
It can be shown that the quantum effective action is also related to \(W\left[J\right]\) by 

\begin{equation}
    \label{W-QEA}
    i\,W\left[J\right]=\int\limits_{\substack{\textrm{CONNECTED}\\ \textrm{TREE}}} \mathcal{D}\chi \exp\left\{i\Gamma\left[\chi\right]+i\int{\rm d}^4 x \chi^n\left(x\right)J_{n}\left(x\right)\right\}.
\end{equation}
This means that the quantum effective action
contains all the information about the quantum corrections of the theory. Every amplitude can be obtained from tree-level calculations using \(\Gamma\left[\chi\right]\) instead of \(I\left[\chi\right]\). From these notions, it can be shown that the quantum effective action can be computed as

\begin{equation}
    \label{Calculation_QEA}
    i\Gamma\left[\chi_0\right]=\int\limits_{\substack{1\textrm{PI} \\ \textrm{CONNECTED}}}\mathcal{D}\chi\exp\left\{iI\left[\chi+\chi_0\right]\right\},
\end{equation}

\noindent where a \(1\) Particle Irreducible (1PI) diagram is one that remains connected when we cut any one of its internal lines. In words, the quantum effective action collects the results of 1PI connected diagrams.

We would like to know how a symmetry of the action \(I\left[\chi\right]\) translates into properties of the quantum effective action \(\Gamma\left[\chi\right]\). Let us consider a transformation of the fields

\begin{equation}
    \label{Field_Transformation}
    \chi^n\left(x\right)\rightarrow\chi^n\left(x\right)+\widetilde{\varepsilon} F^n\left[x;\chi\right],
\end{equation}

\noindent where \(\widetilde\varepsilon\) is a commuting or anticommuting infinitesimal number (introduced with a tilde to distinguish it from the parameter from dimensional regularization). At the quantum level, the product \(\mathcal{D}\chi\exp\left\{iI\left[\chi\right]\right\}\) has to remain invariant under the transformation in Eq.~\eqref{Field_Transformation},
otherwise we would have a classical symmetry broken at the quantum level, a so-called anomaly. If the theory is not anomalous then the quantum effective action satisfies the well-known Slavnov-Taylor identity

\begin{equation}
    \label{Slavnov_Taylor}
    \int{\rm d}^4 x \left<F^n\left(x\right)\right>_{J_X}\frac{\delta_L\Gamma\left[X\right]}{\delta X^n\left(x\right)}=0,
\end{equation}

\noindent where the vacuum expectation value \( \left<F^n\left(x\right)\right>_{J_X}\) is given by

\begin{equation}
    \label{Vacuum_Expectation_Value}
        \left<F^n\left(y\right)\right>_{J}\coloneqq Z^{-1}\left[J\right]\int\mathcal{D}\chi F^n\left[y;\chi\right]\exp\left\{iI\left[\chi\right]+i\int{\rm d}^4x\chi^n\left(x\right)J_{n}\left(x\right)\right\}.
\end{equation}
If the functions \(F^n\left[x;\chi\right]\) are linear on the fields then the Slavnov-Taylor identity implies that the quantum effective action satisfies the same linear symmetries as the classical action.
In the case of non-linear functions, as in the case of the BRST transformation, of course a symmetry is still respected by the quantum effective action, but not necessarily the one satisfied by the classical action.
In order to work with non-linear transformations, introduce
further external sources \(K_n\left(x\right)\).
We define the vacuum-vacuum amplitude in presence of the classical currents \(J_n\left(x\right)\) and
\(K_n\left(x\right)\) as:

\begin{equation}
    \label{Path_Integral_Antifield}
    Z\left[J,K\right]\coloneqq e^{iW\left[J,K\right]}=\int\mathcal{D}\chi\exp\left\{iI\left[\chi\right]+i\int{\rm d}^4xF^n\left(x\right)K_n\left(x\right)+i\int{\rm d}^4x\chi^n\left(x\right)J_{n}\left(x\right)\right\},
\end{equation}

\noindent where the sources \(K_n\left(x\right)\) have the opposite statistics with respect to the fields \(\chi^n\left(x\right)\). The new quantum effective action is defined as

\begin{equation}
    \label{QEA_Antifields}
    \Gamma\left[X,K\right]\coloneqq-\int{\rm d}^4xX^n\left(x\right)\left(J_{X,K}\right)_n\left(x\right)+W\left[J_{X,K},K\right],
\end{equation}

\noindent where \(\left(J_{X,K}\right)_n\left(x\right)\) is the value of the current \(J_n\left(x\right)\) for which the classical field

\begin{equation}
    \label{Classical_Field_Antifield}
    X_{J,K}^n\left(x\right)\coloneqq\frac{\delta_R}{\delta J_n\left(x\right)}W\left[J,K\right]
\end{equation}

\noindent takes the particular value \(X^n\left(x\right)\).
The Slavnov-Taylor identity can alternatively be written as

\begin{equation}
    \label{Zinn_Justin}
    \int{\rm d}^4 x \frac{\delta_R\Gamma\left[\chi,K\right]}{\delta K_n\left(x\right)}\frac{\delta_L\Gamma\left[\chi,K\right]}{\delta\chi^n\left(x\right)}=0,
\end{equation}

\noindent which is the Zinn-Justin equation. If we define the antibracket of two quantities \(A\) and \(B\) as

\begin{equation}
    \label{antibracket}
    \left(A,B\right)\coloneqq\int{\rm d}^4 x \left(\frac{\delta_R A\left[\chi,K\right]}{\delta\chi^n\left(x\right)}\frac{\delta_L B\left[\chi,K\right]}{\delta K_n\left(x\right)}-\frac{\delta_R A\left[\chi,K\right]}{\delta K_n\left(x\right)}\frac{\delta_L B\left[\chi,K\right]}{\delta \chi^n\left(x\right)}\right),
\end{equation}

\noindent then the Zinn-Justin equation can also be written as

\begin{equation}
    \label{Zinn_Justin_Antibracket}
    \left(\Gamma,\Gamma\right)=0.
\end{equation}

BRST invariance at the quantum level implies that Eq.~\eqref{Zinn_Justin_Antibracket} is satisfied. From the Zinn-Justin equation it can be shown that, if the regularization procedure used preserves the BRST invariance, then all the singular terms that can appear in the quantum effective action satisfy the (renormalized version of the) BRST symmetry. Consequently, they already appeared in the original action Eq.~\eqref{Action_BRST} and can be eliminated via a redefinition of the
coefficients and fields, including when non-power counting renormalizable operators are present. This is equivalent to saying that the theory is renormalizable. Indeed,
as we will later discuss, in dimensional regularization BRST invariance at the tree-level can be ensured for non-chiral gauge theories like QED $\times$ QCD,
so the terms that are proportional to \(\varepsilon^{-n}\), with \(\varepsilon\coloneqq 4-D\) and \(D\) the new dimension of the spacetime, can be reabsorbed by symmetry invariant counterterms. The problem arises with chiral theories.

\subsection{Dimensional regularization in the BMHV scheme}\label{sec:BMHV}

In dimensional regularization, one performs calculations in a spacetime of dimension \(D\) and subsequently makes the analytic continuation to  \(D\to4\); see e.g. Ref.~\cite{Collins:1984xc} for an introduction. This requires defining elements of the Clifford algebra with infinite dimensionality. Therefore, we need to extend the field theory to an arbitrary dimension, which means that we will have to give a prescription for the properties of the Lorentz covariants in this case. The problem is that some objects like \(\varepsilon_{\mu\nu\rho\sigma}\) or \(\gamma_5\coloneqq i\gamma^0\gamma^1\gamma^2\gamma^3\)
are intrinsically four-dimensional. Some important properties of these objects rely on this fact. For instance, we have the contraction of two Levi-Civita tensors in \(D=4\):

\begin{equation}
    \label{Levi_Civita_Dim4}
    \varepsilon_{\mu_1\dots\mu_4}\varepsilon_{\nu_1\dots\nu_4}=-\sum_{\pi\in S_4}\mathrm{sign}\left(\pi\right)\prod_{i=1}^4\eta_{\mu_i\nu_{\pi\left(i\right)}},
\end{equation}

\noindent where \(S_n\) is the group of permutations of \(n\) objects, \(\mathrm{sign}\left(\pi\right)\) is the signature of the permutation and we are taking the prescription \(\varepsilon^{0123}=1\).
This property is the one that enables us to reduce products of Levi-Civita tensors
and relies on the fact that every possible value of a Lorentz index corresponds to a value of the indices of \(\varepsilon_{\mu\nu\rho\sigma}\) if it is non-vanishing in \(D=4\).
Furthermore, if we assume that the relation \(\left\{\gamma_{\mu},\gamma_5\right\}=0\) is maintained in an arbitrary dimension, then we can prove that:

\begin{equation}
    \label{Trace_4gamma_gamma5}
    \left(D-2\right)\left(D-4\right)\mathrm{tr}\left\{\gamma_{\mu}\gamma_{\nu}\gamma_{\rho}\gamma_{\sigma}\gamma_5\right\}=0.
\end{equation}
Consequently, there is no continuous limit matching
the value of that trace in dimension \(D\) with its value in \(D=4\): \(\mathrm{tr}\left\{\gamma_{\mu}\gamma_{\nu}\gamma_{\rho}\gamma_{\sigma}\gamma_5\right\}=-4i\varepsilon_{\mu\nu\rho\sigma}\).
Then, if we want to have an algebraically consistent scheme in dimensional regularization for \(\gamma_5\) we need \(\left\{\gamma_{\mu},\gamma_5\right\}\neq0\). The only known example of such a scheme is the Breitenlohner-Maison-’t Hooft-Veltman (BMHV) scheme. It is based on the definition of \(4\)- and \(\left(D-4\right)\)-dimensional objects. We have

\begin{equation}
    \label{BMHV_Metrics}
    D-\textrm{dim:}\hspace{0.2cm} \eta_{\mu\nu}, \hspace{1cm} 4-\textrm{dim:}\hspace{0.2cm} \bar{\eta}_{\mu\nu}, \hspace{1cm} \left(D-4\right)-\textrm{dim:}\hspace{0.2cm} \hat{\eta}_{\mu\nu}, 
\end{equation}

\noindent where we define

\begin{equation}
    \label{Metric_Projectors}
    \hat{\eta}_{\mu\nu}\coloneqq\left\{
    \begin{array}{ll}
       \eta_{\mu\nu}, &\textrm{if} \,\,\mu\geq4\,\, \textrm{or}\,\, \nu\geq 4  \\
        0,  &  \textrm{otherwise}
    \end{array}\right., \hspace{2cm} \bar{\eta}_{\mu\nu}\coloneqq\eta_{\mu\nu}-\hat{\eta}_{\mu\nu}.
\end{equation}
These matrices behave as projectors into the \(\left(D-4\right)\)-dimensional and the \(4\)-dimensional spaces, respectively. Therefore, they satisfy properties like:

\begin{equation}
    \label{Projectors}
    \begin{array}{lll}
        \eta^{\mu\alpha}\hat{\eta}_{\alpha\nu}=\hat{\eta}^{\mu\alpha}\hat{\eta}_{\alpha\nu}=\hat{\eta}^{\mu}_{\nu}, &  \hspace{1cm}  \eta^{\mu\alpha}\bar{\eta}_{\alpha\nu}=\bar{\eta}^{\mu\alpha}\bar{\eta}_{\alpha\nu}=\bar{\eta}^{\mu}_{\nu}, &  \hspace{1cm} \hat{\eta}^{\mu\alpha}\bar{\eta}_{\alpha\nu}=0,  \\ & & \\
         \eta^{\mu}_{\mu}=D, & \hspace{1cm} \hat{\eta}^{\mu}_{\mu}=D-4,  & \hspace{1cm} \bar{\eta}^{\mu}_{\mu}=4.
    \end{array} 
\end{equation}

\noindent We can also define the projections of vectors and Dirac matrices:

\begin{equation}
    \label{Projections_Vectors_Matrices}
    \begin{array}{lll}
        p^{\mu}\coloneqq\eta^{\mu}_{\nu}\,p^{\nu}, &  \hspace{1cm}  \hat{p}^{\mu}\coloneqq\hat{\eta}^{\mu}_{\nu}\,p^{\nu}, &  \hspace{1cm} \bar{p}^{\mu}\coloneqq\bar{\eta}^{\mu}_{\nu}\,p^{\nu},  \\ & & \\
         \gamma^{\mu}\coloneqq\eta^{\mu}_{\nu}\,\gamma^{\nu}, &  \hspace{1cm}  \hat{\gamma}^{\mu}\coloneqq\hat{\eta}^{\mu}_{\nu}\,\gamma^{\nu}, &  \hspace{1cm} \bar{\gamma}^{\mu}\coloneqq\bar{\eta}^{\mu}_{\nu}\,\gamma^{\nu}. 
    \end{array} 
\end{equation}

\noindent We still impose the properties

\begin{equation}
    \label{Commutator_Trace}
    \left\{\gamma_{\mu},\gamma_{\nu}\right\}\coloneqq2\eta_{\mu\nu}\mathbb{I}, \hspace{1cm} \textrm{tr}\left\{\mathbb{I}\right\}\coloneqq4.
\end{equation}

\noindent With these definitions, we can prove that

\begin{equation}
    \label{Trace_Gamma}
    \textrm{tr}\left\{\gamma^{\mu}\right\}=0\,\,\textrm{if}\,\,D\neq1,
\end{equation}

\begin{equation}
    \label{Projectors-Commutators}
    \begin{array}{ll}
        \eta^{\mu}_{\nu}\,\hat{\gamma}^{\nu}=\hat{\gamma}^{\mu}, &  \hspace{1cm}  \eta^{\mu}_{\nu}\,\bar{\gamma}^{\nu}=\bar{\gamma}^{\mu}, \\ & \\  \hat{\eta}^{\mu}_{\nu}\,\hat{\gamma}^{\nu}=\hat{\gamma}^{\mu}, & \hspace{1cm}  \bar{\eta}^{\mu}_{\nu}\,\bar{\gamma}^{\nu}=\bar{\gamma}^{\mu}, \\ & \\
         \hat{\eta}^{\mu}_{\nu}\,\bar{\gamma}^{\nu}=0,  & \hspace{1cm} \bar{\eta}^{\mu}_{\nu}\,\hat{\gamma}^{\nu}=0, 
    \end{array} \hspace{2cm}
    \begin{array}{ll}
         \left\{\gamma_{\mu},\hat{\gamma}_{\nu}\right\}\coloneqq2\hat{\eta}_{\mu\nu}\mathbb{I}, &  \hspace{0.5cm}   \left\{\hat{\gamma}_{\mu},\hat{\gamma}_{\nu}\right\}\coloneqq2\hat{\eta}_{\mu\nu}\mathbb{I}, \\ & \\  \left\{\gamma_{\mu},\bar{\gamma}_{\nu}\right\}\coloneqq2\bar{\eta}_{\mu\nu}\mathbb{I}, &  \hspace{0.5cm}   \left\{\bar{\gamma}_{\mu},\bar{\gamma}_{\nu}\right\}\coloneqq2\bar{\eta}_{\mu\nu}\mathbb{I}, \\ & \\
          \left\{\hat{\gamma}_{\mu},\bar{\gamma}_{\nu}\right\}=0.  &  
    \end{array} 
\end{equation}

Now, we define the Levi-Civita tensor. It will be kept  as a four-dimensional--like object in order to maintain all of its properties:

\begin{eqnarray}
    \label{Levi_Civita_Definition}
    \varepsilon^{\mu\nu\rho\sigma}\coloneqq\left\{
    \begin{array}{rl}
        1 & \textrm{if} \,\, \left(\mu\nu\rho\sigma\right) \,\,\textrm{is an even permutation of}\,\, \left(0123\right)  \\
        -1 & \textrm{if} \,\, \left(\mu\nu\rho\sigma\right) \,\,\textrm{is an odd permutation of}\,\, \left(0123\right)  \\
        0 & \textrm{otherwise}
    \end{array}\right..
\end{eqnarray}

\noindent With this definition, we can prove that

\begin{equation}
    \label{Levi_Civita_Properties}
    \begin{array}{ll}
        \varepsilon_{\mu\nu\rho\sigma}\,\hat{\eta}^{\sigma}_{\tau}=0, & \hspace{1cm} \varepsilon_{\mu_1\dots\mu_4}=\textrm{sign}\left(\pi\right)\varepsilon_{\mu_{\pi\left(1\right)}\dots\mu_{\pi\left(4\right)}},   \\ & \\
        \varepsilon_{\mu\nu\rho\sigma}\,\bar{\eta}^{\sigma}_{\tau}=\varepsilon_{\mu\nu\rho\tau}, & \hspace{1cm} \varepsilon_{\mu_1\dots\mu_4}\varepsilon_{\nu_1\dots\nu_4}=-\sum\limits_{\pi\in S_4}\mathrm{sign}\left(\pi\right)\prod\limits_{i=1}^4\bar{\eta}_{\mu_i\nu_{\pi\left(i\right)}}, \\ & \\
        \varepsilon_{\mu\nu\rho\sigma}\,\eta^{\sigma}_{\tau}=\varepsilon_{\mu\nu\rho\tau}. &
    \end{array}
\end{equation}

Finally, we do the same with the matrix \(\gamma_5\) and give a four-dimensional--like definition for it:

\begin{equation}
    \label{gamma_5_definition}
    \gamma_5\coloneqq i\,\gamma^0\gamma^1\gamma^2\gamma^3=-\frac{i}{4!}\varepsilon_{\mu\nu\rho\sigma}\gamma^{\mu}\gamma^{\nu}\gamma^{\rho}\gamma^{\sigma}.
\end{equation}

\noindent Then, it satisfies the properties

\begin{equation}
    \label{Gamma_5_Properties}
    \begin{array}{ll}
        \textrm{tr}\left\{\gamma_5\right\}=0, & \hspace{1cm} \gamma_5^2=\mathbb{I},   \\ & \\
        \left\{\gamma_5,\bar{\gamma}^{\mu}\right\}=0, & \hspace{1cm} \left\{\gamma_5,\gamma^{\mu}\right\}=\left\{\gamma_5,\hat{\gamma}^{\mu}\right\}=2\gamma_5\hat{\gamma}^{\mu}, \\ & \\
        \left[\gamma_5,\hat{\gamma}^{\mu}\right]=0, & \hspace{1cm} \left[\gamma_5,\gamma^{\mu}\right]=\left[\gamma_5,\bar{\gamma}^{\mu}\right]=2\gamma_5\bar{\gamma}^{\mu}.
    \end{array}
\end{equation}

As we can see, in the BMHV scheme \(\gamma_5\) does not anticommute with all of the \(\gamma^{\mu}\) anymore.
This scheme breaks Lorentz invariance in $D$ dimensions \cite{Collins:1984xc},
but the 4 and $D-4$ dimensional sectors separately respect a reduced set of symmetries.
This is an algebraically consistent scheme: we can univocally calculate whatever expression with Lorentz covariants and chiral objects and the result will have a continuous limit \(D\to 4\). In other words, it will lead to an unambigous result, not requiring any further prescription. Nonetheless, we will show now that having a non-anticommuting \(\gamma_5\) induces a BRST breaking of the regularized Lagrangian already at the classical level and, as a consequence, we will need symmetry-breaking counterterms in order to renormalize the theory. The quantum effective action will in this way satisfy basic symmetry requirements, after the suitable finite renormalization is performed.

\subsection{Chiral symmetry breaking in the BMHV scheme}\label{sec:BMHV_breaking}

The breaking of the chiral gauge symmetry due to taking a non-anticommuting scheme for \(\gamma_5\) can only be induced by a change in the chiral structure of the Lagrangian. For definiteness, let us consider the four-dimensional kinetic term of a right-handed Dirac field \(\bar{\xi}\gamma^{\mu}\partial_{\mu}P_R\xi\), with \(P_{L,R}\coloneqq\left(1\mp\gamma_5\right)/2\). In the BMHV scheme there are at least three inequivalent ways to extend this operator to \(D\) dimensions because \(\gamma^{\mu}P_R\neq P_L\gamma^{\mu}\):

\begin{equation}
    \label{Chiral_Kinetic_Terms}
    \bar{\xi}\gamma^{\mu}\partial_{\mu}P_R\xi, \hspace{1.5cm} \bar{\xi}P_L\gamma^{\mu}\partial_{\mu}\xi, \hspace{1.5cm} \bar{\xi}P_L\gamma^{\mu}\partial_{\mu}P_R\xi.
\end{equation}
The usual convention is to take the most symmetric scenario, which is \(\bar{\xi}P_L\gamma^{\mu}\partial_{\mu}P_R\xi\), see e.g. Ref.~\cite{Ferrari:1994ct}. This choice also guarantees hermiticity. Nonetheless, even in this case we will face a problem. It is easy to show that this term is a four-dimensional object, i.e., \(\bar{\xi}P_L\gamma^{\mu}\partial_{\mu}P_R\xi=\bar{\xi}P_L\bar{\gamma}^{\mu}\partial_{\bar \mu}P_R\xi\), which means that the associated propagator will be \(iP_R\bar{\slashed{p}}P_L/\bar{p}^2\) (suppressing the usual imaginary infinitesimal piece from Feynman's prescription).
This leads
to unregularized amplitudes when integrating over the full \(D\)-dimensional space of momenta. See Ref.~\cite{Jegerlehner:2000dz} for a previous discussion.

The usual solution for this kind of chiral theories is to
introduce a
fermion with opposite chirality, but imposing that the same fermion is invariant under BRST transformations
so
that there is no interaction term in the Lagrangian density with these extra fields,
see Refs.~\cite{Weinberg:1996kr,Jegerlehner:2000dz,Belusca-Maito:2020ala} (and Refs.~\cite{Kuhler:2024fak,Ebert:2024xpy} for alternative formulations). For this reason these mathematical devices will be called ``fictitious fields'' hereafter, and not included among the ``matter fields''; in the example of the previous right-handed model, it will simply be denoted $ \xi_L^f $, not to be confused with the left-handed fields $ \xi_L $ introduced later in the text:
under the full SM gauge group, $ \xi_L^f $ is a singlet under $SU(2)_L$ just like its right-handed counterpart, while the physical $ \xi_L $ is the usual doublet under gauged $SU(2)_L$. Conversely, the fictitious field $\xi_R^f$ has the necessary sterile degrees of freedom to counterpart the upper and lower degrees of freedom under global $SU(2)_L$ transformations of $\xi_L$.\footnote{In four dimensions where the evanescent symmetry-breaking term is absent, these fictitious fields can be integrated over exactly in the path integral formulation, resulting in a field-independent determinant that does not contribute to the connected parts of the $S$-matrix for the remaining field variables, see e.g. Ref.~\cite{Weinberg:1996kr}. The case in $D$ dimensions is the main topic of interest of Sec.~\ref{sec:Restoring_Symmetry}.}
In this way, with the introduction of fictitious fields we preserve global symmetry transformations, that we exploit in App.~\ref{sec:list_CTs} when enumerating possible counterterm candidates.
Gauge bosons couple to the chiral fields, and thus interactions are purely four dimensional, which specifies the gauge sector properties for the purpose of our calculations.

We can now write the kinetic term of the two fermions in the same way as we would do in a theory where they were already present from the very beginning

\begin{equation}
    \label{Kinetic_Term}
    i\bar{\xi}\slashed{\partial}\xi=i\bar{\xi}_L^f\bar{\slashed{\partial}}\xi_L^f+i\bar{\xi}_R\bar{\slashed{\partial}}\xi_R+i\bar{\xi}\hat{\slashed{\partial}}\xi,
\end{equation}

\noindent with \(\xi_{R}\coloneqq P_{R}\xi\), \(\bar{\xi}_{R}=\bar{\xi}P_{L}\), \(\xi_{L}^f\coloneqq P_{L}\xi\) and \(\bar{\xi}_{L}^f=\bar{\xi}P_{R}\). It means that we have added an evanescent operator \(i\bar{\xi}\hat{\slashed{\partial}}\xi\) to the original Lagrangian. It can also be written as \(i\bar{\xi}\hat{\slashed{\partial}}\xi=i\bar{\xi}^f_L\hat{\slashed{\partial}}\xi_R+i\bar{\xi}_R\hat{\slashed{\partial}}\xi_L^f\), i.e., mixing different chiralities. Now, let us consider a term of the form \(i\bar{\xi}\hat{\slashed{\partial}}\xi\) where the \(\xi\) have the BRST transformation indicated in Eq.~\eqref{BRST_Transformation}. It is straightforward to show that

\begin{equation}
    \label{Kinetic_BRST_Breaking}
    %\begin{split}
        \delta_{\theta}\left(i\bar{\xi}\hat{\slashed{\partial}}\xi\right)=\theta c_{\alpha} %&
        \left[\left(t_{\alpha}^R\right)_{ji}\bar{\xi}_{R,j} \overrightarrow{\hat{\slashed{\partial}}} \xi_{L,i}^f-\left(t_{\alpha}^R\right)_{ij}\bar{\xi}_{L,i}^f \overleftarrow{\hat{\slashed{\partial}}} \xi_{R,j}\right]
        %\right. \\ & \left.+\left(t_{\alpha}^L\right)_{ji}\bar{\xi}_{L,j}\hat{\slashed{\partial}}\xi_{R,i}-\left(t_{\alpha}^L\right)_{ij}\bar{\xi}_{R,i}\hat{\slashed{\partial}}\xi_{L,j}\right].
    %\end{split}
\end{equation}

\noindent in the previous example where $\xi_R$ is physical and $\xi_L^f$ fictitious; another way to write this term is

\begin{equation}
    \label{Kinetic_BRST_Breaking_1}
    \begin{split}
       s\left(i\bar{\xi}\hat{\slashed{\partial}}\xi\right)= c^{\alpha}\left[\bar{\xi}_i\left(t_{\alpha}^R\right)_{ij}\left(\overrightarrow{\hat{\slashed{\partial}}}P_L+\overleftarrow{\hat{\slashed{\partial}}}P_R\right)
       %+\left(t_{\alpha}^L\right)_{ij}\left(\overrightarrow{\hat{\slashed{\partial}}}\mathcal{P}_R+\overleftarrow{\hat{\slashed{\partial}}}\mathcal{P}_L\right)
       \xi_j\right].
    \end{split}
\end{equation}

Therefore, the evanescent operator that we have added to the regularized Lagrangian is not BRST invariant because it contains a chiral structure of the form \(\bar{\xi}_{L,R}\hat{\gamma}_{\mu}\xi_{R,L}\).
We stress that the addition of the new chiral structure is needed because the appearance of a four-dimensional momentum in the denominator of the propagator leads
to unregularized amplitudes. There is no problem if the four-dimensional momentum appears in the numerator of the integrand while keeping the \(D\)-dimensional analogue in the denominator (i.e., \(i\bar{\slashed{p}}/p^2\)), because this does not produce undefined results.
In the same way,
we can preserve the chiral structure of the interaction terms (whose effects also appear in the numerators of the integrands of Feynman integrals), so they
are not responsible for producing extra
symmetry-breaking effects.

As it will become clear later in the text, due to the chiral nature of the full SM gauge symmetry QCD interactions can also lead to (spurious) symmetry-breaking effects. This is so because physical fields of different chiralities have distinct quantum numbers.
It is worth mentioning that for a non-chiral theory, where fermion masses are consistent with the local symmetries, as in the case of the low-energy EFT invariant under QED $\times$ QCD, a different
scheme employs only the left- and right-handed physical fields of the model, thus not introducing any fictitious fields \cite{Cornella:2022hkc,Naterop:2023dek}.
In this case of a non-chiral theory, there is no symmetry breaking at the tree level, because $s ( i\bar{\xi} \hat{\slashed{\partial}} \xi ) = 0$ (compare to Eq.~\eqref{Kinetic_BRST_Breaking_1} above, where fictitious fields are introduced), which is an advantage of this approach.
Obviously, in the case of a vector theory like QED $\times$ QCD there is no breaking of their corresponding global symmetry transformations neither.
However, this scheme leads to the breaking of both global and local symmetries in the case of chiral theories, that must be then cured with the addition of local counterterms.\footnote{Ref.~\cite{Naterop:2023dek} is interested in global chiral symmetries,
which are broken by fermion mass terms and non-renormalizable interactions.}

The addition of symmetry-breaking evanescent terms in the Lagrangian has no important consequences at the tree level because all of the diagrams in which insertions of these operators are present are going to vanish in the limit \(D\to4\). Nonetheless, we face some difficulties at the loop level because we find three kinds of symmetry-breaking contributions to the quantum effective action \(\Gamma\) when calculating amplitudes at one loop: singular evanescent terms of the form \(E/\varepsilon\) ($E$ being an evanescent operator),
evanescent terms proportional to \(\varepsilon\) (i.e., $O \times \varepsilon$, where $O$ is any operator), and terms for which the symmetry-breaking evanescent operator ``hits'' the divergence, of the form \(\varepsilon/\varepsilon\) (i.e., $O \times 1$, where $O$ is not evanescent). We will have to add counterterms in such a way that, in the end, all of the symmetry breaking disappears when \(D\to4\), thus enforcing the quantum effective action
is BRST invariant:

\begin{equation}
    \label{4-dimensional_Limit}
    \lim_{D\to4}\mathcal{S}\left(\Gamma+I_{\textrm{sct}}+I_{\textrm{fct}}\right)=0,
\end{equation}

\noindent where \(\mathcal{S}\) is the Slavnov operator \(\mathcal{S}\left(\Gamma\right)\coloneqq\left(\Gamma,\Gamma\right)\),
which flags symmetry breaking when it acquires non-vanishing values, \(I_{\textrm{sct}}\) is the action for the singular counterterms that we have to add to cancel in particular the symmetry breaking, and \(I_{\textrm{fct}}\) the action for the finite ones which is the main focus of this work. We need to see how to identify these two contributions.

\subsection{Quantum effective action in presence of symmetry-breaking evanescent terms}\label{sec:Restoring_Symmetry}

Following a similar presentation as in Sec.~\ref{sec:Usual_Renormalization}, we give a heuristic proof of which diagrams contribute to the spurious symmetry-breaking of the quantum effective action in \(D\)-dimensions. Let us consider again the transformation \(\chi^n\left(x\right)\rightarrow\chi^n\left(x\right)+\widetilde\varepsilon F^n\left[x;\chi\right]\) of Eq.~\eqref{Field_Transformation}. It produces a change in the action due to the addition of the symmetry-breaking evanescent operators

\begin{equation}
    \label{Action_Transformation}
    I\left[\chi\right]\rightarrow I\left[\chi+\widetilde\varepsilon F\right]=I\left[\chi\right]+\widetilde\varepsilon\int{\rm d}^D x G\left[x;\chi\right].
\end{equation}

\noindent
For the case of the gauge symmetry-breaking that appears in the BMHV scheme, \(G[x;\chi]=s(\mathcal{L})=s(i\bar{\xi}\hat{\slashed{\partial}}\xi)\) at tree level.
In the same way that we added the sources \(K_n\left(x\right)\) in order to study the transformation of the fields \(\chi_n\left(x\right)\), we introduce now the classical source \(\mathcal{K}\left(x\right)\) in order to study the transformation of the action:

\begin{equation}
    \label{Complete_Path_Integral}
    \begin{split}
        Z\left[J,K,\mathcal{K}\right]\coloneqq e^{iW\left[J,K,\mathcal{K}\right]}=\int\mathcal{D}\chi\textrm{exp}&\left\{iI\left[\chi\right]+ i\int{\rm d}^DxF^n\left(x\right)K_n\left(x\right)\right.\\&\left.+i\int{\rm d}^Dx\chi^n\left(x\right)J_{n}\left(x\right)+i\int{\rm d}^DxG\left(x\right)\mathcal{K}\left(x\right)\right\}.
    \end{split}
\end{equation}

\noindent Again, we define the classical fields in presence of these sources as 

\begin{equation}
    \label{Classical_Field_Symmetry_Breaking}
    X_{J,K,\mathcal{K}}^n\left(x\right)\coloneqq\frac{\delta_R}{\delta J_n\left(x\right)}W\left[J,K,\mathcal{K}\right]
\end{equation}

\noindent and \(\left(J_{X,K,\mathcal{K}}\right)_n\left(x\right)\) is the value of \(J_n\left(x\right)\) for which the classical field \( X_{J,K,\mathcal{K}}^n\left(x\right)\) takes the particular value \(X_n\left(x\right)\). Then, the quantum effective action in presence of \(\mathcal{K}\left(x\right)\) is defined as the Legendre transformation

\begin{equation}
    \label{QEA_Symmetry_Breaking}
    \Gamma\left[X,K,\mathcal{K}\right]\coloneqq-\int{\rm d}^DxX^n\left(x\right)\left(J_{X,K,\mathcal{K}}\right)_n\left(x\right)+W\left[J_{X,K,\mathcal{K}},K,\mathcal{K}\right].
\end{equation}

\noindent It is easy to show that

\begin{equation}
    \label{Derivatives_QEA_1}
    \frac{\delta_L\Gamma\left[X,K,\mathcal{K}\right]}{\delta X^n\left(x\right)}= -\left(J_{X,K,\mathcal{K}}\right)_n\left(x\right),
\end{equation}

\begin{equation}
    \label{Derivatives_QEA_2}
    \frac{\delta_R\Gamma\left[X,K,\mathcal{K}\right]}{\delta K_n\left(x\right)}= \left<F^n\left(x\right)\right>_{J_{X,K,\mathcal{K}},K,\mathcal{K}},
\end{equation}

\begin{equation}
    \label{Derivatives_QEA_3}
    \frac{\delta_R\Gamma\left[X,K,\mathcal{K}\right]}{\delta \mathcal{K}\left(x\right)}= \left<G\left(x\right)\right>_{J_{X,K,\mathcal{K}},K,\mathcal{K}},
\end{equation}

\noindent where the vacuum expectation value of a collection of fields $H$ is given by

\begin{equation}
    \label{Vacuum_Expectation_Value_Symmetry_Breaking}
    \begin{split}
        \left<H\left(y\right)\right>_{J,K,\mathcal{K}}\coloneqq Z^{-1}\left[J,K,\mathcal{K}\right]
        \int\mathcal{D}\chi& H\left[y;\chi\right]\textrm{exp}\left\{iI\left[\chi\right]+ i\int{\rm d}^DxF^n\left(x\right)K_n\left(x\right)\right.\\&\left.+i\int{\rm d}^Dx\chi^n\left(x\right)J_{n}\left(x\right)+i\int{\rm d}^DxG\left(x\right)\mathcal{K}\left(x\right)\right\}.
    \end{split}
\end{equation}

Now, we make the change of variables \(\chi^n\left(x\right)\rightarrow\chi^n\left(x\right)+\widetilde\varepsilon F^n\left[x;\chi\right]\) in the path integral of Eq.~\eqref{Complete_Path_Integral}.
We are assuming that the measure \(\mathcal{D}\chi\) is invariant;\footnote{Going beyond this heuristic presentation, the Quantum Action Principle, valid in particular for dimensional regularization \cite{Breitenlohner:1975hg,Breitenlohner:1976te,Breitenlohner:1977hr}, states which Green's functions must be calculated in order to study the effect of the symmetry transformation that is broken by the regularization procedure; see also e.g. Refs.~\cite{Lowenstein:1971jk,Lam:1972mb,Lam:1973qa,Piguet:1980nr}. This is the main expression we aim at obtaining in the present section. Importantly, this principle also applies to the case of true, unremovable anomalies, while in the heuristic proof they are assumed to be absent.}
the parameter \(\widetilde\varepsilon\) is infinitesimal and the BRST transformation is nilpotent, so \(F^{n}\left[x;\chi+\widetilde\varepsilon F\right]=F^n\left[x;\chi\right]\) and \(G\left[x;\chi+\widetilde\varepsilon F\right]=G\left[x;\chi\right]\). Therefore, after the change of variables we arrive at

\begin{equation}
    \label{Change_of_Variables}
    \begin{split}
        &Z\left[J,K,\mathcal{K}\right]=Z\left[J,K,\mathcal{K}\right]+i\widetilde\varepsilon\int\mathcal{D}\chi\textrm{exp}\left\{iI\left[\chi\right]+ i\int{\rm d}^DxF^n\left(x\right)K_n\left(x\right)\right.\\&\quad\left.+i\int{\rm d}^Dx\chi^n\left(x\right)J_{n}\left(x\right)+i\int{\rm d}^DxG\left(x\right)\mathcal{K}\left(x\right)\right\}\int{\rm d}^Dy\left\{G\left[y;\chi\right]+F^n\left[y;\chi\right]J_n\left(y\right)\right\}.
    \end{split}
\end{equation}

\noindent Consequently, it is clear that 

\begin{equation}
    \label{Slavnov_Taylor_Expectation_Values}
    \int{\rm d}^Dx\left<F^n\left(x\right)\right>_{J,K,\mathcal{K}}J_n\left(x\right)=-\int{\rm d}^Dx\left<G\left(x\right)\right>_{J,K,\mathcal{K}}.
\end{equation}

\noindent Using Eqs.~\eqref{Derivatives_QEA_1}-\eqref{Derivatives_QEA_3} we can write Eq.~\eqref{Slavnov_Taylor_Expectation_Values} as

\begin{equation}
    \label{Slavnov_Taylor_Symmetry_Breaking}
    \int{\rm d}^Dx \frac{\delta_R\Gamma\left[X,K,\mathcal{K}\right]}{\delta K_n\left(x\right)}\frac{\delta_L\Gamma\left[X,K,\mathcal{K}\right]}{\delta X^n\left(x\right)}=\int{\rm d}^Dx\frac{\delta_R\Gamma\left[X,K,\mathcal{K}\right]}{\delta \mathcal{K}\left(x\right)}.
\end{equation}

If we take the limit \(\mathcal{K}\left(x\right)\to 0\) at the left-hand side we find the usual \(D\)-dimensional Slavnov operator up to a factor of \(-1/2\), so

\begin{equation}
    \label{AntiBracket_Symmetry_Breaking}
    \mathcal{S}\left(\Gamma\right)=\left(\Gamma,\Gamma\right)=-2\lim_{\mathcal{K}\left(x\right)\to0}\int{\rm d}^Dx\frac{\delta_R\Gamma\left[X,K,\mathcal{K}\right]}{\delta \mathcal{K}\left(x\right)},
\end{equation}
where now the Slavnov operator is defined from spacetime integration in $D$ dimensions.
Considering that the classical source \(\mathcal{K}\left(x\right)\) appears in Eq.~\eqref{Complete_Path_Integral}
as \(\int{\rm d}^DxG\left(x\right)\mathcal{K}\left(x\right)\), it acts as a coupling among the fields encoded in the operator \(G\left(x\right)\) (which is nothing but \(s\left(\mathcal{L}\right)\)). The only surviving terms in the limit \(\mathcal{K}\left(x\right)\to0\) are the ones where \(G\left(x\right)\) only appears once. Consequently, taking into account Eq.~\eqref{Calculation_QEA}, the Feynman diagrams that contribute to the symmetry breaking of the chiral theory in \(D\) dimensions as expressed by the Slavnov operator above are
1PI diagrams
with single insertions of the operator \(s\left(\mathcal{L}\right)\), which carry ghost number $1$. Physical amplitudes do not carry ghosts as external fields: this however serves as a convenient device to compute the symmetry-restoring finite counterterms.

When calculating such diagrams, there are divergent parts for which we have to find appropriate counterterms (at one loop, of the form evanescent operator over \(\varepsilon\); their discussion is beyond the scope of the present paper).
There are also contributions that are proportional to \(\varepsilon\) which are completely irrelevant since they vanish in the limit \(D\to4\) when defining the quantum effective action. The problematic parts are the ones proportional to \(\varepsilon/\varepsilon\), which do not cancel in the limit \(\varepsilon\to0\) and are not eliminated by the usual renormalization program based on minimal subtraction (or its often-used modified version). We have then to find finite counterterms to cancel them.  Of course, one option is to calculate all of the diagrams with single insertions
of \(s\left(\mathcal{L}\right)\)
and find the counterterms that eliminate symmetry-breaking contributions to the Slavnov-Taylor identities, but this is inefficient. Instead, we consider the method by Bonneau \cite{Bonneau:1979jx,Bonneau:1980zp}, which allows us to calculate the \(\varepsilon/\varepsilon\) parts from the computation of the divergent parts of a related set of diagrams (namely, the ones with insertions of the $\check{\Delta}$ operator introduced in the following section).

We now exploit the antibracket formalism to render the previous discussion clearer.
At one-loop order, we expand $\Gamma$ as $I + \Gamma_1$, replacing $ (\Gamma, \Gamma) $ by the expression $ 2 (I^{(4)}, \Gamma_1^{(4)}) $, since $ (I^{(4)}, I^{(4)}) = 0 $, when taking the limit in four dimensions (and when neglecting the $ \mathcal{O} (\hbar^2)$ term $ (\Gamma_1, \Gamma_1) $, once $\hbar$ is reintroduced everywhere). This means that when the right-hand side of Eq.~\eqref{AntiBracket_Symmetry_Breaking} can be expressed as $ 2 (I^{(4)}, F_1) $ where $F_1$ is a local functional,
one can perform the finite renormalization $ \Gamma_1^{(4)} \to \Gamma_1^{(4)}-F_1 $ such that $(I^{(4)}, \Gamma_1^{(4)}-F_1) = 0$, thus restoring the Zinn-Justin equation. Using the same antibracket formalism, we also arrive at the following formulation of the Wess-Zumino consistency condition: $(I^{(4)}, (I^{(4)}, \Gamma_1^{(4)}))=-(\Gamma_1^{(4)}, (I^{(4)}, I^{(4)})) / 2=0$, which must be respected whether in presence of true anomalies or not (i.e., whether a local functional $F_1$ can be introduced or not).

A final comment is in order regarding Batalin-Vilkovisky antifields.
When a functional $F$ is built exclusively from the available fields of the theory, $ (I^{(4)}, F) = - s ( \chi^n ) \, \delta_L F / \delta \chi^n = - s ( F ) $. We will find below that this simpler formulation is enough in order to identify the finite counterterms needed to reestablish the Slavnov-Taylor identities. In general, however, a dependence on antifields could also be expected \cite{Weinberg:1996kr}.

\subsection{Bonneau method}\label{sec:Restoring_Symmetry_Bonneau}

We illustrate how the Bonneau method \cite{Bonneau:1979jx,Bonneau:1980zp} works particularizing for the one-loop case.
The BRST breaking contributions in the BMHV scheme
come from the fact that the commutator of \(\gamma_5\) with \(\gamma^{\mu}\) is not zero: we have instead \(\left\{\gamma_5,\gamma_{\mu}\right\}=2\hat{\eta}_{\mu\nu}\gamma_5\gamma^{\nu}\). Thus, generally speaking, the BRST-breaking operators \(\Delta\) can be written as \(\Delta=-\hat{\eta}_{\mu\nu}\Delta^{\mu\nu}\) where \(\Delta^{\mu\nu}\) does not vanish in the limit \(\varepsilon\to0\). At tree level this \(\Delta^{\mu\nu}\) operator is the Fourier transform of \(-c^{\alpha}[\bar{\xi}_i\left(t_{\alpha}^R\right)_{ij}\gamma^{\mu}(\overrightarrow{\partial^{\nu}}P_L+\overleftarrow{\partial^{\nu}}P_R)\xi_j]\) (when the physical degree of freedom is right-handed; see Eq.~\eqref{Kinetic_BRST_Breaking_1}). Then, let us consider a one-loop graph \(G\) with one special vertex
\(V\)
associated with the operator \(\Delta^{\mu\nu}\). The regularized amplitude \(I_{\varepsilon}^{G^{\Delta_{\mu\nu}}}\) can be written as

\begin{equation}
    \label{Amplitude_Tensor}
   ( I_{\varepsilon}^{G^{\Delta_{\mu\nu}}})_{\mu\nu}=M_{\mu\nu}^{G^{\Delta_{\mu\nu}}}+\tilde{M}_{\mu\nu}^{G^{\Delta_{\mu\nu}}},
\end{equation}

\noindent  where \(M_{\mu\nu}^{G^{\Delta_{\mu\nu}}}\) is the part of the amplitude that does not produce an \(\varepsilon\) term when contracted with \(-\hat{\eta}^{\mu\nu}\) and \(\tilde{M}_{\mu\nu}^{G^{\Delta_{\mu\nu}}}\) the part that does. In the first case, the Lorentz indices can be associated with products of external momenta \(p_{\mu}^{i}p_{\nu}^{j}\),
where \(i\) and \(j\) represent the different external lines of the Feynman diagrams; or with products of one external momentum and a \(\gamma\)-matrix. On the other hand, the second term can contain the metric tensor \(\eta_{\mu\nu}\) or the product of two \(\gamma\)-matrices, since \(-\hat{\eta}^{\mu\nu}\gamma_{\mu}\gamma_{\nu}=-\hat{\eta}^{\mu\nu}\eta_{\mu\nu}=4-D=\varepsilon\).
Now, in order to obtain the amplitude with one single insertion of the BRST-breaking operator \(\Delta\), we contract Eq.~\eqref{Amplitude_Tensor} with \(-\hat{\eta}^{\mu\nu}\):

\begin{equation}
    \label{Amplitud_1Insertion}
    I_{\varepsilon}^{G^{\Delta}}=-\hat{\eta}^{\mu\nu}( I_{\varepsilon}^{G^{\Delta_{\mu\nu}}})_{\mu\nu}=-M^{G^{\Delta_{\mu\nu}}}+\varepsilon \tilde{M}^{G^{\Delta_{\mu\nu}}},
\end{equation}

\noindent where we have defined \(M^{G^{\Delta_{\mu\nu}}}\coloneqq\hat{\eta}^{\mu\nu}M_{\mu\nu}^{G^{\Delta_{\mu\nu}}}\) and \(\tilde{M}^{G^{\Delta_{\mu\nu}}}\coloneqq-\hat{\eta}^{\mu\nu}\tilde{M}_{\mu\nu}^{G^{\Delta_{\mu\nu}}}/\varepsilon\). By assumption, the contraction \(\hat{\eta}^{\mu\nu}M_{\mu\nu}^{G^{\Delta_{\mu\nu}}}\) does not add a factor \(\varepsilon\) in the numerator, so
\(\varepsilon/\varepsilon\) contributions, with the \(\varepsilon\) in the numerator resulting from the evanescent nature of the symmetry-breaking operator, only appear in the second term of Eq.~\eqref{Amplitud_1Insertion}.
(Note that we still did not derive precisely the result, which will come in the following, that \(\varepsilon/\varepsilon\) is the only concern to worry about, although qualitatively this result was already announced at the end of Sec.~\ref{sec:BMHV_breaking} and restated at the end of Sec.~\ref{sec:Restoring_Symmetry}. Also note that \(\varepsilon/\varepsilon\) contributions also happen outside the present context, e.g. when evanescent symmetry-conserving operators are considered.)

Consider the Minimal Subtraction (MS) scheme when subtracting divergences,\footnote{Since we will eventually subtract finite pieces, the scheme we employ later is not an MS(-bar) scheme, but a modified version of it instead that subtracts finite pieces beyond the usual combination $\mu^2 4 \pi e^{- \gamma_E}$.} where we add the appropriate counterterms
to cancel only the $1/\varepsilon^n$
part, $n$ being a positive integer. At the one-loop level of interest in this section, these are terms proportional to \(1/\varepsilon\). Then, we want to identify those \(1/\varepsilon\) contributions:

\begin{equation}
    \label{divergent_parts_1}
    \begin{split}
         -\textrm{p.p.}(I_{\varepsilon}^{G^{\Delta}})&=-\textrm{p.p.}\left(-\hat{\eta}^{\mu\nu}M_{\mu\nu}^{G^{\Delta_{\mu\nu}}}\right)-\textrm{p.p.}\left(\varepsilon \tilde{M}^{G^{\Delta_{\mu\nu}}}\right)\\&=\left(-\hat{\eta}^{\mu\nu}\right)\left[-\textrm{p.p.}\left(M_{\mu\nu}^{G^{\Delta_{\mu\nu}}}\right)\right]-\textrm{p.p.}\left(\varepsilon \tilde{M}^{G^{\Delta_{\mu\nu}}}\right).
    \end{split}
\end{equation}

\noindent
We have used the fact that the contraction of \(\hat{\eta}^{\mu\nu}\) with \(M_{\mu\nu}^{G^{\Delta_{\mu\nu}}}\) does not change the structure in \(1/\varepsilon\) (the divergent part) of the amplitude. Now, we will study the second term in the right-hand side of Eq.~\eqref{divergent_parts_1}. For this purpose, we use a simple theorem proved in App.~\ref{sec:Proof_Theorem}, which relates the singular part ``p.p.'' of a meromorphic function \(f\left(\varepsilon\right)\) that has a pole at \(\varepsilon=0\) with its residue at the simple pole ``r.s.p.'':

\begin{equation}
    \label{Meromorphic_Functions}
    -\textrm{p.p.}\left(\varepsilon f\left(\varepsilon\right)\right)=-\varepsilon\,\textrm{p.p.}\left(f\left(\varepsilon\right)\right)+\textrm{r.s.p.}\left(f\left(\varepsilon\right)\right).
\end{equation}

\noindent Therefore, we can write

\begin{equation}
    \label{divergent_parts_2}
        -\textrm{p.p.}\left(\varepsilon \tilde{M}^{G^{\Delta_{\mu\nu}}}\right)=-\varepsilon\,\textrm{p.p.}\left( \tilde{M}^{G^{\Delta_{\mu\nu}}}\right)+\textrm{r.s.p.}\left( \tilde{M}^{G^{\Delta_{\mu\nu}}}\right) \,.
\end{equation}

\noindent Now, we decompose \(\varepsilon\) into the original structures \(\varepsilon=-\hat{\eta}^{\mu\nu}S_{\mu\nu}\) with \(S_{\mu\nu}\) being \(\eta_{\mu\nu}=\left(\gamma_{\mu}\gamma_{\nu}+\gamma_{\nu}\gamma_{\mu}\right)/2\) in such a way that 

\begin{equation}
    \label{varepsilon_decomposition}
    -\varepsilon\,\textrm{p.p.}\left( \tilde{M}^{G^{\Delta_{\mu\nu}}}\right)=\left(-\hat{\eta}^{\mu\nu}\right)\left[-{\rm p.p.}\left(\tilde{M}_{\mu\nu}^{G^{\Delta_{\mu\nu}}}\right)\right],
\end{equation}
 
\noindent where we have introduced again the structure \(S_{\mu\nu}\) into the principal part of \(\tilde{M}^{G^{\Delta_{\mu\nu}}}\), and rearranged the terms using the symmetry of \(\hat{\eta}^{\mu\nu}\) in order to reconstruct the original amplitude \(\tilde{M}_{\mu\nu}^{G^{\Delta_{\mu\nu}}}\)
Consequently, the complete divergent part reads

\begin{equation}
    \label{divergent_parts_3}
    \begin{split}
         -\textrm{p.p.}(I_{\varepsilon}^{G^{\Delta}})&=\left(-\hat{\eta}^{\mu\nu}\right)\left[-\textrm{p.p.}\left(M_{\mu\nu}^{G^{\Delta_{\mu\nu}}}+\tilde{M}_{\mu\nu}^{G^{\Delta_{\mu\nu}}}\right)\right]+\textrm{r.s.p.}\left( \tilde{M}^{G^{\Delta_{\mu\nu}}}\right)\\&=\left(-\hat{\eta}^{\mu\nu}\right)\left[-\textrm{p.p.}(I^{G^{\Delta_{\mu\nu}}}_{\varepsilon})_{\mu\nu}\right]+\textrm{r.s.p.}\left( \tilde{M}^{G^{\Delta_{\mu\nu}}}\right).
    \end{split}
\end{equation}

Introduce now the renormalized amplitudes \(\mathcal{R}\) which, at one loop in the MS scheme, are just the regularized amplitudes after subtracting the divergent parts (we will take the limit in four dimensions shortly). Then, it is easy to see from Eq.~\eqref{divergent_parts_3} by adding $I_{\varepsilon}^{G^{\Delta}}$ on both sides that

\begin{equation}
    \label{Renormalized_Amplitudes}
    \mathcal{R}^{G^{\Delta}}_{\varepsilon}=\left(-\hat{\eta}^{\mu\nu}\right)(\mathcal{R}_{\varepsilon}^{G^{\Delta_{\mu\nu}}})_{\mu\nu}+\textrm{r.s.p.}\left( \tilde{M}^{G^{\Delta_{\mu\nu}}}\right).
\end{equation}
(Note that we are thus subtracting away the $E/\varepsilon$ part mentioned towards the end of Secs.~\ref{sec:BMHV_breaking} and \ref{sec:Restoring_Symmetry}; their detailed discussion is beyond the scope of this paper.)
Since \((\mathcal{R}_{\varepsilon}^{G^{\Delta_{\mu\nu}}})_{\mu\nu}\) is finite, $\textrm{r.s.p.}\left( \tilde{M}^{G^{\Delta_{\mu\nu}}}\right)$ is the only part of the renormalized amplitude that survives in the limit \(D\to4\):

\begin{equation}
    \label{Renormalized_Amplitudes_D=4}
    \lim_{D\to4}\mathcal{R}^{G^{\Delta}}_{\varepsilon}=\lim_{D\to4}\textrm{r.s.p.}\left( \tilde{M}^{G^{\Delta_{\mu\nu}}}\right).
\end{equation}
Therefore, (minus) \(\lim_{D\to4}\textrm{r.s.p.}\left( \tilde{M}^{G^{\Delta_{\mu\nu}}}\right)\) are the finite counterterms
that we need in order to restore the BRST symmetry of the quantum effective action (further divided by $i$, since the Feynman rules are derived from $i$ times the Lagrangian density).
It is interesting to notice that \(\varepsilon/\varepsilon\), with the \(\varepsilon\) in the numerator resulting from the fact that the symmetry-breaking operator is evanescent, is the contribution we want to calculate, as already announced at the end of Secs.~\ref{sec:BMHV_breaking} and \ref{sec:Restoring_Symmetry}: since \(\varepsilon \tilde{M}^{G^{\Delta_{\mu\nu}}}\) is where the \(\varepsilon/\varepsilon\) parts appear, then the residue at the simple pole of \(\tilde{M}^{G^{\Delta_{\mu\nu}}}\) (the part proportional to \(1/\varepsilon\)) is precisely this \(\varepsilon/\varepsilon\) contribution.

In order to calculate these finite counterterms, we introduce the operator \(\check{\Delta}\coloneqq\check{\eta}^{\mu\nu}\Delta_{\mu\nu}\), where the metric \(\check{\eta}^{\mu\nu}\) is defined by the properties

\begin{equation}
    \label{check_metric}
    \check{\eta}^{\mu\nu}=\check{\eta}^{\nu\mu},\hspace{1cm}\check{\eta}^{\mu\alpha}\eta_{\alpha\nu}=\check{\eta}^{\mu\alpha}\hat{\eta}_{\alpha\nu}=\check{\eta}^{\mu}_{\nu},\hspace{1cm} \check{\eta}^{\mu}_{\mu}=\check{\eta}^{\mu\nu}\gamma_{\mu}\gamma_{\nu}=1,
\end{equation}

\noindent so it behaves like \(\check{\eta}^{\mu\nu}=-\hat{\eta}^{\mu\nu}/\varepsilon\). Then, calculate the one-loop amplitude with a single insertion of this operator

\begin{equation}
    \label{sigle_insertion_check_operator}
    I_{\varepsilon}^{G^{\check{\Delta}}}=\check{\eta}^{\mu\nu}( I_{\varepsilon}^{G^{\Delta_{\mu\nu}}})_{\mu\nu}=\check{\eta}^{\mu\nu}M_{\mu\nu}^{G^{\Delta_{\mu\nu}}}+\tilde{M}^{G^{\Delta_{\mu\nu}}}.
\end{equation}

\noindent Therefore, it is clear that

\begin{equation}
    \label{check_amplitude}
    \left.I_{\varepsilon}^{G^{\check{\Delta}}}\right|_{\check{\eta}=0}=\tilde{M}^{G^{\Delta_{\mu\nu}}}
\end{equation}

\noindent and we can write

\begin{equation}
    \label{Calculation_Renormalized_Amplitudes_D=4}
    \lim_{D\to4}\mathcal{R}^{G^{\Delta}}_{\varepsilon}=\lim_{D\to4}\textrm{r.s.p.}\left( \tilde{M}^{G^{\Delta_{\mu\nu}}}\right)=\lim_{D\to4}\textrm{r.s.p.}\left(\left.I_{\varepsilon}^{G^{\check{\Delta}}}\right|_{\check{\eta}=0}\right).
\end{equation}
Thus, in order to find the \(\varepsilon/\varepsilon\) finite counterterms we only need to calculate the 1PI amplitudes with one insertion of the operator \(\check{\Delta}\),
set
\(\check{\eta}=0\) and find its divergent parts in the limit \(D\to4\).
Since these terms, together with possible true anomalies, are calculated from $1/\varepsilon$ terms, they are forcefully local following the proof of Weinberg's theorem \cite{Weinberg:1959nj}.

At this point, we mention an
ingenious method discussed recently in Ref.~\cite{OlgosoRuiz:2024dzq}, where a spurion field is introduced to establish
invariance in $D$-dimensions (in short, the evanescent kinetic term in Eq.~\eqref{Kinetic_Term} gives raise to a symmetric term linear in the external spurion field).
Counterterms are found more directly therein (and shown to be quadratic, quartic, etc., in the external spurion field or the evanescent symmetry-breaking term, contrarily to the discussion presented in Sec.~\ref{sec:Restoring_Symmetry}), without the need to identify the operator whose BRST transformation leads to the result that one obtains from calculating Feynman diagrams carrying insertions of the symmetry-breaking vertex (in our case, the latter have an external ghost; instead, the analogous Feynman diagrams in the other method have background spurion fields attached).
Thus, this method apparently circumvents the need for discussing the cohomology in ghost number $1$.
Ref.~\cite{OlgosoRuiz:2024dzq} does not discuss non-renormalizable interactions. In contrast to the latter reference, we introduce the fictitious fields discussed above, with the advantage of preserving global symmetry transformations from the beginning.
(We also do not consider the background field gauge \cite{Abbott:1981ke}, which is not central to their spurion method.)
This method has been recently applied to SMEFT in Ref.~\cite{Fuentes-Martin:2025meq}; further comments follow in Sec.~\ref{sec:final_discussion}.

\subsection{Renormalizable interactions and dimension-6 operator basis}\label{sec:operators}

We consider the SM Lagrangian $ \mathcal{L}_{\rm SM} $, added by right-handed neutrinos to enlarge the scope of our analysis, and dimension-6 operators, $ \mathcal{L} = \mathcal{L}_{{\rm SM} + \nu} + \sum_i C_i Q_i $, where\footnote{One employs here the convention $ D_\mu = \partial_\mu - i g_3 T^A G^A_\mu - i g_2 T^I_L W^I_\mu - i g_1 \text{y} B_\mu $ for the covariant derivative, opposite to the convention in use in Refs.~\cite{Jenkins:2013zja,Jenkins:2013wua,Alonso:2013hga}, which means that gauge couplings $g_1, g_2, g_3$ carry a relative minus sign between the two sets of references.}

\begin{eqnarray}\label{eq:SM_nuR_Lagrangian}
	\mathcal{L}_{{\rm SM} + \nu} &=& - \frac{1}{4} G^{A}_{\mu \nu} G^{A \mu \nu}_{} - \frac{1}{4} W^{I}_{\mu \nu} W^{I \mu \nu}_{} - \frac{1}{4} B^{}_{\mu \nu} B^{\mu \nu}_{} \\
    &+& (D_\mu H^\dagger) \, (D^\mu H) - \lambda \left( H^\dagger H - \frac{1}{2} v^2 \right)^2 + \sum_{\psi = q, u, d, \ell, e, \nu} \bar{\psi} i \slashed{D} \psi \nonumber\\
	&-& \left[ \frac{1}{2} (\nu C M_\nu \nu) + H^{\dagger j} \bar{d} Y_d q_j + \tilde{H}^{\dagger j} \bar{u} Y_u q_j + H^{\dagger j} \bar{e} Y_e \ell_j + \tilde{H}^{\dagger j} \bar{\nu} Y_\nu \ell_j + {\rm h.c.} \right] \,; \nonumber
\end{eqnarray}

\begin{table}[]
    \centering
    \renewcommand{\arraystretch}{1.2}
    \begin{tabular}{l}
        $ ( \bar{L} L ) ( \bar{L} L ) $ \\
        \hline
        $ Q^{}_{\ell \ell} = (\bar{\ell} \gamma_\mu \ell) (\bar{\ell} \gamma^\mu \ell) $ \\
        $ Q^{(1)}_{q q} = (\bar{q} \gamma_\mu q) (\bar{q} \gamma^\mu q) $ \\
        $ Q^{(3)}_{q q} = (\bar{q} \gamma_\mu \tau^I q) (\bar{q} \gamma^\mu \tau^I q) $ \\
        $ Q^{(1)}_{\ell q} = (\bar{\ell} \gamma_\mu \ell) (\bar{q} \gamma^\mu q) $ \\
        $ Q^{(3)}_{\ell q} = (\bar{\ell} \gamma_\mu \tau^I \ell) (\bar{q} \gamma^\mu \tau^I q) $ \\
    %\end{tabular}
    %
    %\begin{tabular}{l}
        \phantom{} \\
        $ ( \bar{R} R ) ( \bar{R} R ) $ \\
        \hline
        %$ Q^{}_{e e} = (\bar{e} \gamma_\mu e) (\bar{e} \gamma^\mu e) $ \\
        $ Q^{}_{y y} = (\bar{y} \gamma_\mu y) (\bar{y} \gamma^\mu y) $ \\
        %$ Q_{u u} = (\bar{u} \gamma_\mu u) (\bar{u} \gamma^\mu u) $ \\
        %$ Q_{d d} = (\bar{d} \gamma_\mu d) (\bar{d} \gamma^\mu d) $ \\
        $ Q^{}_{x x} = (\bar{x} \gamma_\mu x) (\bar{x} \gamma^\mu x) $ \\
        %$ Q_{e u} = (\bar{e} \gamma_\mu e) (\bar{u} \gamma^\mu u) $ \\
        %$ Q_{e d} = (\bar{e} \gamma_\mu e) (\bar{d} \gamma^\mu d) $ \\
        %$ Q^{}_{e x} = (\bar{e} \gamma_\mu e) (\bar{x} \gamma^\mu x) $ \\
        $ Q^{}_{y x} = (\bar{y} \gamma_\mu y) (\bar{x} \gamma^\mu x) $ \\
        $ Q_{\nu e} = (\bar{\nu} \gamma_\mu \nu) (\bar{e} \gamma^\mu e) $ \\
        $ Q^{(1)}_{u d} = (\bar{u} \gamma_\mu u) (\bar{d} \gamma^\mu d) $ \\
        $ Q^{(8)}_{u d} = (\bar{u} \gamma_\mu T^A u) (\bar{d} \gamma^\mu T^A d) $ \\
        $ Q_{d u \nu e} = (\bar{d} \gamma_\mu u) (\bar{\nu} \gamma^\mu e) $ \\
        $[ Q_{d u \nu e}^\dagger = (\bar{u} \gamma_\mu d) (\bar{e} \gamma^\mu \nu) ]$ \\
    \end{tabular}
    \hspace{10mm}
    \begin{tabular}{l}
        \phantom{} \\
        $ ( \bar{L} L ) ( \bar{R} R ) $ \\
        \hline
        %$ Q^{}_{\ell e} = (\bar{\ell} \gamma_\mu \ell) (\bar{e} \gamma^\mu e) $ \\
        $ Q^{}_{\ell y} = (\bar{\ell} \gamma_\mu \ell) (\bar{y} \gamma^\mu y) $ \\
        %$ Q_{\ell u} = (\bar{\ell} \gamma_\mu \ell) (\bar{u} \gamma^\mu u) $ \\
        %$ Q_{\ell d} = (\bar{\ell} \gamma_\mu \ell) (\bar{d} \gamma^\mu d) $ \\
        $ Q^{}_{\ell x} = (\bar{\ell} \gamma_\mu \ell) (\bar{x} \gamma^\mu x) $ \\
        %$ Q^{}_{q e} = (\bar{q} \gamma_\mu q) (\bar{e} \gamma^\mu e) $ \\
        $ Q^{}_{q y} = (\bar{q} \gamma_\mu q) (\bar{y} \gamma^\mu y) $ \\
        %$ Q^{(1)}_{q u} = (\bar{q} \gamma_\mu q) (\bar{u} \gamma^\mu u) $ \\
        %$ Q^{(8)}_{q u} = (\bar{q} \gamma_\mu T^A q) (\bar{u} \gamma^\mu T^A u) $ \\
        %$ Q^{(1)}_{q d} = (\bar{q} \gamma_\mu q) (\bar{d} \gamma^\mu d) $ \\
        %$ Q^{(8)}_{q d} = (\bar{q} \gamma_\mu T^A q) (\bar{d} \gamma^\mu T^A d) $ \\
        $ Q^{(1)}_{q x} = (\bar{q} \gamma_\mu q) (\bar{x} \gamma^\mu x) $ \\
        $ Q^{(8)}_{q x} = (\bar{q} \gamma_\mu T^A q) (\bar{x} \gamma^\mu T^A x) $ \\
    %\end{tabular}
    %
    %\begin{tabular}{l}
        \phantom{} \\
        $ ( \bar{L} R ) ( \bar{R} L ) $ + h.c. \\
        \hline
        $ Q^{}_{\ell e d q} = (\bar{\ell}^m e) (\bar{d} q_m) $ \\
        $ Q^{}_{\ell \nu u q} = (\bar{\ell}^m \nu) (\bar{u} q_m) $ \\
    %\end{tabular}
    %
    %\begin{tabular}{l}
        \phantom{} \\
        $ ( \bar{L} R ) ( \bar{L} R ) $ + h.c. \\
        \hline
        $ Q_{\ell \nu \ell e} = (\bar{\ell}^m \nu) \epsilon_{mn} (\bar{\ell}^n e) $ \\
        $ Q^{(1)}_{q u q d} = (\bar{q}^m u) \epsilon_{mn} (\bar{q}^n d) $ \\
        $ Q^{(8)}_{q u q d} = (\bar{q}^m T^A u) \epsilon_{mn} (\bar{q}^n T^A d) $ \\
        $ Q^{(1)}_{\ell e q u} = (\bar{\ell}^m e) \epsilon_{mn} (\bar{q}^n u) $ \\
        $ Q^{(3)}_{\ell e q u} = (\bar{\ell}^m \sigma_{\mu \nu} e) \epsilon_{mn} (\bar{q}^n \sigma^{\mu \nu} u) $ \\
        $ Q^{(1)}_{\ell \nu q d} = (\bar{\ell}^m \nu) \epsilon_{mn} (\bar{q}^n d) $ \\
        $ Q^{(3)}_{\ell \nu q d} = (\bar{\ell}^m \sigma_{\mu \nu} \nu) \epsilon_{mn} (\bar{q}^n \sigma^{\mu \nu} d) $ \\
    \end{tabular}
    \caption{Four-fermion operators of the so-called Warsaw basis extended to include right-handed neutrinos $\nu$, where $ x = u, d $ and $ y = e, \nu $; $q$ ($\ell$) are weak-isospin doublet quarks (leptons), and $u, d$ ($e, \nu$) are weak-isospin singlet quarks (leptons). Flavour or generation indices are omitted; when indicated in the text (e.g., as in the Wilson coefficient $ C_{\ell e d q; f i j k} $) they correspond to the fields above in that same ordering (i.e., $ (\bar{\ell}^m_f e_i) (\bar{d}_j q_{m; k}) $).}
    \label{tab:dimension_6}
\end{table}

\noindent
we are not writing explicitly the gauge fixing functional $\Psi$ of Eq.~\eqref{Action_BRST}, see e.g. Ref.~\cite{Weinberg:1996kr}.
We introduce the set of non-renormalizable operators in Tab.~\ref{tab:dimension_6}.
For completeness, and since the calculation is very similar, we extend the basis of operators of Ref.~\cite{Grzadkowski:2010es} to include right-handed neutrinos.
To avoid confusion, we stress that such a physical degree of freedom is not needed in the computations carrying only physical left-handed neutrinos.
Operators that violate total baryon or lepton numbers are found in Refs.~\cite{Alonso:2014zka,Liao:2016qyd}.
With these operators it is not possible to contract together two fermions, which is the only possibly non-vanishing topology as further discussed in App.~\ref{sec:properties_Greens_functions}; therefore, at one loop, this category of operators does not require the introduction of finite counterterms, see also Refs.~\cite{Naterop:2025lzc,Banik:2025wpi}.

At times
it is more convenient to consider operators in a different basis with respect to the Warsaw basis \cite{Grzadkowski:2010es}, following the use of Fierz identities
\begin{eqnarray}\label{eq:Fierz_1}
    && (\gamma_\mu P_L)_{i j} (\gamma^\mu P_L)_{k l} = - (\gamma_\mu P_L)_{i l} (\gamma^\mu P_L)_{k j} \,, \\
    && (\gamma_\mu P_L)_{i j} (\gamma^\mu P_R)_{k l} = 2 (P_R)_{i l} (P_L)_{k j} \,, \\
    && (P_L)_{i j} (P_L)_{k l} = \frac{1}{2} (P_L)_{i l} (P_L)_{k j} + \frac{1}{8} (\sigma_{\mu \nu} P_L)_{i l} (\sigma^{\mu \nu} P_L)_{k j} \,, \\
    && (\sigma_{\mu \nu} P_L)_{i j} (\sigma^{\mu \nu} P_L)_{k l} = 6 (P_L)_{i l} (P_L)_{k j} - \frac{1}{2} (\sigma_{\mu \nu} P_L)_{i l} (\sigma^{\mu \nu} P_L)_{k j} \,, \label{eq:Fierz_4}
\end{eqnarray}
where $i, j, k, l$ are spinorial indices, and
$ \sigma^{\mu \nu} = i [\gamma^\mu, \gamma^\nu] / 2 $,
while other identities are found by interchanging the chiral projectors $ P_L \leftrightarrow P_R $ (of course, fermions anticommute, and this leads to an additional sign when presenting these identities along with fermionic operators).
When employing these identities, it is useful to consider the following ones for the generators of $SU(3)$
\begin{eqnarray}\label{eq:GellMann_identity_1}
    && \delta_{\alpha \tilde{\alpha}} \delta_{\beta \tilde{\beta}} = \frac{1}{N_c} \delta_{\alpha \tilde{\beta}} \delta_{\beta \tilde{\alpha}} + 2 T^A_{\alpha \tilde{\beta}} T^A_{\beta \tilde{\alpha}} \,, \\
    && T^A_{\alpha \tilde{\alpha}} T^A_{\beta \tilde{\beta}} = \frac{C_A}{N_c} \delta_{\alpha \tilde{\beta}} \delta_{\beta \tilde{\alpha}} - \frac{1}{N_c} T^A_{\alpha \tilde{\beta}} T^A_{\beta \tilde{\alpha}} \,, \label{eq:GellMann_identity_2}
\end{eqnarray}
where $ \alpha, \tilde{\alpha}, \beta, \tilde{\beta} $ are color indices, and $ C_A = \frac{1}{2} \left( N_c - \frac{1}{N_c} \right) $. Analogous expressions hold for the generators $T_L^I$ of $SU(2)_L$, with $N_c$ replaced by $2$.

The two bases differ in general by evanescent operators. Qualitatively, since the symmetry-breaking operator is itself an evanescent operator, and since we are looking for the infinite parts of the diagrams introduced above, of ghost number $1$ that are not evanescent, the choice of the basis has no effect on the results that will be displayed hereafter (apart from overall constant factors resulting from the use of the Fierz identities).

\section{Green's functions and counterterms}\label{sec:results_CTs}

In computing possible contributions to the quantum effective action we consider connected 1PI diagrams.
We divide the Green's functions with one insertion of the symmetry-breaking operator according to the matter fields: none, one, and two scalar fields, all having two fermion fields. The Green's functions having four fermion fields vanish.
See App.~\ref{sec:properties_Greens_functions} for a detailed discussion.
As it will be made explicit in the following, given the basis of operators already discussed, cases of interest involve closed (that require the computation of a trace of Dirac matrices) and/or open fermion lines,\footnote{In theories with renormalizable interactions as in the SM, the analogous case of having an open fermion line (with an internal vector or scalar being exchanged) and more than one external vector boson or scalar legs does not require any finite renormalization. This can be easily understood based on identifying the superficial degree of divergence, and using the same arguments as presented in App.~\ref{sec:properties_Greens_functions}. See also Ref.~\cite{ValeSilva:2025}.}
although this distinction is not important at this order due to possible Fierz reshuffling.
The full list of possible counterterms is discussed in App.~\ref{sec:list_CTs};
BRST transformations of the elementary fields are found in Eq.~\eqref{BRST_Transformation}. Hereafter the counterterms will be given a more explicit form.

We now provide a solution for the counterterms required in the finite renormalization program, that eliminate the obstructions for the Zinn-Justin equation to hold.
Such a solution is not unique: it is defined up to an arbitrary combination of operators that vanish under the BRST transformation; one can also employ the usual integration by parts to move around partial derivatives.
We have not been able to design a strategy to prescribe symmetry-restoring counterterms up to these BRST-invariant ambiguities, aside from requiring the ``simplicity'' of the counterterms; there are specific cases in which BRST-invariant ambiguities are absent, e.g., there are no gauge-invariant operators of dimension 6 coupling leptons to chromodynamic fields.
We stress that gauge invariant structures, or BRST-exact ones, alone cannot provide the needed renormalization, since they both vanish under the BRST transformation.
Also, the results for the Green's functions and the corresponding counterterms are given in 4 dimensions: it is then made manifest
that such counterterms are not evanescent.\footnote{Evanescent pieces to the following discussion could be added at will. Here we make the choice of not including them.}

We perform our calculations in the BMHV scheme using \texttt{FeynCalc} \cite{Mertig:1990an,Shtabovenko:2016sxi,Shtabovenko:2020gxv,Shtabovenko:2023idz}.
When indicated, the momenta $ k_g, k_h, k_v $ are all incoming.

\subsection{No scalars}\label{sec:no_scalars_results}

\begin{figure}
    \centering
    \hspace{6mm} \includegraphics[width=0.18\linewidth]{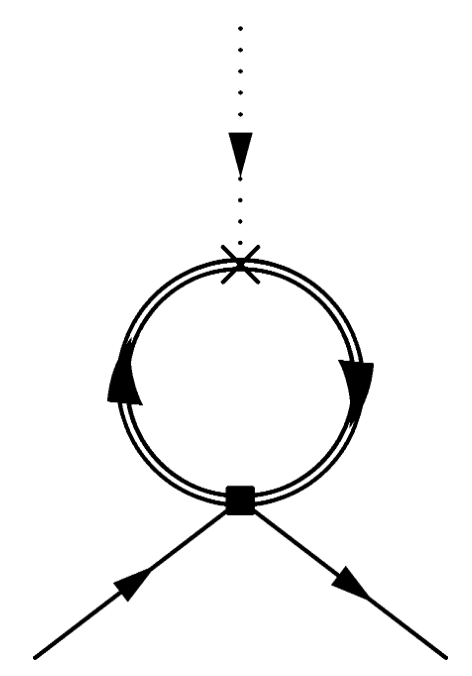} \hspace{11mm}
    \includegraphics[width=0.27\linewidth]{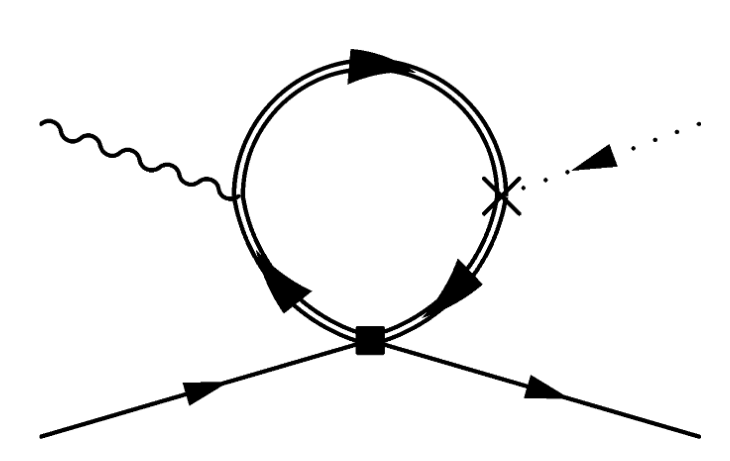} \hspace{5mm}
    \includegraphics[width=0.2\linewidth]{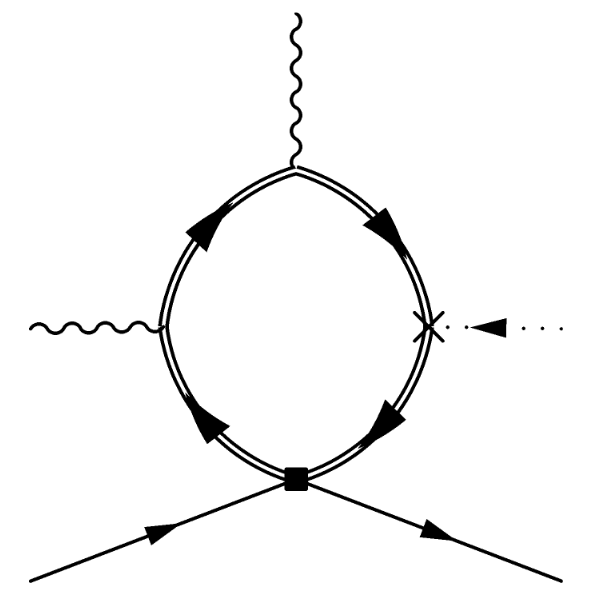}
    \caption{Examples of 1PI diagrams with single insertions of SMEFT four-fermion operators (represented by filled squares) at one loop. No external scalars are present in these examples. Insertions of the symmetry-breaking operator are indicated by a cross, and ghosts are represented by a dotted line. Gauge bosons are indicated by wiggly lines. The single (double) solid lines represent physical right-handed (left-handed) fermions; analogous diagrams with opposite chiralities are also possible.}
    \label{fig:diagrams_no_scalars}
\end{figure}

We first discuss Green's functions having no scalars, see Fig.~\ref{fig:diagrams_no_scalars}.
With little effort, the expressions for the different amplitudes can be adapted or directly taken from those presented in App.~\ref{app:example}.
For instance, the discussion of $\gamma_\mu \tau^I$ structures can be easily adapted from there, the only important difference being that the (fully) symmetric constant is now absent.

\subsubsection{$\gamma_\mu$-structure counterterms}\label{sec:gamma_structure_CTs}

We start by discussing couterterm structures having a fermion bilinear of the form $ \bar{\xi} \bar{\gamma}_\mu \xi $, where $\xi$ is any left- or right-handed physical fermion, being a quark or a lepton of possibly different generations; the structure $ \bar{\xi} \bar{\gamma}_\alpha \bar{\gamma}_\beta \bar{\gamma}_\mu \xi $ can also be cast in such a way (for instance, as displayed below, it comes contracted with three gauge boson fields allowing further simplification), but we find more appropriate keeping the bilinear under such a Lorentz form in some instances.
In order to obtain non-vanishing amplitudes, zero or at least two $SU(2)_L$ or $SU(3)$ gauge bosons are needed.
A possible basis of counterterm operators includes

\begin{equation}
    \mathcal{Q}_1 = g_1 (\bar{\xi} \bar{\gamma}_\mu \xi) \bar{\partial}^2 B^\mu \,, \label{eq:T1} %T^{(1)}
\end{equation}
which has a single gauge boson; then, the following cases having a term with two gauge bosons (which consist of the Chern-Simons form of QCD, and analogously for $U(1)_Y$ and $SU(2)_L$, contracted with a fermion bilinear) are also included

\begin{eqnarray}
    && \mathcal{Q}_2 = g_1^2 (\bar{\xi} \bar{\gamma}_\mu \xi) \epsilon^{\mu \nu \rho \sigma} B_{\sigma} \bar{\partial}_\rho B_{\nu} \,, \label{eq:T1CS}\\ %T^{(\rm CS)}_1
    && \mathcal{Q}_3 = (\bar{\xi} \bar{\gamma}_\mu \xi) \epsilon^{\mu \nu \rho \sigma} \left( g_2^2 W^b_\sigma \bar{\partial}_\rho W^b_\nu - g_2^3 \frac{1}{3} C^{SU(2)}_{b e f} W^b_\sigma W^e_\nu W^f_\rho \right) \,, \label{eq:T2CS}\\ %T^{(\rm CS)}_2
    && \mathcal{Q}_4 = (\bar{\xi} \bar{\gamma}_\mu \xi) \epsilon^{\mu \nu \rho \sigma} \left( g_3^2 G^B_\sigma \bar{\partial}_\rho G^B_\nu - g_3^3 \frac{1}{3} C_{B E F} G^B_\sigma G^E_\nu G^F_\rho \right) \,; \label{eq:T3CS} %T^{(\rm CS)}_3
\end{eqnarray}
finally, there are also various cases having three gauge bosons

\begin{eqnarray}
    \mathcal{Q}_5 = g_1^3 B_\alpha B_\mu B_\beta ( \bar{\xi} \bar{\gamma}^\alpha \bar{\gamma}^\mu \bar{\gamma}^\beta \xi ) \,, \label{eq:T3A} %T^{(3A)}
\end{eqnarray}

\begin{eqnarray}
    && \mathcal{Q}_6 = g_1 g_2^2 W^b_\alpha W^b_\mu B_\beta \bar{\xi} ( \bar{\gamma}^\beta \bar{\gamma}^\alpha \bar{\gamma}^\mu + \bar{\gamma}^\alpha \bar{\gamma}^\beta \bar{\gamma}^\mu + \bar{\gamma}^\alpha \bar{\gamma}^\mu \bar{\gamma}^\beta ) \xi \,, \label{eq:T23A}\\ %T^{(3A)}_2
    && \mathcal{Q}_7 = g_1 g_3^2 G_\alpha^B G_\mu^B B_\beta \bar{\xi} ( \bar{\gamma}^\beta \bar{\gamma}^\alpha \bar{\gamma}^\mu + \bar{\gamma}^\alpha \bar{\gamma}^\beta \bar{\gamma}^\mu + \bar{\gamma}^\alpha \bar{\gamma}^\mu \bar{\gamma}^\beta ) \xi \,, \label{eq:T33A} %T^{(3A)}_3
\end{eqnarray}

\begin{eqnarray}
    \mathcal{Q}_8 = g_3^3 D_{B C D} G^B_\alpha G^C_\mu G^D_\beta (\bar{\xi} \bar{\gamma}^\alpha \bar{\gamma}^\mu \bar{\gamma}^\beta \xi) \,, \label{eq:T3d} %T^{(3d)}
\end{eqnarray}

\begin{eqnarray}
    && \mathcal{Q}_9 = ( \bar{\xi} \bar{\gamma}_\mu \xi ) g_2^3 W^b_\nu W^c_\rho W^d_\sigma \epsilon^{\mu \nu \rho \sigma} C^{SU(2)}_{b c d} \,, \label{eq:T23f}\\ %T^{(3f)}_2
    && \mathcal{Q}_{10} = ( \bar{\xi} \bar{\gamma}_\mu \xi ) g_3^3 G^B_\nu G^C_\rho G^D_\sigma \epsilon^{\mu \nu \rho \sigma} C_{B C D} \,. \label{eq:T33f} %T^{(3f)}_3
\end{eqnarray}

The finite counterterm Lagrangian is

\begin{comment}
\begin{eqnarray}
    \mathcal{L}_{\rm fct} \supset && - \frac{C^{\mathcal{O}}_{\xi \xi}}{3 \cdot 16 \pi^2} \left( h^{\mathcal{O}}_{(1)} \, T^{(1)} + h^{\mathcal{O}}_{(3A)} \, T^{(3A)} + h^{\mathcal{O}}_{(3d)} \, T^{(3d)} \right. \\
    && \qquad \left. + h^{\mathcal{O}}_{({\rm CS}),i} \, T^{(\rm CS)}_i + h^{\mathcal{O}}_{(3A),j} \, T^{(3A)}_j + h^{\mathcal{O}}_{(3f),j} \, T^{(3f)}_j \right) + {\rm h.c.} \,, \nonumber
\end{eqnarray}
\end{comment}

\begin{equation}
    \mathcal{L}_{\rm fct} \supset - \frac{C^{\mathcal{O}}_{\xi \xi}}{3 \cdot 16 \pi^2} \sum^{10}_{i=1} h^{\mathcal{O}}_{i} \, \mathcal{Q}_i + {\rm h.c.}
\end{equation}
%where $i=1,2,3$ and $j=2,3$
Hermitian conjugation has to be included only for cases where the fermion bilinear involves two different flavors. The coefficients $h^{\mathcal{O}}$ are shown in Tab.~\ref{tab:gamma_gamma}.
This table is organized as follows: the first, second, third and fourth sectors of the first half correspond to $\ell$, $q$, $x=u, d$ and $y=e, \nu$ loops, respectively, when the Feynman diagrams are obtained by contracting the fermion fields that appear within the same bilinear of the four-fermion operators, following the convention of Tab.~\ref{tab:dimension_6}; the second half of the table corresponds to those other topologies obtained when contracting fermions in different bilinears of the four-fermion operators, again when following the Warsaw basis. These two halves can be related by the Fierz identities shown in Eqs.~\eqref{eq:Fierz_1}-\eqref{eq:Fierz_4}. As can be understood from Eq.~\eqref{eq:GellMann_identity_1}, $SU(2)_L$ triplet and $SU(3)$ octet structures can be obtained from $SU(2)_L$ and $SU(3)$ singlet cases, and the converse is also true following Eq.~\eqref{eq:GellMann_identity_2}; the resulting counterterms are discussed in Secs.~\ref{sec:gamma_Pauli_structure_CTs}, \ref{sec:gamma_GellMann_structure_CTs}, \ref{sec:gamma_Pauli_GellMann_structure_CTs}.

It turns out that some of the different counterterm structures of Eqs.~\eqref{eq:T1}-\eqref{eq:T33f} appear in specific combinations

\begin{comment}
\begin{eqnarray}
    && y_L^q T^{(1)} - (y_L^q)^3 T^{(3A)} \,, \; \text{or} \;\; y_L^\ell T^{(1)} - (y_L^\ell)^3 T^{(3A)} \,, \nonumber\\
    && \qquad \text{or} \;\; y_R^x T^{(1)} - (y_R^x)^3 T^{(3A)} \,, \; \text{or} \;\; y_R^y T^{(1)} - (y_R^y)^3 T^{(3A)} \,, \\
    && T^{(\rm CS)}_2 - \frac{1}{4} y_L^q T^{(3A)}_2 - \frac{1}{24} T^{(3f)}_2 \,, \; \text{or} \;\; T^{(\rm CS)}_2 - \frac{1}{4} y_L^\ell T^{(3A)}_2 - \frac{1}{24} T^{(3f)}_2 \,, \\
    && T^{(\rm CS)}_3 - \frac{1}{24} T^{(3f)}_3 \,, \\
    && T^{(3d)} + 2 y_L^q T^{(3A)}_3 \,, \; \text{or} \;\; T^{(3d)} + 2 y_L^\ell T^{(3A)}_3 \,.
\end{eqnarray}
\end{comment}

\begin{eqnarray}
    && y_L^q \mathcal{Q}_1 - (y_L^q)^3 \mathcal{Q}_5 \;\; \text{or} \;\; y_L^\ell \mathcal{Q}_1 - (y_L^\ell)^3 \mathcal{Q}_5 \nonumber\\
    && \qquad \text{or} \;\; y_R^x \mathcal{Q}_1 - (y_R^x)^3 \mathcal{Q}_5 \;\; \text{or} \;\; y_R^y \mathcal{Q}_1 - (y_R^y)^3 \mathcal{Q}_5 \,, \\
    && \mathcal{Q}_3 - \frac{1}{4} y_L^q \mathcal{Q}_6 - \frac{1}{24} \mathcal{Q}_9 \;\; \text{or} \;\; \mathcal{Q}_3 - \frac{1}{4} y_L^\ell \mathcal{Q}_6 - \frac{1}{24} \mathcal{Q}_9 \,, \\
    && \mathcal{Q}_4 - \frac{1}{24} \mathcal{Q}_{10} \,, \\
    && \mathcal{Q}_8 + 2 y_L^q \mathcal{Q}_7 \;\; \text{or} \;\; \mathcal{Q}_8 + 2 y_L^\ell \mathcal{Q}_7 \,.
\end{eqnarray}

To provide a concrete example of a particular structure in the finite counterterm Lagrangian, we have:

\begin{eqnarray}
    && \mathcal{L}_{\rm fct} \supset \frac{1}{3 \cdot 16 \pi^2} \left( g_3^2 G^B_\nu \bar{\partial}_\rho G^B_\sigma + g_3^3 \frac{1}{3} C_{B E F} G^B_\nu G^E_\rho G^F_\sigma \right) \epsilon^{\mu \nu \rho \sigma} \\
    && \left[ C^{q q (1)}_{f i j k} 4 \delta_{fi} \bar{q}_j \bar{\gamma}_\mu q_k + C^{q q (1)}_{f i j k} 4 \delta_{jk} \bar{q}_f \bar{\gamma}_\mu q_i + C^{\ell q (1)}_{f i j k} 4 \delta_{jk} \bar{\ell}_f \bar{\gamma}_\mu \ell_i \right. \nonumber\\
    && + C^{q \nu}_{f i j k} 4 \delta_{fi} \bar{\nu}_j \bar{\gamma}_\mu \nu_k + C^{q e}_{f i j k} 4 \delta_{fi} \bar{e}_j \bar{\gamma}_\mu e_k \nonumber\\
    && + C^{q u (1)}_{f i j k} 4 \delta_{fi} \bar{u}_j \bar{\gamma}_\mu u_k + C^{q d (1)}_{f i j k} 4 \delta_{fi} \bar{d}_j \bar{\gamma}_\mu d_k \nonumber\\
    %%%%%%%%%%%%%%%%%%%%%%%%%%%%%%%%%%%%%%%%%%%%%%%%%%%%%%%%%%%%%%%%%%%%%%%%%%%%%%%%%%%%%%%%%
    && - C^{u u}_{f i j k} 2 \delta_{fi} \bar{u}_j \bar{\gamma}_\mu u_k - C^{d d}_{f i j k} 2 \delta_{fi} \bar{d}_j \bar{\gamma}_\mu d_k \nonumber\\
    && - C^{u u}_{f i j k} 2 \delta_{jk} \bar{u}_f \bar{\gamma}_\mu u_i - C^{d d}_{f i j k} 2 \delta_{jk} \bar{d}_f \bar{\gamma}_\mu d_i \nonumber\\
    && - C^{\nu u}_{f i j k} 2 \delta_{jk} \bar{\nu}_f \bar{\gamma}_\mu \nu_i - C^{e u}_{f i j k} 2 \delta_{jk} \bar{e}_f \bar{\gamma}_\mu e_i \nonumber\\
    && - C^{\nu d}_{f i j k} 2 \delta_{jk} \bar{\nu}_f \bar{\gamma}_\mu \nu_i - C^{e d}_{f i j k} 2 \delta_{jk} \bar{e}_f \bar{\gamma}_\mu e_i \nonumber\\
    && - C^{u d (1)}_{f i j k} 2 \delta_{fi} \bar{d}_j \bar{\gamma}_\mu d_k - C^{u d (1)}_{f i j k} 2 \delta_{jk} \bar{u}_f \bar{\gamma}_\mu u_i \nonumber\\
    && - C^{\ell u}_{f i j k} 2 \delta_{jk} \bar{\ell}_f \bar{\gamma}_\mu \ell_i - C^{\ell d}_{f i j k} 2 \delta_{jk} \bar{\ell}_f \bar{\gamma}_\mu \ell_i \nonumber\\
    && - C^{q u (1)}_{f i j k} 2 \delta_{jk} \bar{q}_f \bar{\gamma}_\mu q_i - C^{q d (1)}_{f i j k} 2 \delta_{jk} \bar{q}_f \bar{\gamma}_\mu q_i \nonumber\\
    %%%%%%%%%%%%%%%%%%%%%%%%%%%%%%%%%%%%%%%%%%%%%%%%%%%%%%%%%%%%%%%%%%%%%%%%%%%%%%%%%%%%%%%%%
    && + C^{q q (1)}_{f i j k} \frac{2}{N_c} \delta_{fk} \bar{q}_j \bar{\gamma}_\mu q_i + C^{q q (1)}_{f i j k} \frac{2}{N_c} \delta_{ji} \bar{q}_f \bar{\gamma}_\mu q_k \nonumber\\
    %%%%%%%%%%%%%%%%%%%%%%%%%%%%%%%%%%%%%%%%%%%%%%%%%%%%%%%%%%%%%%%%%%%%%%%%%%%%%%%%%%%%%%%%%
    && + C^{q q (3)}_{f i j k} \frac{6}{N_c} \delta_{fk} \bar{q}_j \bar{\gamma}_\mu q_i + C^{q q (3)}_{f i j k} \frac{6}{N_c} \delta_{ji} \bar{q}_f \bar{\gamma}_\mu q_k \nonumber\\
    %%%%%%%%%%%%%%%%%%%%%%%%%%%%%%%%%%%%%%%%%%%%%%%%%%%%%%%%%%%%%%%%%%%%%%%%%%%%%%%%%%%%%%%%%
    && - C^{u u}_{f i j k} \frac{2}{N_c} \delta_{fk} \bar{u}_j \bar{\gamma}_\mu u_i - C^{d d}_{f i j k} \frac{2}{N_c} \delta_{fk} \bar{d}_j \bar{\gamma}_\mu d_i \nonumber\\
    && \left. - C^{u u}_{f i j k} \frac{2}{N_c} \delta_{ji} \bar{u}_f \bar{\gamma}_\mu u_k - C^{d d}_{f i j k} \frac{2}{N_c} \delta_{ji} \bar{d}_f \bar{\gamma}_\mu d_k \right] + {\rm h.c.} \nonumber
\end{eqnarray}
As this example illustrates, there are for instance lepton-gluonic finite counterterm structures, which are not possible when considering dimension-6 gauge-invariant operators.

Some generation assignments are identical: $ Q^{}_{e e; f i j k} = Q^{}_{e e; j k f i} $, to give an example. We do not use this symmetry to abbreviate our results,
i.e., Tab.~\ref{tab:gamma_gamma} includes $ \delta_{f i} $ and $ \delta_{j k} $ in such cases.
Eliminating redundancies leads to symmetry factors that can be absorbed into the Wilson coefficients.

\begin{sidewaystable}
%\begin{table}[]
    %{\scriptsize
    \centering
    \begin{tabular}{|cc||c|c|c|c|c|c|c|c|c|c|}
        \hline
        $\mathcal{O}$ & & $ h^{\mathcal{O}}_{1} $ & $ h^{\mathcal{O}}_{2} $ & $ h^{\mathcal{O}}_{3} $ & $ h^{\mathcal{O}}_{4} $ & $ h^{\mathcal{O}}_{5} $ & $ h^{\mathcal{O}}_{6} $ & $ h^{\mathcal{O}}_{7} $ & $ h^{\mathcal{O}}_{8} $ & $ h^{\mathcal{O}}_{9} $ & $ h^{\mathcal{O}}_{10} $ \\
        \hline
        \hline
        $ Q_{\ell \ell} $ & $\delta_{fi}, \delta_{jk}$ & \multirow{4}{*}{$2 y^\ell_L$} & \multirow{4}{*}{$8 (y^\ell_L)^2$} & \multirow{4}{*}{$2$} & \multirow{4}{*}{$0$} & \multirow{4}{*}{$- 2 (y^\ell_L)^3$} & \multirow{4}{*}{$-\frac{1}{2} y^\ell_L$} & \multirow{4}{*}{$0$} & \multirow{4}{*}{$0$} & \multirow{4}{*}{$-\frac{1}{12}$} & \multirow{4}{*}{$0$} \\
        $ Q^{(1)}_{\ell q} $ & $\delta_{fi}$ & & & & & & & & & & \\
        $ Q_{\ell y} $ & $\delta_{fi}$ & & & & & & & & & & \\
        $ Q_{\ell x} $ & $\delta_{fi}$ & & & & & & & & & & \\
        \hline
        $ Q^{(1)}_{q q} $ & $\delta_{fi}, \delta_{jk}$ & \multirow{4}{*}{$2 N_c y^q_L$} & \multirow{4}{*}{$8 N_c (y^q_L)^2$} & \multirow{4}{*}{$2 N_c$} & \multirow{4}{*}{$4$} & \multirow{4}{*}{$- 2 N_c (y^q_L)^3$} & \multirow{4}{*}{$-\frac{N_c}{2} y^q_L$} & \multirow{4}{*}{$- y^q_L$} & \multirow{4}{*}{$-\frac{1}{2}$} & \multirow{4}{*}{$-\frac{N_c}{12}$} & \multirow{4}{*}{$-\frac{1}{6}$} \\
        $ Q^{(1)}_{\ell q} $ & $\delta_{jk}$ & & & & & & & & & & \\
        $ Q_{q y} $ & $\delta_{fi}$ & & & & & & & & & & \\
        $ Q^{(1)}_{q x} $ & $\delta_{fi}$ & & & & & & & & & & \\
        \hline
        $ Q_{x x} $ & $\delta_{fi}, \delta_{jk}$ & \multirow{5}{*}{$N_c y^x_R$} & \multirow{5}{*}{$- 4 N_c (y^x_R)^2$} & \multirow{5}{*}{$0$} & \multirow{5}{*}{$- 2$} & \multirow{5}{*}{$- N_c (y^x_R)^3$} & \multirow{5}{*}{$0$} & \multirow{5}{*}{$-\frac{1}{2} y^x_R$} & \multirow{5}{*}{$-\frac{1}{4}$} & \multirow{5}{*}{$0$} & \multirow{5}{*}{$\frac{1}{12}$} \\
        $ Q_{y x} $ & $\delta_{jk}$ & & & & & & & & & & \\
        $ Q^{(1)}_{u d} $ & $\delta_{fi}, \delta_{jk}$ & & & & & & & & & & \\
        $ Q_{\ell x} $ & $\delta_{jk}$ & & & & & & & & & & \\
        $ Q^{(1)}_{q x} $ & $\delta_{jk}$ & & & & & & & & & & \\
        \hline
        $ Q_{y y} $ & $\delta_{fi}, \delta_{jk}$ & \multirow{5}{*}{$y^y_R$} & \multirow{5}{*}{$- 4 (y^y_R)^2$} & \multirow{5}{*}{$0$} & \multirow{5}{*}{$0$} & \multirow{5}{*}{$- (y^y_R)^3$} & \multirow{5}{*}{$0$} & \multirow{5}{*}{$0$} & \multirow{5}{*}{$0$} & \multirow{5}{*}{$0$} & \multirow{5}{*}{$0$} \\
        $ Q_{y x} $ & $\delta_{fi}$ & & & & & & & & & & \\
        $ Q_{\nu e} $ & $\delta_{fi}, \delta_{jk}$ & & & & & & & & & & \\
        $ Q_{\ell y} $ & $\delta_{jk}$ & & & & & & & & & & \\
        $ Q_{q y} $ & $\delta_{jk}$ & & & & & & & & & & \\
        \hline
        \hline
        $ Q_{\ell \ell} $ & $\delta_{fk}, \delta_{ji}$ & $y^\ell_L$ & $4 (y^\ell_L)^2$ & $1$ & $0$ & $- (y^\ell_L)^3$ & $-\frac{1}{4} y^\ell_L$ & $0$ & $0$ & $-\frac{1}{24}$ & $0$ \\
        \hline
        $ Q^{(1)}_{q q} $ & $\delta_{fk}, \delta_{ji}$ & $y^q_L$ & $4 (y^q_L)^2$ & $1$ & $\frac{2}{N_c}$ & $- (y^q_L)^3$ & $-\frac{1}{4} y^q_L$ & $-\frac{1}{2 N_c} y^q_L$ & $-\frac{1}{4 N_c}$ & $-\frac{1}{24}$ & $-\frac{1}{12 N_c}$ \\
        \hline
        $ Q^{(3)}_{q q} $ & $\delta_{fk}, \delta_{ji}$ & $3 y^q_L$ & $12 (y^q_L)^2$ & $3$ & $\frac{6}{N_c}$ & $-3 (y^q_L)^3$ & $-\frac{3}{4} y^q_L$ & $-\frac{3}{2 N_c} y^q_L$ & $-\frac{3}{4 N_c}$ & $-\frac{1}{8}$ & $-\frac{1}{4 N_c}$ \\
        \hline
        $ Q_{x x} $ & $\delta_{fk}, \delta_{ji}$ & $y^x_R$ & $- 4 (y^x_R)^2$ & $0$ & $- \frac{2}{N_c}$ & $- (y^x_R)^3$ & $0$ & $-\frac{1}{2 N_c} y^x_R$ & $-\frac{1}{4 N_c}$ & $0$ & $\frac{1}{12 N_c}$ \\
        \hline
        $ Q_{y y} $ & $\delta_{fk}, \delta_{ji}$ & $y^y_R$ & $- 4 (y^y_R)^2$ & $0$ & $0$ & $- (y^y_R)^3$ & $0$ & $0$ & $0$ & $0$ & $0$ \\
        \hline
    \end{tabular}
    %}
    \caption{
    Coefficients in the finite counterterm Lagrangian, which present a fermion bilinear of structure $ \bar{\xi} \bar{\gamma}_\mu \xi $.
    The fermion generation indices of the four-fermion operators of the Warsaw basis are $f i j k$, in this same ordering. The operators and the fermions contracted in the loop can be read from the first two columns.
    In the case of $ Q^{(1)}_{u d} $ ($ Q_{\nu e} $), since the indicated hypercharge $y^x_R$ ($y^y_R$, respectively) is the one of the fermion running inside the loop, $x = u$ for $\delta_{fi}$ and $x = d$ for $\delta_{jk}$ ($y = \nu$ for $\delta_{fi}$ and $y = e$ for $\delta_{jk}$, respectively).
    }
    \label{tab:gamma_gamma}
%\end{table}
\end{sidewaystable}

\subsubsection{$\gamma_\mu \tau^I$-structure counterterms}\label{sec:gamma_Pauli_structure_CTs}

We now move to those structures that have the fermion bilinear $ \bar{\xi} \bar{\gamma}_\mu T_L^I \xi $, where $T_L^I$ is a generator of $SU(2)_L$; it is also convenient at times not to simplify structures with three $\gamma$-matrices (e.g., by employing the Chisholm identity). Non-vanishing obstructions are only possible when at least one $SU(2)_L$ gauge boson is radiated from the fermion loop. We have the following structure with a single gauge boson

\begin{eqnarray}
    \mathcal{Q}_{11} = - g_2 \frac{1}{2} \bar{\partial}^2 W^I_\mu ( \bar{\xi} \bar{\gamma}^\mu T^I_L \xi ) + g_2 \bar{\partial}_\mu \bar{\partial}^\nu W^I_\nu ( \bar{\xi} \bar{\gamma}^\mu T^I_L \xi ) \,; \label{eq:Ftilde21} %\widetilde{F}^{(1)}_2
\end{eqnarray}
we also have the following cases with three gauge bosons

\begin{eqnarray}
    \mathcal{Q}_{12} = g_2^3 {\rm Tr} \{ T^I_L T^J_L T^K_L T^L_L \} W^J_\alpha W^K_\mu W^L_\beta ( \bar{\xi} \bar{\gamma}^\alpha \bar{\gamma}^\mu \bar{\gamma}^\beta T^I_L \xi ) \,, \label{eq:Ftilde13} %\widetilde{F}^{(3)}_1
\end{eqnarray}

\begin{comment}
\begin{equation}
    \mathcal{Q}'_{13} = i g_2^3 ( \bar{\xi} \bar{\gamma}_\mu T^I_L \xi ) W^J_\nu W^K_\rho W^L_\sigma \epsilon^{\mu \nu \rho \sigma} \mathcal{A}_{I,SU(2)}^{J K L} \,, \label{eq:Ftilde23} %\widetilde{F}^{(3)}_2
\end{equation}
where

\begin{eqnarray}
    && \mathcal{A}_{I,SU(2)}^{J K L} = {\rm Tr} \{ T^I_L T^L_L ( T^J_L T^K_L - T^K_L T^J_L ) \} \\
    && - {\rm Tr} \{ T^I_L ( T^J_L T^L_L T^K_L - T^K_L T^L_L T^J_L ) \} + {\rm Tr} \{ T^I_L ( T^J_L T^K_L - T^K_L T^J_L ) T^L_L \} \nonumber\\
    && = \frac{1}{4} i \left( C_{JKE} D_{ILE} - C_{LKE} D_{IJE} - C_{JLE} D_{IKE} \right) \,, \nonumber
\end{eqnarray}
which is totally antisymmetric
in $J, K, L$, and
\end{comment}

\begin{eqnarray}
    && \mathcal{Q}_{13} = g_1^2 g_2 B_\alpha B_\mu W_\beta^I \bar{\xi} ( \bar{\gamma}^\beta \bar{\gamma}^\alpha \bar{\gamma}^\mu + \bar{\gamma}^\alpha \bar{\gamma}^\beta \bar{\gamma}^\mu + \bar{\gamma}^\alpha \bar{\gamma}^\mu \bar{\gamma}^\beta ) T^I_L \xi \,, \label{eq:Ftilde1AG}\\ %\widetilde{F}^{(AG)}_1
    && \mathcal{Q}_{14} = g_2 g_3^2 G^B_\alpha G^B_\mu W_\beta^I \bar{\xi} ( \bar{\gamma}^\beta \bar{\gamma}^\alpha \bar{\gamma}^\mu + \bar{\gamma}^\alpha \bar{\gamma}^\beta \bar{\gamma}^\mu + \bar{\gamma}^\alpha \bar{\gamma}^\mu \bar{\gamma}^\beta ) T^I_L \xi \,. \label{eq:Ftilde2AG} %\widetilde{F}^{(AG)}_2
\end{eqnarray}

In terms of this basis of structures, the finite counterterm Lagrangian is

\begin{comment}
\begin{eqnarray}
    \mathcal{L}_{\rm fct} \supset && - \frac{C^{\mathcal{O}}_{\xi \xi}}{3 \cdot 16 \pi^2} \left( \tilde{f}^{\mathcal{O} (1)}_2 \widetilde{F}^{(1)}_2 + \tilde{f}^{\mathcal{O} (3)}_1 \widetilde{F}^{(3)}_1 + \tilde{f}^{\mathcal{O} (3)}_2 \widetilde{F}^{(3)}_2 \right. \\
    && \qquad \left. + \tilde{f}^{\mathcal{O} (AG)}_1 \widetilde{F}^{(AG)}_1 + \tilde{f}^{\mathcal{O} (AG)}_2 \widetilde{F}^{(AG)}_2 \right) + {\rm h.c.} \nonumber
\end{eqnarray}
\end{comment}

\begin{equation}
    \mathcal{L}_{\rm fct} \supset - \frac{C^{\mathcal{O}}_{\xi \xi}}{3 \cdot 16 \pi^2} \sum^{14}_{i=11} h^{\mathcal{O}}_{i} \, \mathcal{Q}_i + {\rm h.c.}
\end{equation}
Coefficients are given in Tab.~\ref{tab:gammaTL_gammaTL}.
The structure of this table is as follows: the first and second sectors of the first half correspond to $q$ and $\ell$ loops, respectively, when contracting the fermion fields that appear within the same bilinear, following the convention of Tab.~\ref{tab:dimension_6}; the second half of the table corresponds to contracting fermions in different bilinears, again when following the Warsaw basis.
As seen from Tab.~\ref{tab:gammaTL_gammaTL}, operators other than $ Q^{(3)}_{q q}, Q^{(3)}_{\ell q} $ from the Warsaw basis also require finite renormalization by the structures shown above (cf. $SU(2)_L$-analogous identities of Eqs.~\eqref{eq:GellMann_identity_1} and \eqref{eq:GellMann_identity_2}).

We observe that some of the different counterterm structures of Eqs.~\eqref{eq:Ftilde21}-\eqref{eq:Ftilde2AG} appear in specific combinations

\begin{comment}
\begin{equation}
    \widetilde{F}^{(1)}_2 - \widetilde{F}^{(3)}_1 + \frac{1}{6} \widetilde{F}^{(3)}_2 - \frac{1}{2} (y_L^q)^2 \widetilde{F}^{(AG)}_1 \,, \; \text{or} \;\; \widetilde{F}^{(1)}_2 - \widetilde{F}^{(3)}_1 + \frac{1}{6} \widetilde{F}^{(3)}_2 - \frac{1}{2} (y_L^\ell)^2 \widetilde{F}^{(AG)}_1 \,.
\end{equation}
\end{comment}

\begin{comment}
\begin{equation}
    \mathcal{Q}_{11} - \mathcal{Q}_{12} + \frac{1}{6} \mathcal{Q}_{13} - \frac{1}{2} (y_L^q)^2 \mathcal{Q}_{14} \;\; \text{or} \;\; \mathcal{Q}_{11} - \mathcal{Q}_{12} + \frac{1}{6} \mathcal{Q}_{13} - \frac{1}{2} (y_L^\ell)^2 \mathcal{Q}_{14} \,.
\end{equation}
    
\end{comment}

\begin{equation}
    \mathcal{Q}_{11} - \mathcal{Q}_{12} - \frac{1}{2} (y_L^q)^2 \mathcal{Q}_{13} \;\; \text{or} \;\; \mathcal{Q}_{11} - \mathcal{Q}_{12} - \frac{1}{2} (y_L^\ell)^2 \mathcal{Q}_{13} \,.
\end{equation}

To provide a concrete example of a term present in the finite renormalization Lagrangian, we have

\begin{eqnarray}\label{eq:Lfct_gammamutauI}
    \mathcal{L}_{\rm fct} \supset && g_2 g_3^2 \frac{1}{6 \cdot 16 \pi^2} G^B_\alpha G^B_\mu W_\beta^I \\
    && \left[ C^{q q (3)}_{f i j k} \delta_{fi} \bar{q}_j S^{\alpha \mu \beta} T^I_L q_k + C^{q q (3)}_{f i j k} \delta_{jk} \bar{q}_f S^{\alpha \mu \beta} T^I_L q_i + C^{\ell q (3)}_{f i j k} \delta_{jk} \bar{\ell}_f S^{\alpha \mu \beta} T^I_L \ell_i \right. \nonumber\\
    && + C^{q q (1)}_{f i j k} \frac{1}{2 N_c} \delta_{fk} \bar{q}_j S^{\alpha \mu \beta} T^I_L q_i + C^{q q (1)}_{f i j k} \frac{1}{2 N_c} \delta_{ji} \bar{q}_f S^{\alpha \mu \beta} T^I_L q_k \nonumber\\
    && \left. - C^{q q (3)}_{f i j k} \frac{1}{2 N_c} \delta_{fk} \bar{q}_j S^{\alpha \mu \beta} T^I_L q_i - C^{q q (3)}_{f i j k} \frac{1}{2 N_c} \delta_{ji} \bar{q}_f S^{\alpha \mu \beta} T^I_L q_k \right] + {\rm h.c.} \,, \nonumber
\end{eqnarray}
where $ S^{\alpha \mu \beta} = \bar{\gamma}^\alpha (\bar{\gamma}^\mu \bar{\gamma}^\beta + \bar{\gamma}^\beta \bar{\gamma}^\mu) + (\bar{\gamma}^\mu \bar{\gamma}^\alpha \bar{\gamma}^\beta + \bar{\gamma}^\beta \bar{\gamma}^\alpha \bar{\gamma}^\mu) + (\bar{\gamma}^\mu \bar{\gamma}^\beta + \bar{\gamma}^\beta \bar{\gamma}^\mu) \bar{\gamma}^\alpha $.
This example also displays a lepton-gluonic counterterm.

\begin{comment}
        \begin{tabular}{|cc||c|c|c|c|c|}
        \hline
        $ \mathcal{O} $ & & $h^{\mathcal{O}}_{11}$ & $h^{\mathcal{O}}_{12}$ & $h^{\mathcal{O}}_{13}$ & $h^{\mathcal{O}}_{14}$ & $h^{\mathcal{O}}_{15}$ \\
        \hline
        \hline
        $ Q^{(3)}_{q q} $ & $ \delta_{fi}, \delta_{jk} $ & \multirow{2}{*}{$ 4 N_c $} & \multirow{2}{*}{$ - 4 N_c $} & \multirow{2}{*}{$ \frac{2}{3} N_c $} & \multirow{2}{*}{$ -2 N_c (y_L^q)^2 $} & \multirow{2}{*}{$ -1 $} \\
        $ Q^{(3)}_{\ell q} $ & $ \delta_{jk} $ & & & & & \\
        \hline
        $ Q^{(3)}_{\ell q} $ & $ \delta_{fi} $ & $ 4 $ & $ -4 $ & $ \frac{2}{3} $ & $ -2 (y_L^\ell)^2 $ & $ 0 $ \\
        \hline
        \hline
        $ Q^{(1)}_{q q} $ & $ \delta_{fk}, \delta_{ji} $ & $ 2 $ & $ -2 $ & $ \frac{1}{3} $ & $ - (y_L^q)^2 $ & $ - \frac{1}{2 N_c} $ \\
        \hline
        $ Q^{(3)}_{q q} $ & $ \delta_{fk}, \delta_{ji} $ & $ -2 $ & $ 2 $ & $ -\frac{1}{3} $ & $ (y_L^q)^2 $ & $ \frac{1}{2 N_c} $ \\
        \hline
        $ Q_{\ell \ell} $ & $ \delta_{fk}, \delta_{ji} $ & $ 2 $ & $ -2 $ & $ \frac{1}{3} $ & $ - (y_L^\ell)^2 $ & $ 0 $ \\
        \hline
    \end{tabular}
\end{comment}

\begin{table}[]
    \centering
    \begin{tabular}{|cc||c|c|c|c|}
    \hline
    $ \mathcal{O} $ & 
    & $h^{\mathcal{O}}_{11}$ 
    & $h^{\mathcal{O}}_{12}$ 
    & $h^{\mathcal{O}}_{13}$ 
    & $h^{\mathcal{O}}_{14}$ \\
    \hline
    \hline
    $ Q^{(3)}_{q q} $ 
    & $ \delta_{fi}, \delta_{jk} $ 
    & \multirow{2}{*}{$ 4 N_c $} 
    & \multirow{2}{*}{$ - 4 N_c $} 
    & \multirow{2}{*}{$ -2 N_c (y_L^q)^2 $} 
    & \multirow{2}{*}{$ -1 $} \\
    $ Q^{(3)}_{\ell q} $ 
    & $ \delta_{jk} $ 
    &  &  &  &  \\
    \hline
    $ Q^{(3)}_{\ell q} $ 
    & $ \delta_{fi} $ 
    & $ 4 $ 
    & $ -4 $ 
    & $ -2 (y_L^\ell)^2 $ 
    & $ 0 $ \\
    \hline
    \hline
    $ Q^{(1)}_{q q} $ 
    & $ \delta_{fk}, \delta_{ji} $ 
    & $ 2 $ 
    & $ -2 $ 
    & $ - (y_L^q)^2 $ 
    & $ - \frac{1}{2 N_c} $ \\
    \hline
    $ Q^{(3)}_{q q} $ 
    & $ \delta_{fk}, \delta_{ji} $ 
    & $ -2 $ 
    & $ 2 $ 
    & $ (y_L^q)^2 $ 
    & $ \frac{1}{2 N_c} $ \\
    \hline
    $ Q_{\ell \ell} $ 
    & $ \delta_{fk}, \delta_{ji} $ 
    & $ 2 $ 
    & $ -2 $ 
    & $ - (y_L^\ell)^2 $ 
    & $ 0 $ \\
    \hline
\end{tabular}
    \caption{
    Coefficients multiplying the counterterms in the finite counterterm Lagrangian, which present a fermion bilinear of structure $ \bar{\xi} \bar{\gamma}_\mu T_L^I \xi $.
    The fermion generation indices of the four-fermion operators of the Warsaw basis are $f i j k$, in this same order. The operators and the fermions contracted in the loop can be read from the first two columns.
    }
    \label{tab:gammaTL_gammaTL}
\end{table}

\subsubsection{$\gamma_\mu T^A$-structure counterterms}\label{sec:gamma_GellMann_structure_CTs}

Analogously to Sec.~\ref{sec:gamma_Pauli_structure_CTs}, we now discuss fermion bilinears of structure $\bar{\xi} \bar{\gamma}^\mu T^A \xi$ (while cases with three $\gamma$-matrices are not simplified for convenience).
Obstructions are only possible when at least one $SU(3)$ gauge boson is radiated.
We define the following operator with a single gauge boson
\begin{eqnarray}
    \mathcal{Q}_{15} = - g_3 \frac{1}{2} \bar{\partial}^2 G^A_\mu ( \bar{\xi} \bar{\gamma}^\mu T^A \xi )
    + g_3 \bar{\partial}_\mu \bar{\partial}^\nu G^A_\nu ( \bar{\xi} \bar{\gamma}^\mu T^A \xi ) \,, \label{eq:F21} %F^{(1)}_2
\end{eqnarray}
followed by an operator with two gauge bosons

\begin{eqnarray}
    \mathcal{Q}_{16} = \frac{1}{2} g_3^2 D_{A B C} \, [\bar{\xi} \bar{\gamma}_\mu T^A \xi] G^C_\sigma \bar{\partial}_\rho G^B_\nu \epsilon^{\mu \nu \rho \sigma} \,; \label{eq:F42} %F^{(2)}_4
\end{eqnarray}
there are many more structures with three gauge bosons

\begin{eqnarray}
    \mathcal{Q}_{17} = g_3^3 {\rm Tr} \{ T^A T^B T^C T^D \} G^B_\alpha G^C_\mu G^D_\beta ( \bar{\xi} \bar{\gamma}^\alpha \bar{\gamma}^\mu \bar{\gamma}^\beta T^A \xi ) \,, \label{eq:F13} %F^{(3)}_1
\end{eqnarray}

\begin{equation}
    \mathcal{Q}_{18} = i g_3^3 ( \bar{\xi} \bar{\gamma}_\mu T^A \xi ) G^B_\nu G^C_\rho G^D_\sigma \epsilon^{\mu \nu \rho \sigma} \mathcal{A}_{A}^{B C D} \,, \label{eq:F23} %F^{(3)}_2
\end{equation}
where

\begin{eqnarray}\label{eq:ASU3_definition}
    && \mathcal{A}_{A}^{B C D} = {\rm Tr} \{ T^A T^D ( T^B T^C - T^C T^B ) \} \\
    && - {\rm Tr} \{ T^A ( T^B T^D T^C - T^C T^D T^B ) \} + {\rm Tr} \{ T^A ( T^B T^C - T^C T^B ) T^D \} \nonumber\\
    && = \frac{1}{4} i \left( C_{BCE} D_{ADE} - C_{DCE} D_{ABE} - C_{BDE} D_{ACE} \right) \,, \nonumber
\end{eqnarray}
which is totally antisymmetric
in $B, C, D$, and

\begin{eqnarray}
    \mathcal{Q}_{19} = - g_3^3 \frac{i}{8} ( \bar{\xi} \bar{\gamma}^\alpha \bar{\gamma}^\mu \bar{\gamma}^\beta T^A \xi ) \left( G^B_\alpha G^C_\mu G^D_\beta + G^C_\alpha G^B_\mu G^D_\beta + G^C_\alpha G^D_\mu G^B_\beta \right) C_{C D E} D_{A B E} \,, \nonumber\\
    \label{eq:F13DC} %F^{(3DC)}_1
\end{eqnarray}

\begin{eqnarray}
    \mathcal{Q}_{20} = g_3^3 ( \bar{\xi} \bar{\gamma}_\mu T^A \xi ) G^B_\nu G^C_\rho G^D_\sigma \epsilon^{\mu \nu \rho \sigma} D_{A D E} C_{B C E} \,,\label{eq:F33} %F^{(3)}_3
\end{eqnarray}

\begin{eqnarray}
    && \mathcal{Q}_{21} = g_1^2 g_3 B_\alpha B_\mu G_\beta^C \bar{\xi} ( \bar{\gamma}^\beta \bar{\gamma}^\alpha \bar{\gamma}^\mu + \bar{\gamma}^\alpha \bar{\gamma}^\beta \bar{\gamma}^\mu + \bar{\gamma}^\alpha \bar{\gamma}^\mu \bar{\gamma}^\beta ) T^C \xi \,, \label{eq:F1AG}\\ %F^{(AG)}_1
    && \mathcal{Q}_{22} = g_2^3 g_3 W^b_\alpha W^b_\mu G_\beta^C \bar{\xi} ( \bar{\gamma}^\beta \bar{\gamma}^\alpha \bar{\gamma}^\mu + \bar{\gamma}^\alpha \bar{\gamma}^\beta \bar{\gamma}^\mu + \bar{\gamma}^\alpha \bar{\gamma}^\mu \bar{\gamma}^\beta ) T^C \xi \,. \label{eq:F2AG} %F^{(AG)}_2
\end{eqnarray}
The operator $ \mathcal{Q}_{19} $ can be simplified with the use of the Chisholm identity, and only the term proportional to $i$ times the Levi-Civita symbol remains, with an overall sign that depends on the chirality of $\xi$; in this way, it can be seen that this operator self-conjugates when the omitted generation indices of the fermion bilinear are the same. Using the same identity, the operator $\mathcal{Q}_{17}$ (and the analogous operator $\mathcal{Q}_{12}$ for $SU(2)_L$) can also be shown to self-conjugate when the two fermions carry the same flavour (with the help of the identity Eq.~\eqref{eq:identity_Haber} shown in App.~\ref{app:example}).

The finite counterterm Lagrangian is

\begin{comment}
\begin{eqnarray}
    \mathcal{L}_{\rm fct} \supset && - \frac{C^{\mathcal{O}}_{\xi \xi}}{3 \cdot 16 \pi^2} \left( f^{\mathcal{O} (1)}_2 F^{(1)}_2 + f^{\mathcal{O} (3)}_1 F^{(3)}_1 + f^{\mathcal{O} (3)}_2 F^{(3)}_2 + f^{\mathcal{O} (3DC)}_1 F^{(3DC)}_1 \right. \\
    && \qquad \left. + f^{\mathcal{O} (2)}_4 F^{(2)}_4 + f^{\mathcal{O} (3)}_3 F^{(3)}_3 + f^{\mathcal{O} (AG)}_1 F^{(AG)}_1 + f^{\mathcal{O} (AG)}_2 F^{(AG)}_2 \right) + {\rm h.c.} \nonumber
\end{eqnarray}
\end{comment}

\begin{equation}
    \mathcal{L}_{\rm fct} \supset - \frac{C^{\mathcal{O}}_{\xi \xi}}{3 \cdot 16 \pi^2} \sum^{22}_{i=15} h^{\mathcal{O}}_{i} \, \mathcal{Q}_i + {\rm h.c.}
\end{equation}

\noindent
Coefficients are given in Tab.~\ref{tab:gammaT_gammaT}.
In this table, the first and second sectors of the first half correspond to $q$ and $x = u, d$ loops, respectively, when contracting the fermion fields that appear within the same bilinear, following the convention of Tab.~\ref{tab:dimension_6}; the second half of the table corresponds to contracting fermions in different bilinears, again following the convention of Tab.~\ref{tab:dimension_6}.
Due to the identities Eqs.~\eqref{eq:GellMann_identity_1} and \eqref{eq:GellMann_identity_2},
operators other than $ Q^{(8)}_{q x}, Q^{(8)}_{u d} $ from the Warsaw basis also require finite renormalization by $ \bar{\gamma}_\mu T^A $ structures.
We find that some of the counterterms previously defined in Eqs.~\eqref{eq:F21}-\eqref{eq:F2AG} appear in specific combinations in the renormalization

\begin{comment}
\begin{eqnarray}
    && F^{(2)}_4 + \frac{1}{12} F^{(3)}_2 - \frac{1}{8} F^{(3)}_3  \,, \\
    && F^{(1)}_2 - F^{(3)}_1 - F^{(3DC)}_1 - \frac{1}{2} (y_L^q)^2 F^{(AG)}_1 \,, \; \text{or} \;\; F^{(1)}_2 - F^{(3)}_1 - F^{(3DC)}_1 - \frac{1}{2} (y_R^x)^2 F^{(AG)}_1 \,. \nonumber\\
\end{eqnarray}
\end{comment}

\begin{eqnarray}
    && \mathcal{Q}_{16} + \frac{1}{12} \mathcal{Q}_{18} - \frac{1}{8} \mathcal{Q}_{20}  \,, \\
    && \mathcal{Q}_{15} - \mathcal{Q}_{17} - \mathcal{Q}_{19} - \frac{1}{2} (y_L^q)^2 \mathcal{Q}_{21} \,, \; \text{or} \;\; \mathcal{Q}_{15} - \mathcal{Q}_{17} - \mathcal{Q}_{19} - \frac{1}{2} (y_R^x)^2 \mathcal{Q}_{21} \,. \nonumber\\
\end{eqnarray}

The following term provides a concrete example:

\begin{eqnarray}
    \mathcal{L}_{\rm fct} \supset && \frac{1}{3 \cdot 16 \pi^2} \left[ g_3^2 D_{A B C} G^B_\nu \bar{\partial}_\rho G^C_\sigma + g_3^3 G^B_\nu G^C_\rho G^D_\sigma \left( \frac{1}{4} D_{A D E} C_{B C E} - \frac{1}{6} i \mathcal{A}_{A}^{B C D} \right) \right] \nonumber\\
    && \epsilon^{\mu \nu \rho \sigma} \left[ C_{f i j k}^{q u (8)} \delta_{fi} \bar{u}_j \bar{\gamma}_\mu T^A u_k + C_{f i j k}^{q d (8)} \delta_{fi} \bar{d}_j \bar{\gamma}_\mu T^A d_k \right. \nonumber\\
    %%%%%%%
    && - C_{f i j k}^{q u (8)} \delta_{jk} \bar{q}_f \bar{\gamma}_\mu T^A q_i - C_{f i j k}^{q d (8)} \delta_{jk} \bar{q}_f \bar{\gamma}_\mu T^A q_i \nonumber\\
    %%%%%%%
    && - C_{f i j k}^{u d (8)} \delta_{fi} \bar{d}_j \bar{\gamma}_\mu T^A d_k - C_{f i j k}^{u d (8)} \delta_{jk} \bar{u}_f \bar{\gamma}_\mu T^A u_i \nonumber\\
    %%%%%%%
    && + C_{f i j k}^{q q (1)} \delta_{fk} \bar{q}_j \bar{\gamma}_\mu T^A q_i + C_{f i j k}^{q q (1)} \delta_{ji} \bar{q}_f \bar{\gamma}_\mu T^A q_k \nonumber\\
    %%%%%%%
    && + C_{f i j k}^{q q (3)} 3 \delta_{fk} \bar{q}_j \bar{\gamma}_\mu T^A q_i + C_{f i j k}^{q q (3)} 3 \delta_{ji} \bar{q}_f \bar{\gamma}_\mu T^A q_k \nonumber\\
    %%%%%%%
    && - C^{uu}_{f i j k} 2 \delta_{fk} \bar{u}_j \bar{\gamma}_\mu T^A u_i - C^{dd}_{f i j k} 2 \delta_{fk} \bar{d}_j \bar{\gamma}_\mu T^A d_i \nonumber\\
    && \left. - C^{uu}_{f i j k} 2 \delta_{ji} \bar{u}_f \bar{\gamma}_\mu T^A u_k - C^{dd}_{f i j k} 2 \delta_{ji} \bar{d}_f \bar{\gamma}_\mu T^A d_k \right] + {\rm h.c.}
\end{eqnarray}
This structure is similar to the one of lower mass dimensionality found when computing the allowed possible forms of true anomalies in the SM. Although similar, a crucial difference is that in SMEFT the obstruction can be compensated for by finite counterterms.

\begin{comment}
        \begin{tabular}{|cc||c|c|c|c|c|c|c|c|}
        \hline
        $ \mathcal{O} $ & & $h^{\mathcal{O}}_{16}$ & $h^{\mathcal{O}}_{18}$ & $h^{\mathcal{O}}_{19}$ & $h^{\mathcal{O}}_{20}$ & $h^{\mathcal{O}}_{17}$ & $h^{\mathcal{O}}_{21}$ & $h^{\mathcal{O}}_{22}$ & $h^{\mathcal{O}}_{23}$ \\
        \hline
        \hline
        $Q_{qx}^{(8)}$ & $\delta_{fi}$ & $2$ & $-2$ & $\frac{1}{3}$ & $-2$ & $4$ & $-\frac{1}{2}$ & $- (y_L^q)^2$ & $-\frac{1}{4}$ \\
        \hline
        $Q_{qx}^{(8)}$ & $\delta_{jk}$ & \multirow{2}{*}{$1$} & \multirow{2}{*}{$-1$} & \multirow{2}{*}{$-\frac{1}{6}$} & \multirow{2}{*}{$-1$} & \multirow{2}{*}{$-2$} & \multirow{2}{*}{$\frac{1}{4}$} & \multirow{2}{*}{$-\frac{1}{2} (y_R^x)^2$} & \multirow{2}{*}{${\color{red}0}$} \\
        $Q_{ud}^{(8)}$ & $\delta_{fi}, \delta_{jk}$ & & & & & & & & \\
        \hline
        \hline
        $Q_{qq}^{(1)}$ & $\delta_{fk}, \delta_{ji}$ & $2$ & $-2$ & $\frac{1}{3}$ & $-2$ & $4$ & $-\frac{1}{2}$ & $- (y_L^q)^2$ & $-\frac{1}{4}$ \\
        \hline
        $Q_{qq}^{(3)}$ & $\delta_{fk}, \delta_{ji}$ & $6$ & $-6$ & $1$ & $-6$ & $12$ & $-\frac{3}{2}$ & $-3 (y_L^q)^2$ & $-\frac{3}{4}$ \\
        \hline
        $Q_{xx}$ & $\delta_{fk}, \delta_{ji}$ & $2$ & $-2$ & $-\frac{1}{3}$ & $-2$ & $-4$ & $\frac{1}{2}$ & $- (y_R^x)^2$ & ${\color{red}0}$ \\
        \hline
    \end{tabular}
\end{comment}

\begin{table}[]
    \centering
    \begin{tabular}{|cc||c|c|c|c|c|c|c|c|}
    \hline
    $ \mathcal{O} $ & 
    & $h^{\mathcal{O}}_{15}$ 
    & $h^{\mathcal{O}}_{16}$ 
    & $h^{\mathcal{O}}_{17}$ 
    & $h^{\mathcal{O}}_{18}$ 
    & $h^{\mathcal{O}}_{19}$ 
    & $h^{\mathcal{O}}_{20}$ 
    & $h^{\mathcal{O}}_{21}$ 
    & $h^{\mathcal{O}}_{22}$ \\
    \hline
    \hline
    $Q_{qx}^{(8)}$ & $\delta_{fi}$ 
    & $2$ & $4$ & $-2$ & $\frac{1}{3}$ & $-2$ 
    & $-\frac{1}{2}$ & $- (y_L^q)^2$ & $-\frac{1}{4}$ \\
    \hline
    $Q_{qx}^{(8)}$ & $\delta_{jk}$ 
    & \multirow{2}{*}{$1$} 
    & \multirow{2}{*}{$-2$} 
    & \multirow{2}{*}{$-1$} 
    & \multirow{2}{*}{$-\frac{1}{6}$} 
    & \multirow{2}{*}{$-1$} 
    & \multirow{2}{*}{$\frac{1}{4}$} 
    & \multirow{2}{*}{$-\frac{1}{2} (y_R^x)^2$} 
    & \multirow{2}{*}{$0$} \\
    $Q_{ud}^{(8)}$ & $\delta_{fi}, \delta_{jk}$ 
    & & & & & & & & \\
    \hline
    \hline
    $Q_{qq}^{(1)}$ & $\delta_{fk}, \delta_{ji}$ 
    & $2$ & $4$ & $-2$ & $\frac{1}{3}$ & $-2$ 
    & $-\frac{1}{2}$ & $- (y_L^q)^2$ & $-\frac{1}{4}$ \\
    \hline
    $Q_{qq}^{(3)}$ & $\delta_{fk}, \delta_{ji}$ 
    & $6$ & $12$ & $-6$ & $1$ & $-6$ 
    & $-\frac{3}{2}$ & $-3 (y_L^q)^2$ & $-\frac{3}{4}$ \\
    \hline
    $Q_{xx}$ & $\delta_{fk}, \delta_{ji}$ 
    & $2$ & $-4$ & $-2$ & $-\frac{1}{3}$ & $-2$ 
    & $\frac{1}{2}$ & $- (y_R^x)^2$ & $0$ \\
    \hline
    \end{tabular}
    \caption{
    Coefficients of the finite counterterm Lagrangian, for fermion bilinears of structure $ \bar{\xi} \bar{\gamma}_\mu T^A \xi $.
    The fermion generation indices of the four-fermion operators of the Warsaw basis are $f i j k$, in this same order. The operators and the fermions contracted in the loop can be read from the first two columns.
    In the case of $ Q^{(8)}_{u d} $, $x = u$ for $\delta_{fi}$ and $x = d$ for $\delta_{jk}$.
    }
    \label{tab:gammaT_gammaT}
\end{table}

\subsubsection{$\gamma_\mu \tau^I T^A$-structure counterterms}\label{sec:gamma_Pauli_GellMann_structure_CTs}

Finally, there are special structures of the form $\bar{\gamma}_\mu \tau^I T^A$, which thus require at least one $SU(2)_L$, and at least one $SU(3)$ external gauge boson.
The required counterterms, having two or one gauge bosons, are

\begin{eqnarray}
    && \mathcal{Q}_{23} = g_1 g_2 g_3 W^a_\alpha G^B_\mu B_\beta \bar{\xi} S^{\beta \alpha \mu} T^B T^a_L \xi \,, \label{eq:TS111}\\ %T^{(111)}_S
    && \mathcal{Q}_{24} = g_2 g_3^2 {\rm Tr} \{ T^A (T^B T^C + T^C T^B) \} {\rm Tr} \{ T^a_L T^b_L \} \, G^B_\nu \, W_\rho^b \, G^C_\sigma ( \bar{\xi} S^{\sigma \nu \rho} T^A T^a_L \xi ) \,, \label{eq:TS12}\\ %T^{(12)}_S
    && \mathcal{Q}_{25} = g_2 g_3 ( \bar{\xi} \bar{\gamma}_\mu T^A T^a_L \xi ) \bar{\partial}_\rho G^A_\nu W^a_\sigma \epsilon^{\mu \nu \rho \sigma} \,, \label{eq:T1eps11}\\ %T^{(11)}_{1 \epsilon}
    && \mathcal{Q}_{26} = g_2 g_3^2 C_{A B C} ( \bar{\xi} \bar{\gamma}_\mu T^A T^a_L \xi ) W^a_\nu G^C_\rho G^B_\sigma \epsilon^{\mu \nu \rho \sigma} \,, \label{eq:T1eps12}\\ %T^{(12)}_{1 \epsilon}
    && \mathcal{Q}_{27} = g_2 g_3 ( \bar{\xi} \bar{\gamma}_\mu T^A T^a_L \xi ) \bar{\partial}_\rho W^a_\nu G^A_\sigma \epsilon^{\mu \nu \rho \sigma} \,, \label{eq:T2eps11}\\ %T^{(11)}_{2 \epsilon}
    && \mathcal{Q}_{28} = g_2^2 g_3 C^{SU(2)}_{a b c} ( \bar{\xi} \bar{\gamma}_\mu T^A T^a_L \xi ) G^A_\nu W^c_\rho W^b_\sigma \epsilon^{\mu \nu \rho \sigma} \,. \label{eq:T2eps21} %T^{(21)}_{2 \epsilon}
\end{eqnarray}
where $S$ is defined after Eq.~\eqref{eq:Lfct_gammamutauI}.

In terms of these operators, the finite counterterm Lagrangian is

\begin{comment}
\begin{eqnarray}
    \mathcal{L}_{\rm fct} \supset && - \frac{C^{\mathcal{O}}_{\xi \xi}}{3 \cdot 16 \pi^2} \left( g^{\mathcal{O}(111)} T^{(111)}_S + g^{\mathcal{O}(12)} T^{(12)}_S \right. \\
    && \left. + g^{\mathcal{O}(11)}_\epsilon T^{(11)}_{1 \epsilon} + g^{\mathcal{O}(12)}_\epsilon T^{(12)}_{1 \epsilon} + h^{\mathcal{O}(11)}_\epsilon T^{(11)}_{2 \epsilon} + h^{\mathcal{O}(21)}_\epsilon T^{(21)}_{2 \epsilon} \right) + {\rm h.c.} \,, \nonumber
\end{eqnarray}
\end{comment}

\begin{equation}
    \mathcal{L}_{\rm fct} \supset - \frac{C^{\mathcal{O}}_{\xi \xi}}{3 \cdot 16 \pi^2} \sum^{28}_{i=23} h^{\mathcal{O}}_{i} \, \mathcal{Q}_i + {\rm h.c.} \,,
\end{equation}
where the hermitian conjugation, as in all previous cases, applies only in cases where the fermion fields do not carry the same flavour.
Coefficients are given in Tab.~\ref{tab:gammaTTL_gammaTTL}.
It turns out that all the structures previously defined appear in the same combination

\begin{comment}
\begin{equation}
    y_L^q T^{(111)}_S
+ T^{(12)}_S
- 4 \left( T^{(11)}_{1 \epsilon} + T^{(11)}_{2 \epsilon} \right)
- \frac{3}{2} \left( T^{(12)}_{1 \epsilon} + T^{(21)}_{2 \epsilon} \right) \,.
\end{equation}
\end{comment}

\begin{equation}
    y_L^q \mathcal{Q}_{23}
+ \mathcal{Q}_{24}
- 4 \left( \mathcal{Q}_{25} + \mathcal{Q}_{27} \right)
- \frac{3}{2} \left( \mathcal{Q}_{26} + \mathcal{Q}_{28} \right) \,.
\end{equation}

To better explain the notation in use,
we have the following example

\begin{eqnarray}
    \mathcal{L}_{\rm fct} \supset && y^q_L \frac{g_1 g_2 g_3}{3 \cdot 16 \pi^2} W^a_\alpha G^B_\mu B_\beta
    \left[ C^{qq (1)}_{f i j k} \delta_{f k} \bar{q}_j S^{\alpha \mu \beta} T^B T^a_L q_i + C^{qq (1)}_{f i j k} \delta_{j i} \bar{q}_f S^{\alpha \mu \beta} T^B T^a_L q_k \right. \nonumber\\
    %%%%%%%%%%
    && \left. - C^{qq (3)}_{f i j k} \delta_{f k} \bar{q}_j S^{\alpha \mu \beta} T^B T^a_L q_i - C^{qq (3)}_{f i j k} \delta_{j i} \bar{q}_f S^{\alpha \mu \beta} T^B T^a_L q_k \right] + {\rm h.c.} \,,
\end{eqnarray}
which has three different gauge bosons.

\begin{table}[]
    \centering
    \begin{tabular}{|cc||c|c|c|c|c|c|}
        \hline
        $ \mathcal{O} $ & & $h^{\mathcal{O}}_{23}$ & $h^{\mathcal{O}}_{24}$ & $h^{\mathcal{O}}_{25}$ & $h^{\mathcal{O}}_{26}$ & $h^{\mathcal{O}}_{27}$ & $h^{\mathcal{O}}_{28}$ \\
        \hline
        \hline
        $ Q^{(1)}_{qq} $ & $ \delta_{f k}, \delta_{j i} $ & $-y_L^q$ & $-1$ & $4$ & $\frac{3}{2}$ & $4$ & $\frac{3}{2}$ \\
        \hline
        $ Q^{(3)}_{qq} $ & $ \delta_{f k}, \delta_{j i} $ & $y_L^q$ & $1$ & $-4$ & $-\frac{3}{2}$ & $-4$ & $-\frac{3}{2}$ \\
        \hline
    \end{tabular}
    \caption{
    Coefficients of the finite counterterm Lagrangian, for fermion bilinears of generic structure $ \bar{\xi} \bar{\gamma}_\mu T_L^I T^A \xi $.
    The fermion generation indices of the four-fermion operators of the Warsaw basis are $f i j k$, in this same order. The operators and the fermions contracted in the loop can be read from the first two columns.
    }
    \label{tab:gammaTTL_gammaTTL}
\end{table}

\subsection{One scalar}\label{sec:one_scalar_results}

\subsubsection{Amplitudes}

\begin{figure}[H]
    \begin{minipage}{.245\textwidth}
        \centering
        \includegraphics[scale=0.175]{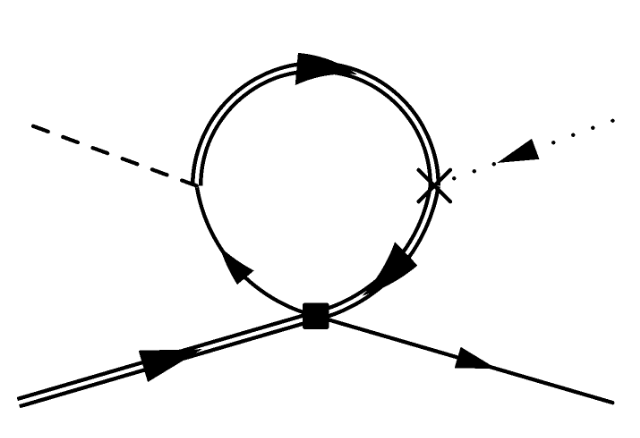} \\
        $ + \ldots $
    \end{minipage}
    \begin{minipage}{.745\textwidth}
        \centering
        \begin{equation}\label{eq:1scalar_0GB}
            = -\frac{1}{3 \cdot 16 \pi^2} C^{\mathcal{O},F}_{\psi \xi} \, Y^\dagger [ \bar{\psi} \{ (C^g_L - C^g_R) C_6^{\mathcal{O}} \}_\mathcal{O} P_X \xi h ] \overline{k}_g \cdot ( \overline{k}_g + 2 \overline{k}_h ) c \,,
        \end{equation}
    \end{minipage}
\end{figure}

\begin{figure}[H]
    \begin{minipage}{.15\textwidth}
        \centering
        \includegraphics[scale=0.15]{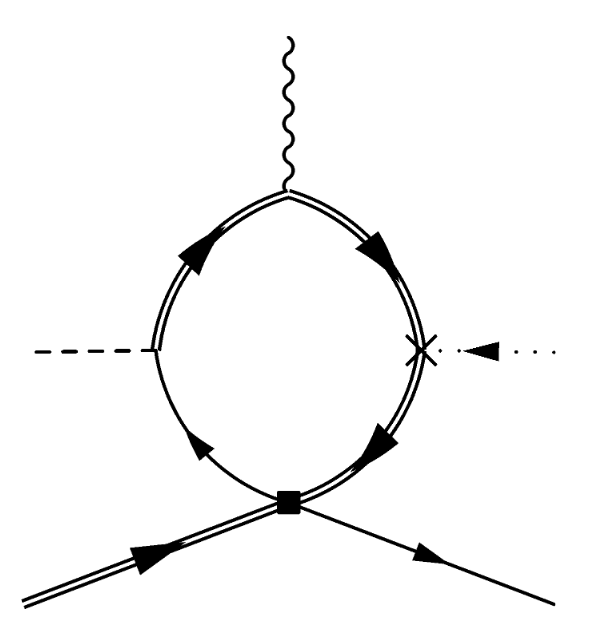} \\
        $ + \ldots $
    \end{minipage}
    \begin{minipage}{.84\textwidth}
        \centering
        \begin{eqnarray}\label{eq:1scalar_1GB}
            && = -\frac{1}{3 \cdot 16 \pi^2} C^{\mathcal{O},F}_{\psi \xi} \, Y^\dagger [ \bar{\psi} \{ (- C_L^g C_R^v - C_L^v C_R^g + 2 C_L^g C_L^v + 2 C_L^v C_L^g \nonumber\\
            && + 2 C_R^g C_R^v + 2 C_R^v C_R^g) C_6^{\mathcal{O}} \}_\mathcal{O} P_X \xi h ] \overline{k}_g^\mu A_\mu c \,,
        \end{eqnarray}
    \end{minipage}
\end{figure}

\begin{figure}[H]
    \begin{minipage}{.15\textwidth}
        \centering
        \includegraphics[scale=0.15]{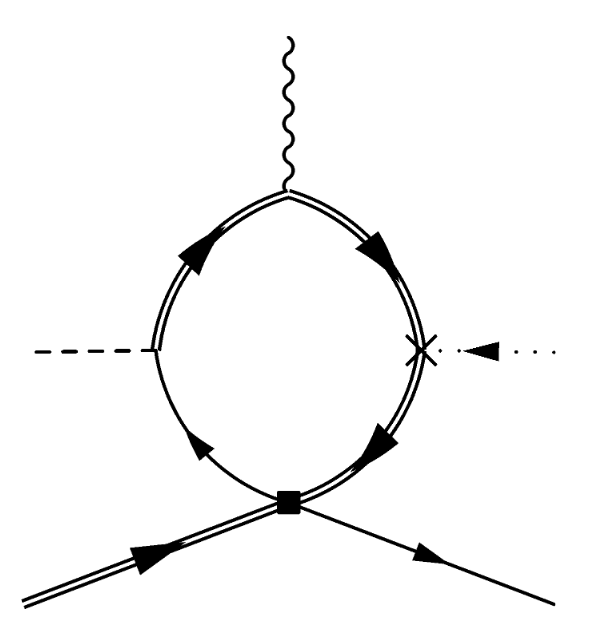} \\
        $ + \ldots $
    \end{minipage}
    \begin{minipage}{.84\textwidth}
        \centering
        \begin{eqnarray}\label{eq:1scalar_sigma_1GB}
            && = i \frac{4}{3 \cdot 16 \pi^2} C^{\mathcal{O},F}_{\psi \xi} \, Y^\dagger [ \bar{\psi} \{ (- C_L^g C_R^v + C_L^v C_R^g + 2 C_L^g C_L^v \nonumber\\
            && - 2 C_R^v C_R^g) C_6^{\mathcal{O}} \}_\mathcal{O} \bar{\sigma}_{\mu \nu} P_X \xi h ] \overline{k}_g^\mu A^\nu c \,.
        \end{eqnarray}
    \end{minipage}
\end{figure}

In the case of Green's functions with a single scalar,
we find the amplitudes in the following Eqs.~\eqref{eq:1scalar_0GB}-\eqref{eq:1scalar_sigma_1GB}. Examples of diagrams are shown, where single (double) fermion lines represent right-handed (respectively, left-handed) fields, which should render the identification of chiralities of the different diagrams easier for the reader; diagrams where the ghost and the gauge boson couple to different internal fermion lines are understood in the ellipses. We have

%%%%%%%%%%%%%%%%%%%%%%%%%%%%%%%%%%%%%%%%%

\noindent
which vanishes when $ C^g_L = C^g_R $, as in the case of $SU(3)$. The incoming four-momenta $ \overline{k}_g $ and $ \overline{k}_h $ are the ones of the ghost and scalar, respectively.
The coefficients $C^{\mathcal{O},F}_{\psi \xi}$ collect the Wilson coefficients and pre-factors from the Fierz transformation used to put all relevant operators into a similar form (i.e., two fermion bilinears, each with a change of chirality), thus leading to a common calculation for the Feynman diagrams of all operator insertions. Then,

%%%%%%%%%%%%%%%%%%%%%%%%%%%%%%%%%%%%%%%%%

\noindent
where $ C_L^v, C_L^g \to y_L, T^I_L, T^A $, and $ C_R^v, C_R^g \to y_R, 0, T^A $ for $ U(1)_Y, SU(2)_L, SU(3) $, respectively; accordingly, $ A_\mu $ ($c$) is the gauge boson ``$v$'' (ghost ``$g$'') with generator index omitted. A trace $ \{ \cdot \}_\mathcal{O} \to {\rm tr} \{ \cdot \} $ over \textit{color} indices applies, which depends on the operator $\mathcal{O}$ being considered, with $C_6^{\mathcal{O}} \to T^A$ for octet structures.

There are also structures in $\bar{\sigma}_{\mu \nu}$:

%%%%%%%%%%%%%%%%%%%%%%%%%%%%%%%%%%%%%%%%%
\begin{figure}[H]
    \begin{minipage}{.245\textwidth}
        \centering
        \includegraphics[scale=0.175]{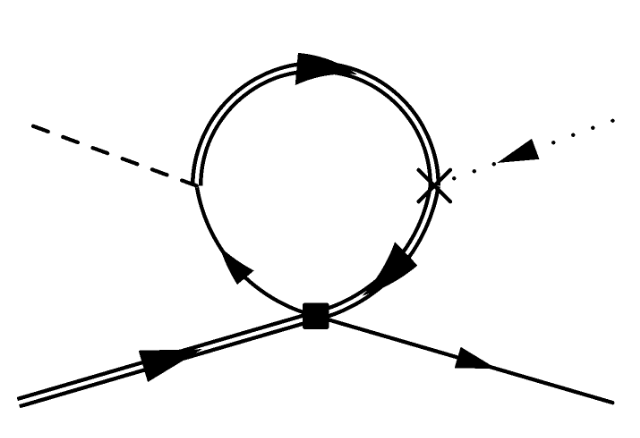} \\
        $ + \ldots $
    \end{minipage}
    \begin{minipage}{.745\textwidth}
        \centering
        \begin{equation}\label{eq:1scalar_sigma_0GB}
            = i \frac{8}{3 \cdot 16 \pi^2} C^{\mathcal{O},F}_{\psi \xi} \, Y^\dagger [ \bar{\psi} \{ (C^g_L + C^g_R) C_6^{\mathcal{O}} \}_\mathcal{O} \bar{\sigma}_{\mu \nu} P_X \xi h ] \overline{k}_g^\mu \overline{k}_h^\nu c \,,
        \end{equation}
    \end{minipage}
\end{figure}

%%%%%%%%%%%%%%%%%%%%%%%%%%%%%%%%%%%%%%%%%

\noindent
Terms with the antisymmetric Levi-Civita symbol contribute in an equal and constructive way to Eqs.~\eqref{eq:1scalar_sigma_0GB}-\eqref{eq:1scalar_sigma_1GB}.\footnote{
When considering cases carrying the anti-symmetric $\sigma_{\mu \nu} = i [\gamma_\mu, \gamma_\nu] / 2$ structure, we found an unexpected problem
while dealing with \texttt{Feyncalc} 9.2.0 (development version): for instance, the function \texttt{EpsChisholm} returns the same value for

\begin{center}
    $ \epsilon_{\mu \nu \alpha \beta} \bar{k}^\alpha_1 \bar{k}^\beta_2 \bar{\gamma}^{\mu} \bar{\gamma}^{\nu} P_R \;\; $ and $ \;\; \epsilon_{\mu \nu \alpha \beta} \bar{k}^\alpha_1 \bar{k}^\beta_2 \bar{\gamma}^{\nu} \bar{\gamma}^{\mu} P_R $,
\end{center}
which obviously should carry instead an opposite relative sign.
This issue has been corrected in subsequent versions, e.g., 10.0.0 (stable version).
Further discussion is found in Ref.~\cite{GitHub_FeynCalc}.
}

When considering the opposite chiralities (i.e., for insertions of the respective hermitian conjugated operators), Eqs.~\eqref{eq:1scalar_0GB} and \eqref{eq:1scalar_sigma_1GB} flip their overall signs.
This fact reflects in the need to introduce ghost number 0 counterterm operators below with an $i$ in the case of the $\bar{\sigma}_{\mu \nu}$ structure contracted with two gauge bosons.
Also consistent with the change of signs of Eqs.~\eqref{eq:1scalar_0GB} and \eqref{eq:1scalar_sigma_1GB}, when taking the conjugate of the operators below, note that the BRST transformation of the scalar and its conjugate carry a relative sign as per Eq.~\eqref{BRST_Transformation}.

\subsubsection{counterterms}

We can consider the following counterterm structures

\begin{equation}\label{eq:CT_Th}
    T_h = Y^\dagger \bar{\psi} P_X \xi \bar{\partial}^2 h \,,
\end{equation}

\begin{eqnarray}
    && T_{h1} = g_1 Y^\dagger \bar{\psi} \bar\sigma^{\mu \nu} P_X \xi B_\mu \bar{\partial}_\nu h \,, \label{eq:Th1}\\
    && T_{h2} = g_2 Y^\dagger \bar{\psi} T^a_L \bar\sigma^{\mu \nu} P_X \xi W^a_\mu \bar{\partial}_\nu h \,, \label{eq:Th2}\\
    && T_{h3} = g_3 Y^\dagger \bar{\psi} T^\alpha \bar\sigma^{\mu \nu} P_X \xi G^\alpha_\mu \bar{\partial}_\nu h \,, \label{eq:Th3}
\end{eqnarray}

\begin{eqnarray}
    && T^{11}_g = g_1^2 Y^\dagger (\bar{\psi} P_X \xi h) \bar{\eta}^{\mu \nu} B_\mu B_\nu \,, \label{eq:Tg11}\\
    && T^{22}_g = g_2^2 Y^\dagger (\bar{\psi} \mathcal{G}^{a b} P_X \xi h) \bar{\eta}^{\mu \nu} W^a_\mu W^b_\nu \,, \label{eq:Tg22}\\
    && T^{33}_g = g_3^2 Y^\dagger (\bar{\psi} \delta^{\alpha \beta} P_X \xi h) \bar{\eta}^{\mu \nu} G^\alpha_\mu G^\beta_\nu \,, \label{eq:Tg33}\\
    && \tilde{T}^{33}_g = g_3^2 Y^\dagger (\bar{\psi} \mathcal{G}^{\alpha \beta} P_X \xi h) \bar{\eta}^{\mu \nu} G^\alpha_\mu G^\beta_\nu \,, \label{eq:Ttildeg33}\\
    && T^{12}_g = g_1 g_2 Y^\dagger (\bar{\psi} T^a_L P_X \xi h) \bar{\eta}^{\mu \nu} B_\mu W^a_\nu \,, \label{eq:Tg12}\\
    && T^{13}_g = g_1 g_3 Y^\dagger (\bar{\psi} T^\alpha P_X \xi h) \bar{\eta}^{\mu \nu} B_\mu G^\alpha_\nu \,, \label{eq:Tg13}\\
    && T^{23}_g = g_2 g_3 Y^\dagger (\bar{\psi} T^a_L T^\alpha P_X \xi h) \bar{\eta}^{\mu \nu} W^a_\mu G^\alpha_\nu \,, \label{eq:Tg23}
\end{eqnarray}

\begin{eqnarray}
    && T^{22}_\sigma = i g_2^2 Y^\dagger (\bar{\psi} T^a_L T^b_L \bar\sigma^{\mu \nu} P_X \xi h) W^a_\mu W^b_\nu \,, \label{eq:T22sigma}\\
    && \tilde{T}^{33}_\sigma = i g_3^2 Y^\dagger (\bar{\psi} T^\alpha T^\beta \bar\sigma^{\mu \nu} P_X \xi h) G^\alpha_\mu G^\beta_\nu \,, \label{eq:Ttilde33sigma}\\
    && T^{12}_\sigma = i g_1 g_2 Y^\dagger (\bar{\psi} T^a_L \bar\sigma^{\mu \nu} P_X \xi h) B_\mu W^a_\nu \,, \label{eq:T12sigma}\\
    && T^{13}_\sigma = i g_1 g_3 Y^\dagger (\bar{\psi} T^\alpha \bar\sigma^{\mu \nu} P_X \xi h) B_\mu G^\alpha_\nu \,, \label{eq:T13sigma}\\
    && T^{23}_\sigma = i g_2 g_3 Y^\dagger (\bar{\psi} T^a_L T^\alpha \bar\sigma^{\mu \nu} P_X \xi h) W^a_\mu G^\alpha_\nu \,, \label{eq:T23sigma}
\end{eqnarray}

\noindent
where $ \mathcal{G}^{a b} = \frac{1}{2} ( T^a_L T^b_L + T^b_L T^a_L ) = \frac{1}{4} \delta^{a b} \mathbb{I}_2 $, with $T^a_L = \tau^a/2$ and where $\tau^a$ are the Pauli matrices;
$ \mathcal{G}^{\alpha \beta} = \frac{1}{2} ( T^\alpha T^\beta + T^\beta T^\alpha ) $, which is not proportional to the identity due to the contribution from the (totally) symmetric symbol $ D^{\alpha \beta \gamma} $.
We take the normalization $ {\rm tr} \{ T^\alpha T^\beta \} = \frac{1}{2} \delta^{\alpha \beta} $.
The BRST transformation of each one of the operators listed in Eqs.~\eqref{eq:CT_Th}-\eqref{eq:T23sigma} consists of a fermion bilinear, a ghost on which one or two derivatives act, one scalar field and up to one gauge boson.

A linear combination of these counterterms is considered to perform the finite renormalization in order to restore the Slavnov-Taylor identities at one-loop order:

\begin{eqnarray}\label{eq:Lfct_one_scalar}
    \mathcal{L}_{\rm fct} \supset && C^{\mathcal{O}}_{\psi \xi} \frac{1}{3 \cdot 16 \pi^2} ( \alpha_h^\mathcal{O} \, T_h + \beta_{ij}^\mathcal{O} \, T^{ij}_g + \tilde{\beta}_{33}^\mathcal{O} \, \tilde{T}^{33}_g ) \nonumber \\
    && \quad + C^{\mathcal{O}}_{\psi \xi} \frac{8}{3 \cdot 16 \pi^2} ( \alpha_{hi}^\mathcal{O} \, T_{hi} + \zeta_{mn}^\mathcal{O} \, T^{mn}_\sigma + \tilde{\zeta}_{33}^\mathcal{O} \, \tilde{T}^{33}_\sigma ) + {\rm h.c.} \,,
\end{eqnarray}
where $i=1,2,3$, and $ij=11, 12, 13, 22, 23, 33$, $mn = 12, 13, 22, 23$; $C^{\mathcal{O}}_{\psi \xi}$ is the Wilson coefficient.
A sum over SMEFT dimension-6 operators and different possible contractions of their fermion fields via the scalar coupling is implicit.
Eq.~\eqref{eq:Lfct_one_scalar} is schematic in the sense that it does not explicitly display the identities of the fermions $\psi, \xi$ (whether left- or right-handed quarks or leptons), nor the dependence on the scalar field $h$ (whether $H$ or $\tilde{H}$, or their complex conjugates) and the Yukawa matrix (whether $Y_d, Y_u, Y_e, Y_\nu$, or their complex conjugates). We will come back shortly to this point.
The values of the coefficients are given in Tabs.~\ref{tab:CTs_one_scalar_1} and \ref{tab:CTs_one_scalar_2}.
It is convenient to define $ y^{i j} = 2 (y^i_L)^2 - y^i_L \, y^j_R + 2 (y^j_R)^2 $.
In deriving these results, we use that $ y_h = y_L - y_R $, where $ y_h = 1/2 $ for $ H $, and $ y_h = -1/2 $ for $ \tilde{H} $.

It turns out that
some of the counterterm operators previously defined always appear in specific combinations:

%before correcting typo
\begin{comment}
\begin{eqnarray}
    && T_h + y^{q x} T^{11}_g + (4 y_L^q - y_R^x) T^{12}_g \;\; \text{or} \;\;
    T_h + y^{\ell y} T^{11}_g + (4 y_L^\ell - y_R^y) T^{12}_g \,, \\
    && (y_L^q + y_R^x) T^{13}_g + T^{23}_g + \tilde{T}^{33}_g \,, \\
    && (y_L^q + y_R^x) T_{h1} + T_{h2} - \frac{3}{2} y_R^x T^{12}_\sigma - T^{22}_\sigma \nonumber\\
    && \qquad \text{or} \;\; (y_L^\ell + y_R^y) T_{h1} + T_{h2} - \frac{3}{2} y_R^y T^{12}_\sigma - T^{22}_\sigma \,, \\
    && T_{h3} + \frac{3}{4} \left[ (y_L^q - y_R^x) T^{13}_\sigma + T^{23}_\sigma \right] \,.
\end{eqnarray}
\end{comment}

\begin{eqnarray}
    && T_h + y^{q x} T^{11}_g \;\; \text{or} \;\;
    T_h + y^{\ell y} T^{11}_g \,, \\
    && (y_L^q + y_R^x) T^{13}_g + \tilde{T}^{33}_g \,, \\
    && T_{h2} - \frac{3}{2} y_R^x T^{12}_\sigma - T^{22}_\sigma \;\; \text{or} \;\; T_{h2} - \frac{3}{2} y_R^y T^{12}_\sigma - T^{22}_\sigma \,, \\
    && T_{h3} + \frac{3}{4} (y_L^q - y_R^x) T^{13}_\sigma \,.
\end{eqnarray}

The fermion bilinear structures and the generation indices for the counterterms can be established straightforwardly from the first two columns in Tabs.~\ref{tab:CTs_one_scalar_1} and \ref{tab:CTs_one_scalar_2}, while the scalar field ($H^{(\dagger)}$ or $\tilde{H}^{(\dagger)}$) is the one needed to ensure global symmetry transformations (see App.~\ref{sec:list_CTs} for further discussion on this point).
To clarify the schematic notation in use, we write explicitly the finite counterterm Lagrangian for the structures $T_h$

\begin{eqnarray}
    \mathcal{L}_{\rm fct} \supset && \frac{1}{3 \cdot 16 \pi^2} \Big[ C^{\ell e d q}_{f i j k} N_c (Y^\dagger_d)_{kj} \bar{\ell}_m^f e^i \bar{\partial}^2 H^m + (C^{\ell e d q}_{f i j k})^\ast (Y^\dagger_e)_{fi} \bar{q}_m^k d^j \bar{\partial}^2 H^m \\
    && + C^{\ell \nu u q}_{f i j k} N_c (Y^\dagger_u)_{kj} \bar{\ell}_m^f \nu^i \bar{\partial}^2 \tilde{H}^m + (C^{\ell \nu u q}_{f i j k})^\ast (Y^\dagger_\nu)_{fi} \bar{q}_m^k u^j \bar{\partial}^2 \tilde{H}^m \nonumber\\
    && - (C^{\ell e q u (1)}_{f i j k})^\ast N_c (Y^\dagger_u)_{jk} \bar{e}^i \ell_m^f \bar{\partial}^2 H^{\dagger m} - (C^{\ell e q u (1)}_{f i j k})^\ast (Y^\dagger_e)_{fi} \bar{u}^k q_m^j \bar{\partial}^2 \tilde{H}^{\dagger m} \nonumber\\
    && + (C^{\ell \nu q d (1)}_{f i j k})^\ast N_c (Y^\dagger_d)_{jk} \bar{\nu}^i \ell_m^f \bar{\partial}^2 \tilde{H}^{\dagger m} + (C^{\ell \nu q d (1)}_{f i j k})^\ast (Y^\dagger_\nu)_{fi} \bar{d}^k q_m^j \bar{\partial}^2 H^{\dagger m} \nonumber\\
    && + (C^{\ell \nu \ell e}_{f i j k})^\ast (Y^\dagger_e)_{jk} \bar{\nu}^i \ell_m^f \bar{\partial}^2 \tilde{H}^{\dagger m} + (C^{\ell \nu \ell e}_{f i j k})^\ast (Y^\dagger_\nu)_{fi} \bar{e}^k \ell_m^j \bar{\partial}^2 H^{\dagger m} \nonumber\\
    && + (C^{\ell \nu \ell e}_{f i j k})^\ast \frac{1}{2} (Y^\dagger_e)_{fk} \bar{\nu}^i \ell_m^j \bar{\partial}^2 \tilde{H}^{\dagger m} + (C^{\ell \nu \ell e}_{f i j k})^\ast \frac{1}{2} (Y^\dagger_\nu)_{ji} \bar{e}^k \ell_m^f \bar{\partial}^2 H^{\dagger m} \nonumber\\
    && + (C^{q u q d (1)}_{f i j k})^\ast N_c (Y^\dagger_d)_{jk} \bar{u}^i q_m^f \bar{\partial}^2 \tilde{H}^{\dagger m} + (C^{q u q d (1)}_{f i j k})^\ast N_c (Y^\dagger_u)_{fi} \bar{d}^k q_m^j \bar{\partial}^2 H^{\dagger m} \nonumber\\
    && + (C^{q u q d (1)}_{f i j k})^\ast \frac{1}{2} (Y^\dagger_d)_{fk} \bar{u}^i q_m^j \bar{\partial}^2 \tilde{H}^{\dagger m} + (C^{q u q d (1)}_{f i j k})^\ast \frac{1}{2} (Y^\dagger_u)_{ji} \bar{d}^k q_m^f \bar{\partial}^2 H^{\dagger m} \nonumber\\
    && + (C^{q u q d (8)}_{f i j k})^\ast \frac{C_A}{2} (Y^\dagger_d)_{fk} \bar{u}^i q_m^j \bar{\partial}^2 \tilde{H}^{\dagger m} + (C^{q u q d (8)}_{f i j k})^\ast \frac{C_A}{2} (Y^\dagger_u)_{ji} \bar{d}^k q_m^f \bar{\partial}^2 H^{\dagger m} \nonumber\\
    && - C^{\ell \nu}_{f i j k} 2 (Y^\dagger_\nu)_{ij} \bar{\ell}_m^f \nu^k \bar{\partial}^2 \tilde{H}^m - (C^{\ell \nu}_{f i j k})^\ast 2 (Y^\dagger_\nu)_{kf} \bar{\ell}_m^i \nu^j \bar{\partial}^2 \tilde{H}^m \nonumber\\
    && - C^{\ell e}_{f i j k} 2 (Y^\dagger_e)_{ij} \bar{\ell}_m^f e^k \bar{\partial}^2 H^m - (C^{\ell e}_{f i j k})^\ast 2 (Y^\dagger_e)_{kf} \bar{\ell}_m^i e^j \bar{\partial}^2 H^m \nonumber\\
    && - C^{q d (1)}_{f i j k} 2 (Y^\dagger_d)_{ij} \bar{q}_m^f d^k \bar{\partial}^2 H^m - (C^{q d (1)}_{f i j k})^\ast 2 (Y^\dagger_d)_{kf} \bar{q}_m^i d^j \bar{\partial}^2 H^m \nonumber\\
    && - C^{q u (1)}_{f i j k} 2 (Y^\dagger_u)_{ij} \bar{q}_m^f u^k \bar{\partial}^2 \tilde{H}^m - (C^{q u (1)}_{f i j k})^\ast 2 (Y^\dagger_u)_{kf} \bar{q}_m^i u^j \bar{\partial}^2 \tilde{H}^m \nonumber\\
    && - C^{q d (8)}_{f i j k} 2 C_A (Y^\dagger_d)_{ij} \bar{q}_m^f d^k \bar{\partial}^2 H^m - (C^{q d (8)}_{f i j k})^\ast 2 C_A (Y^\dagger_d)_{kf} \bar{q}_m^i d^j \bar{\partial}^2 H^m \nonumber\\
    && - C^{q u (8)}_{f i j k} 2 C_A (Y^\dagger_u)_{ij} \bar{q}_m^f u^k \bar{\partial}^2 \tilde{H}^m - (C^{q u (8)}_{f i j k})^\ast 2 C_A (Y^\dagger_u)_{kf} \bar{q}_m^i u^j \bar{\partial}^2 \tilde{H}^m \Big] \nonumber\\
    && + {\rm h.c.} \nonumber
\end{eqnarray}

\begin{sidewaystable}
%\begin{table}[]
    %{\scriptsize
    \centering
    \renewcommand{\arraystretch}{1.0}
    \begin{tabular}{|cc||c|c|c|c|c|c|c|c|}
        \hline
        $\mathcal{O}$ & $Y^{\dagger}$ & $\alpha_h^\mathcal{O}$ & $\beta_{11}^\mathcal{O}$ & $\beta_{12}^\mathcal{O}$ & $\beta_{13}^\mathcal{O}$ & $\beta_{22}^\mathcal{O}$ & $\beta_{23}^\mathcal{O}$ & $\beta_{33}^\mathcal{O}$ & $\tilde{\beta}_{33}^\mathcal{O}$ \\
        \hline
        \hline
        $Q_{\ell e d q}$ & $(Y^{\dagger}_d)_{kj}$ & \multirow{2}{*}{$N_c$} & \multirow{2}{*}{$N_c y^{q x}$} & \multirow{2}{*}{$N_c (4 y^q_L - y^x_R)$} & \multirow{2}{*}{$0$} & \multirow{2}{*}{$2 N_c$} & \multirow{2}{*}{$0$} & \multirow{2}{*}{$\frac{3}{2}$} & \multirow{2}{*}{$0$} \\
        $Q_{\ell \nu u q}$ & $(Y^{\dagger}_u)_{kj}$ & & & & & & & & \\
        \hline
        $Q_{\ell e d q}^\dagger$ & $(Y^{\dagger}_e)_{fi}$ & \multirow{2}{*}{$1$} & \multirow{2}{*}{$y^{\ell y}$} & \multirow{2}{*}{$4 y^\ell_L - y^y_R$} & \multirow{2}{*}{$0$} & \multirow{2}{*}{$2$} & \multirow{2}{*}{$0$} & \multirow{2}{*}{$0$} & \multirow{2}{*}{$0$} \\
        $Q_{\ell \nu u q}^\dagger$ & $(Y^{\dagger}_\nu)_{fi}$ & & & & & & & & \\
        \hline
        $Q_{\ell e q u}^{(1) \dagger}$ & $(Y^{\dagger}_u)_{jk}$ & \multirow{2}{*}{$\mp N_c$} & \multirow{2}{*}{$\mp N_c y^{q x}$} & \multirow{2}{*}{$\pm N_c (4 y^q_L - y^x_R)$} & \multirow{2}{*}{$0$} & \multirow{2}{*}{$\mp 2 N_c$} & \multirow{2}{*}{$0$} & \multirow{2}{*}{$\mp \frac{3}{2}$} & \multirow{2}{*}{$0$} \\
        $Q_{\ell \nu q d}^{(1) \dagger}$ & $(Y^{\dagger}_d)_{jk}$ & & & & & & & & \\
        \hline
        $Q_{\ell e q u}^{(1) \dagger}$ & $(Y^{\dagger}_e)_{fi}$ & \multirow{2}{*}{$\mp 1$} & \multirow{2}{*}{$\mp y^{\ell y}$} & \multirow{2}{*}{$- (4 y^\ell_L - y^y_R)$} & \multirow{2}{*}{$0$} & \multirow{2}{*}{$\mp 2$} & \multirow{2}{*}{$0$} & \multirow{2}{*}{$0$} & \multirow{2}{*}{$0$} \\
        $Q_{\ell \nu q d}^{(1) \dagger}$ & $(Y^{\dagger}_\nu)_{fi}$ & & & & & & & & \\
        \hline
        \multirow{2}{*}{$Q_{\ell \nu \ell e}^\dagger$} & $(Y^{\dagger}_e)_{jk}$ & \multirow{2}{*}{$1$} & \multirow{2}{*}{$y^{\ell y}$} & \multirow{2}{*}{$- (4 y^\ell_L - y^y_R)$} & \multirow{2}{*}{$0$} & \multirow{2}{*}{$2$} & \multirow{2}{*}{$0$} & \multirow{2}{*}{$0$} & \multirow{2}{*}{$0$} \\
         & $(Y^{\dagger}_\nu)_{fi}$ & & & & & & & & \\
        \hline
        \multirow{2}{*}{$Q_{q u q d}^{(1) \dagger}$} & $(Y^{\dagger}_d)_{jk}$ & \multirow{2}{*}{$N_c$} & \multirow{2}{*}{$N_c y^{q x}$} & \multirow{2}{*}{$- N_c (4 y^q_L - y^x_R)$} & \multirow{2}{*}{$0$} & \multirow{2}{*}{$2 N_c$} & \multirow{2}{*}{$0$} & \multirow{2}{*}{$\frac{3}{2}$} & \multirow{2}{*}{$0$} \\
         & $(Y^{\dagger}_u)_{fi}$ & & & & & & & & \\
        \hline
        \multirow{2}{*}{$Q_{q u q d}^{(8) \dagger}$} & $(Y^{\dagger}_d)_{jk}$ & \multirow{2}{*}{$0$} & \multirow{2}{*}{$0$} & \multirow{2}{*}{$0$} & \multirow{2}{*}{$\frac{3}{2} (y^q_L + y^x_R)$} & \multirow{2}{*}{$0$} & \multirow{2}{*}{$- \frac{3}{2}$} & \multirow{2}{*}{$- \frac{3}{4 N_c}$} & \multirow{2}{*}{$\frac{3}{2}$} \\
         & $(Y^{\dagger}_u)_{fi}$ & & & & & & & & \\
        \hline
        \hline
        $Q_{\ell y}$ & $(Y^{\dagger}_y)_{ij}$ & \multirow{2}{*}{$-2$} & \multirow{2}{*}{$-2 y^{\ell y}$} & \multirow{2}{*}{$-2 (4 y^\ell_L - y^y_R)$} & \multirow{2}{*}{$0$} & \multirow{2}{*}{$-4$} & \multirow{2}{*}{$0$} & \multirow{2}{*}{$0$} & \multirow{2}{*}{$0$} \\
        $Q_{\ell y}^\dagger$ & $(Y^{\dagger}_y)_{kf}$ & & & & & & & & \\
        \hline
        $Q_{q x}^{(1)}$ & $(Y^{\dagger}_x)_{ij}$ & \multirow{2}{*}{$-2$} & \multirow{2}{*}{$-2 y^{q x}$} & \multirow{2}{*}{$-2 (4 y^q_L - y^x_R)$} & \multirow{2}{*}{$-6 (y^q_L + y^x_R)$} & \multirow{2}{*}{$-4$} & \multirow{2}{*}{$-6$} & \multirow{2}{*}{$0$} & \multirow{2}{*}{$-6$} \\
        $Q_{q x}^{(1) \dagger}$ & $(Y^{\dagger}_x)_{kf}$ & & & & & & & & \\
        \hline
        $Q_{q x}^{(8)}$ & $(Y^{\dagger}_x)_{ij}$ & \multirow{2}{*}{$-2 C_A$} & \multirow{2}{*}{$-2 C_A y^{q x}$} & \multirow{2}{*}{$-2 C_A (4 y^q_L - y^x_R)$} & \multirow{2}{*}{$\frac{3}{N_c} (y^q_L + y^x_R)$} & \multirow{2}{*}{$-4 C_A$} & \multirow{2}{*}{$\frac{3}{N_c}$} & \multirow{2}{*}{$- \frac{3}{2}$} & \multirow{2}{*}{$\frac{3}{N_c}$} \\
        $Q_{q x}^{(8) \dagger}$ & $(Y^{\dagger}_x)_{kf}$ & & & & & & & & \\
        \hline
        \multirow{2}{*}{$Q_{\ell \nu \ell e}^\dagger$} & $(Y^{\dagger}_e)_{fk}$ & \multirow{2}{*}{$\frac{1}{2}$} & \multirow{2}{*}{$\frac{1}{2} y^{\ell y}$} & \multirow{2}{*}{$\pm \frac{1}{2} (4 y^\ell_L - y^y_R)$} & \multirow{2}{*}{$0$} & \multirow{2}{*}{$1$} & \multirow{2}{*}{$0$} & \multirow{2}{*}{$0$} & \multirow{2}{*}{$0$} \\
         & $(Y^{\dagger}_\nu)_{ji}$ & & & & & & & & \\
        \hline
        \multirow{2}{*}{$ Q_{q u q d}^{(1) \dagger} $} & $(Y^{\dagger}_d)_{fk}$ & \multirow{2}{*}{$\frac{1}{2}$} & \multirow{2}{*}{$\frac{1}{2} y^{q x}$} & \multirow{2}{*}{$\pm \frac{1}{2} (4 y^q_L - y^x_R)$} & \multirow{2}{*}{$\frac{3}{2} (y^q_L + y^x_R)$} & \multirow{2}{*}{$1$} & \multirow{2}{*}{$\pm \frac{3}{2}$} & \multirow{2}{*}{$0$} & \multirow{2}{*}{$\frac{3}{2}$} \\
         & $(Y^{\dagger}_u)_{ji}$ & & & & & & & & \\
        \hline
        \multirow{2}{*}{$ Q_{q u q d}^{(8) \dagger} $} & $(Y^{\dagger}_d)_{fk}$ & \multirow{2}{*}{$\frac{C_A}{2}$} & \multirow{2}{*}{$\frac{C_A}{2} y^{q x}$} & \multirow{2}{*}{$\pm \frac{C_A}{2} (4 y^q_L - y^x_R)$} & \multirow{2}{*}{$- \frac{3}{4 N_c} (y^q_L + y^x_R)$} & \multirow{2}{*}{$C_A$} & \multirow{2}{*}{$\mp \frac{3}{4 N_c}$} & \multirow{2}{*}{$\frac{3}{8}$} & \multirow{2}{*}{$- \frac{3}{4 N_c}$} \\
         & $(Y^{\dagger}_u)_{ji}$ & & & & & & & & \\
        \hline
    \end{tabular}
    %}
    \caption{
    Coefficients of the set of counterterms needed in the finite renormalization of Green's functions carrying a single scalar field.
    We denote $x=u, d$, and $y=\nu, e$.
    The generation indices of the fields composing the operators are $fijk$.
    When two signs are given, the upper (lower) sign corresponds to the upper (lower, respectively) case in the same row.
    }
    \label{tab:CTs_one_scalar_1}
%\end{table}
\end{sidewaystable}

\begin{sidewaystable}
%\begin{table}[]
    %{\scriptsize
    \centering
    \renewcommand{\arraystretch}{1.0}
    \begin{tabular}{|cc||c|c|c|c|c|c|c|c|}
        \hline
        $ \mathcal{O} $ & $Y^{\dagger}$ & $ \alpha^\mathcal{O}_{h1} $ & $ \alpha^\mathcal{O}_{h2} $ & $ \alpha^\mathcal{O}_{h3} $ & $ \zeta^\mathcal{O}_{12} $ & $ \zeta^\mathcal{O}_{13} $ & $ \zeta^\mathcal{O}_{22} $ & $ \zeta^\mathcal{O}_{23} $ & $ \tilde{\zeta}^\mathcal{O}_{33} $ \\
        \hline
        \hline
        $ Q_{\ell e q u}^{(3) \dagger} $ & $(Y^{\dagger}_u)_{jk}$ & \multirow{2}{*}{$\mp N_c(y^q_L + y^x_R)$} & \multirow{2}{*}{$\pm N_c$} & \multirow{2}{*}{$0$} & \multirow{2}{*}{$\mp \frac{3 N_c}{2} y^x_R$} & \multirow{2}{*}{$0$} & \multirow{2}{*}{$\mp N_c$} & \multirow{2}{*}{$0$} & \multirow{2}{*}{$0$} \\ %\multirow{2}{*}{$- \frac{1}{2}$}
        $ Q_{\ell \nu q d}^{(3) \dagger} $ & $(Y^{\dagger}_d)_{jk}$ & & & & & & & & \\
        \hline
        $ Q_{\ell e q u}^{(3) \dagger} $ & $(Y^{\dagger}_e)_{fi}$ & \multirow{2}{*}{$\mp (y^\ell_L + y^y_R)$} & \multirow{2}{*}{$- 1$} & \multirow{2}{*}{$0$} & \multirow{2}{*}{$\frac{3}{2} y^y_R$} & \multirow{2}{*}{$0$} & \multirow{2}{*}{$1$} & \multirow{2}{*}{$0$} & \multirow{2}{*}{$0$} \\
        $ Q_{\ell \nu q d}^{(3) \dagger} $ & $(Y^{\dagger}_\nu)_{fi}$ & & & & & & & & \\
        \hline
        \hline
        \multirow{2}{*}{$Q_{\ell \nu \ell e}^\dagger$} & $(Y^{\dagger}_e)_{fk}$ & \multirow{2}{*}{$\frac{1}{8} (y^\ell_L + y^y_R)$} & \multirow{2}{*}{$\pm \frac{1}{8}$} & \multirow{2}{*}{$0$} & \multirow{2}{*}{$\mp \frac{3}{16} y^y_R$} & \multirow{2}{*}{$0$} & \multirow{2}{*}{$\mp \frac{1}{8}$} & \multirow{2}{*}{$0$} & \multirow{2}{*}{$0$} \\
         & $(Y^{\dagger}_\nu)_{ji}$ & & & & & & & & \\
        \hline
        \multirow{2}{*}{$ Q_{q u q d}^{(1) \dagger} $} & $(Y^{\dagger}_d)_{fk}$ & \multirow{2}{*}{$\frac{1}{8} (y^q_L + y^x_R)$} & \multirow{2}{*}{$\pm \frac{1}{8}$} & \multirow{2}{*}{$\frac{1}{4}$} & \multirow{2}{*}{$\mp \frac{3}{16} y^x_R$} & \multirow{2}{*}{$\frac{3}{16} (y^q_L - y^x_R)$} & \multirow{2}{*}{$\mp \frac{1}{8}$} & \multirow{2}{*}{$\pm \frac{3}{16}$} & \multirow{2}{*}{$- \frac{1}{16}$} \\ %\multirow{2}{*}{$- \frac{1}{16} \frac{1 + N_c}{N_c}$}
         & $(Y^{\dagger}_u)_{ji}$ & & & & & & & & \\
        \hline
        \multirow{2}{*}{$ Q_{q u q d}^{(8) \dagger} $} & $(Y^{\dagger}_d)_{fk}$ & \multirow{2}{*}{$\frac{C_A}{8} ( y^q_L + y^x_R )$} & \multirow{2}{*}{$\pm \frac{C_A}{8}$} & \multirow{2}{*}{$- \frac{1}{8 N_c}$} & \multirow{2}{*}{$\mp \frac{3 C_A}{16} y^x_R$} & \multirow{2}{*}{$- \frac{3}{32 N_c} (y^q_L - y^x_R)$} & \multirow{2}{*}{$\mp \frac{C_A}{8}$} & \multirow{2}{*}{$\mp \frac{3}{32 N_c}$} & \multirow{2}{*}{$\frac{1}{32 N_c}$} \\ %\multirow{2}{*}{$\frac{1}{32} \frac{1 + N_c - N_c^2}{N_c^2}$}
         & $(Y^{\dagger}_u)_{ji}$ & & & & & & & & \\
        \hline
    \end{tabular}
    %}
    \caption{Continuation of Tab.~\ref{tab:CTs_one_scalar_1}.}
    \label{tab:CTs_one_scalar_2}
%\end{table}
\end{sidewaystable}

\subsection{Two scalars}\label{sec:two_scalars_results}

\subsubsection{Amplitudes}

Finally, shifting to Green's functions with two scalars,
an explicit calculation of this case gives

\begin{figure}[H]
    \begin{minipage}{.25\textwidth}
        \centering
        \includegraphics[scale=0.15]{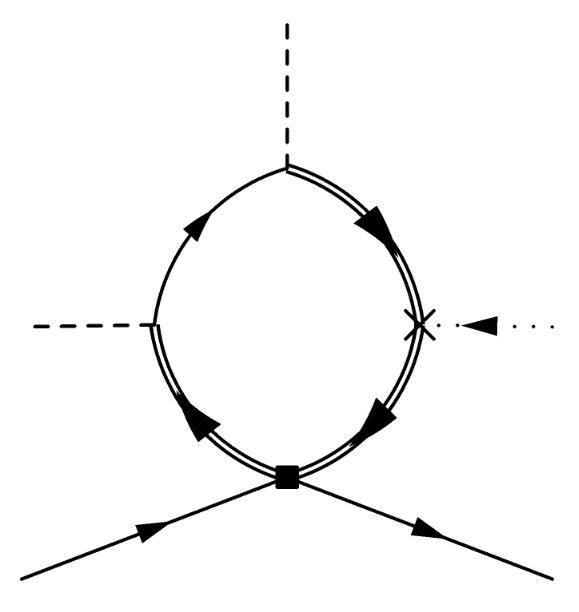} \\
        $ + \ldots $
    \end{minipage}
    \begin{minipage}{.7\textwidth}
        \centering
        \begin{equation}
            = C^{\mathcal{O}}_{\psi \xi} Y^\dagger Y \frac{1}{3 \cdot 16 \pi^2} \, [ \bar{\psi} \{ ( 2 C^g_{Y_1} - C^g_{Y_2} + 2 C^{g'}_{Y_3} ) C_6^\mathcal{O} \}_\mathcal{O} \overline{\slashed{k}}_g P_X \xi ] h^\dagger h c \,,
        \end{equation}
    \end{minipage}
\end{figure}

\noindent
where $Y_1, Y_2, Y_3 = L, R$, with $ Y_1 \neq Y_2 $ and $ Y_1 = Y_3 $; $C^g_{Y_1}$ and $C^{g'}_{Y_3}$ can be in principle different, as it will be later illustrated (e.g., as in the case of the operator $Q_{d u \nu e}$).
We employ the notation previously introduced in the case of Green's functions of a single scalar.

\subsubsection{counterterms}

Possible counterterm operators needed for the finite renormalization are

\begin{eqnarray}\label{eq:Thh1}
    && T^{(1)}_{h^\dagger h} = g_1 [ \bar{\psi} \bar{\gamma}^\mu P_X \xi ] h^\dagger h B_\mu \,, \\
    && T^{(2)}_{h^\dagger h} = g_2 [ \bar{\psi} T_L^a \bar{\gamma}^\mu P_X \xi ] h^\dagger h W_\mu^a \,, \label{eq:Thh2}\\
    && T^{(3)}_{h^\dagger h} = g_3 [ \bar{\psi} T^\alpha \bar{\gamma}^\mu P_X \xi ] h^\dagger h G_\mu^\alpha \,, \label{eq:Thh3}
\end{eqnarray}

\begin{eqnarray}
    && \tilde{T}^{(1)}_{h^\dagger h} = g_1 [ \bar{\psi} T_L^a \bar{\gamma}^\mu P_X \xi ] h^\dagger T_L^a h B_\mu \,, \label{eq:Ttildehh1}\\
    && \tilde{T}^{(2)}_{h^\dagger h} = g_2 [ \bar{\psi} \bar{\gamma}^\mu P_X \xi ] h^\dagger T_L^a h W_\mu^a \,, \label{eq:Ttildehh2}\\
    && \tilde{T}^{(3)}_{h^\dagger h} = g_3 [ \bar{\psi} T_L^a T^\alpha \bar{\gamma}^\mu P_X \xi ] h^\dagger T_L^a h G_\mu^\alpha \,. \label{eq:Ttildehh3}
\end{eqnarray}
The BRST transformation of the operators in Eqs.~\eqref{eq:Thh1}-\eqref{eq:Ttildehh3} all lead to structures carrying a fermion bilinear, a ghost on which a single derivative acts and two scalar fields.

We then include the counterterm Lagrangian

\begin{eqnarray}
    \mathcal{L}_{\rm fct} \supset && - C^{\mathcal{O}}_{\psi \xi} \frac{1}{3 \cdot 16 \pi^2} \left( \gamma^{\mathcal{O}}_{1} T^{(1)}_{h^\dagger h} + \gamma^{\mathcal{O}}_{2} T^{(2)}_{h^\dagger h} + \gamma^{\mathcal{O}}_{3} T^{(3)}_{h^\dagger h} \right. \\
    && \left. \quad \tilde\gamma^{\mathcal{O}}_{1} \tilde{T}^{(1)}_{h^\dagger h} + \tilde\gamma^{\mathcal{O}}_{2} \tilde{T}^{(2)}_{h^\dagger h} + \tilde\gamma^{\mathcal{O}}_{3} \tilde{T}^{(3)}_{h^\dagger h} \right) + {\rm h.c.} \,, \nonumber
\end{eqnarray}
where the coefficients are found in Tabs.~\ref{tab:CTs_two_scalars_1} (when fermions within the same bilinear are contracted, when referring to the Warsaw basis) and \ref{tab:CTs_two_scalars_2} (when fermions in different bilinears are contracted, with respect to the Warsaw basis).
Hermitian conjugation has to be included only for non--self-conjugate cases (such as when two different flavours are present).

As before, the explicit counterterm operator structures can be built from the information contained in the first two columns of Tabs.~\ref{tab:CTs_two_scalars_1} and \ref{tab:CTs_two_scalars_2}.
To better illustrate the notation in use, consider the structures $T^{(3)}_{h^\dagger h}$

\begin{eqnarray}\label{eq:example_2_Higgs}
    \mathcal{L}_{\rm fct} \supset && - \frac{g_3}{16 \pi^2} \Big\{ \frac{1}{2} C^{qu(8)}_{fijk} [(Y^\dagger_u Y_u)_{if} + (Y^\dagger_d Y_d)_{if}] ( \bar{u}^j T^\alpha \bar{\gamma}^\mu u^k ) H^\dagger H G_\mu^\alpha \\
    && + \frac{1}{2} C^{qd(8)}_{fijk} [(Y^\dagger_u Y_u)_{if} + (Y^\dagger_d Y_d)_{if}] ( \bar{d}^j T^\alpha \bar{\gamma}^\mu d^k ) H^\dagger H G_\mu^\alpha \nonumber\\
    %%%%%%%%%%%%%%%%%%%%%%%%%%%%%%%%%%%%%%%%%%%%%%%%%%%%%%%%%%%%%%%%%%%%%%%%%%%%%%%%%%%%%%%%%%%%%%%%%%%%%%%%%%%%%%%%%%%%%%%%%%%%%%%%%%%%%%%%%%%%%%%%%%%%%%%%%%%
    && + \frac{1}{2} C^{qu(8)}_{fijk} (Y_u Y^\dagger_u)_{kj} ( \bar{q}^f T^\alpha \bar{\gamma}^\mu q^i ) H^\dagger H G_\mu^\alpha
       + \frac{1}{2} C^{qd(8)}_{fijk} (Y_d Y^\dagger_d)_{kj} ( \bar{q}^f T^\alpha \bar{\gamma}^\mu q^i ) H^\dagger H G_\mu^\alpha \nonumber\\
    %%%%%%%%%%%%%%%%%%%%%%%%%%%%%%%%%%%%%%%%%%%%%%%%%%%%%%%%%%%%%%%%%%%%%%%%%%%%%%%%%%%%%%%%%%%%%%%%%%%%%%%%%%%%%%%%%%%%%%%%%%%%%%%%%%%%%%%%%%%%%%%%%%%%%%%%%%%
    && + \frac{1}{2} C^{ud(8)}_{fijk} (Y_d Y^\dagger_d)_{kj} ( \bar{u}^f T^\alpha \bar{\gamma}^\mu u^i ) H^\dagger H G_\mu^\alpha
       + \frac{1}{2} C^{ud(8)}_{fijk} (Y_u Y^\dagger_u)_{if} ( \bar{d}^j T^\alpha \bar{\gamma}^\mu d^k ) H^\dagger H G_\mu^\alpha \nonumber\\
    %%%%%%%%%%%%%%%%%%%%%%%%%%%%%%%%%%%%%%%%%%%%%%%%%%%%%%%%%%%%%%%%%%%%%%%%%%%%%%%%%%%%%%%%%%%%%%%%%%%%%%%%%%%%%%%%%%%%%%%%%%%%%%%%%%%%%%%%%%%%%%%%%%%%%%%%%%%
    && + \frac{1}{2} C^{qq(1)}_{fijk} [ (Y^\dagger_u Y_u)_{kf} + (Y^\dagger_d Y_d)_{kf} ] ( \bar{q}^j T^\alpha \bar{\gamma}^\mu q^i ) H^\dagger H G_\mu^\alpha \nonumber\\
    && + \frac{1}{2} C^{qq(1)}_{fijk} [ (Y^\dagger_u Y_u)_{ij} + (Y^\dagger_d Y_d)_{ij} ] ( \bar{q}^f T^\alpha \bar{\gamma}^\mu q^k ) H^\dagger H G_\mu^\alpha \nonumber\\
    %%%%%%%%%%%%%%%%%%%%%%%%%%%%%%%%%%%%%%%%%%%%%%%%%%%%%%%%%%%%%%%%%%%%%%%%%%%%%%%%%%%%%%%%%%%%%%%%%%%%%%%%%%%%%%%%%%%%%%%%%%%%%%%%%%%%%%%%%%%%%%%%%%%%%%%%%%%
    && + \frac{3}{2} C^{qq(3)}_{fijk} [ (Y^\dagger_u Y_u)_{kf} + (Y^\dagger_d Y_d)_{kf} ] ( \bar{q}^j T^\alpha \bar{\gamma}^\mu q^i ) H^\dagger H G_\mu^\alpha \nonumber\\
    && + \frac{3}{2} C^{qq(3)}_{fijk} [ (Y^\dagger_u Y_u)_{ij} + (Y^\dagger_d Y_d)_{ij} ] ( \bar{q}^f T^\alpha \bar{\gamma}^\mu q^k ) H^\dagger H G_\mu^\alpha \nonumber\\
    %%%%%%%%%%%%%%%%%%%%%%%%%%%%%%%%%%%%%%%%%%%%%%%%%%%%%%%%%%%%%%%%%%%%%%%%%%%%%%%%%%%%%%%%%%%%%%%%%%%%%%%%%%%%%%%%%%%%%%%%%%%%%%%%%%%%%%%%%%%%%%%%%%%%%%%%%%%
    && + C^{uu}_{fijk} (Y_u Y_u^\dagger)_{kf} ( \bar{u}^j T^\alpha \bar{\gamma}^\mu u^i ) H^\dagger H G_\mu^\alpha
       + C^{uu}_{fijk} (Y_u Y_u^\dagger)_{ij} ( \bar{u}^f T^\alpha \bar{\gamma}^\mu u^k ) H^\dagger H G_\mu^\alpha \nonumber\\
    && + C^{dd}_{fijk} (Y_d Y_d^\dagger)_{kf} ( \bar{d}^j T^\alpha \bar{\gamma}^\mu d^i ) H^\dagger H G_\mu^\alpha
       + C^{dd}_{fijk} (Y_d Y_d^\dagger)_{ij} ( \bar{d}^f T^\alpha \bar{\gamma}^\mu d^k ) H^\dagger H G_\mu^\alpha \Big\} \nonumber\\
    %%%%%%%%%%%%%%%%%%%%%%%%%%%%%%%%%%%%%%%%%%%%%%%%%%%%%%%%%%%%%%%%%%%%%%%%%%%%%%%%%%%%%%%%%%%%%%%%%%%%%%%%%%%%%%%%%%%%%%%%%%%%%%%%%%%%%%%%%%%%%%%%%%%%%%%%%%%
    %%&& + C^{ud (1)}_{fijk} (Y_u Y^\dagger_d)_{ij} ( \bar{u}^f T^\alpha \bar{\gamma}^\mu d^k ) \tilde{H}^\dagger H G_\mu^\alpha
    %%   + C^{ud (1) \dagger}_{fijk} (Y_u Y^\dagger_d)_{fk} ( \bar{u}^i T^\alpha \bar{\gamma}^\mu d^j ) \tilde{H}^\dagger H G_\mu^\alpha \nonumber\\
    %%%%%%%%%%%%%%%%%%%%%%%%%%%%%%%%%%%%%%%%%%%%%%%%%%%%%%%%%%%%%%%%%%%%%%%%%%%%%%%%%%%%%%%%%%%%%%%%%%%%%%%%%%%%%%%%%%%%%%%%%%%%%%%%%%%%%%%%%%%%%%%%%%%%%%%%%%%
    %%&& - \frac{1}{2 N_c} C^{ud (8)}_{fijk} (Y_u Y^\dagger_d)_{ij} ( \bar{u}^f T^\alpha \bar{\gamma}^\mu d^k ) \tilde{H}^\dagger H G_\mu^\alpha \nonumber\\
    %%&& - \frac{1}{2 N_c} C^{ud (8) \dagger}_{fijk} (Y_u Y^\dagger_d)_{fk} ( \bar{u}^i T^\alpha \bar{\gamma}^\mu d^j ) \tilde{H}^\dagger H G_\mu^\alpha  \Big\} \nonumber\\
    && + {\rm h.c.} \nonumber
\end{eqnarray}

\begin{table}[]
    \centering
    \begin{tabular}{|cc||c|c|c|c|c|c|}
        \hline
        $\mathcal{O}$ & & $\gamma^\mathcal{O}_1$ & $\gamma^\mathcal{O}_2$ & $\gamma^\mathcal{O}_3$ & $\tilde{\gamma}^\mathcal{O}_1$ & $\tilde{\gamma}^\mathcal{O}_2$ & $\tilde{\gamma}^\mathcal{O}_3$ \\
        \hline
        \hline
        \multirow{2}{*}{$ Q_{q q}^{(1)} $} & $ (Y^\dagger_x Y_x)_{kj} $, & \multirow{5}{*}{$ N_c (4 y_L^q - y_R^x) $} & \multirow{5}{*}{$0$} & \multirow{5}{*}{$0$} & \multirow{5}{*}{$0$} & \multirow{5}{*}{$\pm 4 N_c$} & \multirow{5}{*}{$0$} \\
         & $ (Y^\dagger_x Y_x)_{if} $ & & & & & & \\
        $ Q_{\ell q}^{(1)} $ & $ (Y^\dagger_x Y_x)_{kj} $ & & & & & & \\
        $ Q_{q u}^{(1)} $, $ Q_{q d}^{(1)} $ & $ (Y^\dagger_x Y_x)_{if} $ & & & & & & \\
        $ Q_{q y} $ & $ (Y^\dagger_x Y_x)_{if} $ & & & & & & \\
        \hline
        \multirow{2}{*}{$ Q_{\ell \ell} $} & $ (Y^\dagger_y Y_y)_{kj} $, & \multirow{5}{*}{$ 4 y_L^\ell - y_R^y $} & \multirow{5}{*}{$0$} & \multirow{5}{*}{$0$} & \multirow{5}{*}{$0$} & \multirow{5}{*}{$\pm 4$} & \multirow{5}{*}{$0$} \\
         & $ (Y^\dagger_y Y_y)_{if} $ & & & & & & \\
        $ Q_{\ell q}^{(1)} $ & $ (Y^\dagger_y Y_y)_{if} $ & & & & & & \\
        $ Q_{\ell \nu} $, $ Q_{\ell e} $ & $ (Y^\dagger_y Y_y)_{if} $ & & & & & & \\
        $ Q_{\ell x} $ & $ (Y^\dagger_y Y_y)_{if} $ & & & & & & \\
        \hline
        \multirow{2}{*}{$ Q_{x x} $} & $ (Y_x Y^\dagger_x)_{kj} $, & \multirow{7}{*}{$ N_c (4 y_R^x - y_L^q) $} & \multirow{7}{*}{$0$} & \multirow{7}{*}{$0$} & \multirow{7}{*}{$0$} & \multirow{7}{*}{$\mp N_c$} & \multirow{7}{*}{$0$} \\
         & $ (Y_x Y^\dagger_x)_{if} $ & & & & & & \\
        $ Q_{y x} $ & $ (Y_x Y^\dagger_x)_{kj} $ & & & & & & \\
        \multirow{2}{*}{$ Q_{u d}^{(1)} $} & $ (Y_d Y^\dagger_d)_{kj} $, & & & & & & \\
         & $ (Y_u Y^\dagger_u)_{if} $ & & & & & & \\
        $ Q_{\ell x} $ & $ (Y_x Y^\dagger_x)_{kj} $ & & & & & & \\
        $ Q_{q x}^{(1)} $ & $ (Y_x Y^\dagger_x)_{kj} $ & & & & & & \\
        \hline
        \multirow{2}{*}{$ Q_{y y} $} & $ (Y_y Y^\dagger_y)_{kj} $, & \multirow{7}{*}{$4 y_R^y - y_L^\ell$} & \multirow{7}{*}{$0$} & \multirow{7}{*}{$0$} & \multirow{7}{*}{$0$} & \multirow{7}{*}{$\mp1$} & \multirow{7}{*}{$0$} \\
         & $ (Y_y Y^\dagger_y)_{if} $ & & & & & & \\
        $ Q_{y x} $ & $ (Y_y Y^\dagger_y)_{if} $ & & & & & & \\
        \multirow{2}{*}{$ Q_{\nu e} $} & $ (Y_\nu Y^\dagger_\nu)_{if} $, & & & & & & \\
         & $ (Y_e Y^\dagger_e)_{kj} $ & & & & & & \\
        $ Q_{\ell y} $ & $ (Y_y Y^\dagger_y)_{kj} $ & & & & & & \\
        $ Q_{q y} $ & $ (Y_y Y^\dagger_y)_{kj} $ & & & & & & \\
        \hline
        $ Q_{d u \nu e} $ & $ (Y_u Y^\dagger_d)_{if} $ & $0$ & $0$ & $0$ & $0$ & $-N_c$ & $0$ \\ %$N_c (2 y_{R}^u + 2 y_{R}^d - y_L^q) $
        \hline
        $ Q_{d u \nu e}^\dagger $ & $ (Y_\nu Y^\dagger_e)_{jk} $ & $0$ & $0$ & $0$ & $0$ & $-1$ & $0$ \\ %$2 y_{R}^\nu + 2 y_{R}^e - y_L^\ell$
        \hline
        \multirow{2}{*}{$ Q_{q q}^{(3)} $} & $ (Y^\dagger_x Y_x)_{kj} $, & \multirow{3}{*}{$0$} & \multirow{3}{*}{$4 N_c$} & \multirow{3}{*}{$0$} & \multirow{3}{*}{$\pm 4 N_c ( 4 y_L^q - y_R^x )$} & \multirow{3}{*}{$0$} & \multirow{3}{*}{$0$} \\
         & $ (Y^\dagger_x Y_x)_{if} $ & & & & & & \\
        $ Q_{\ell q}^{(3)} $ & $ (Y^\dagger_x Y_x)_{kj} $ & & & & & & \\
        \hline
        $ Q_{\ell q}^{(3)} $ & $ (Y^\dagger_y Y_y)_{if} $ & $0$ & $4$ & $0$ & $ \pm 4 ( 4 y_L^\ell - y_R^y )$ & $0$ & $0$ \\
        \hline
        $ Q_{q u}^{(8)} $, $ Q_{q d}^{(8)} $ & $ (Y^\dagger_x Y_x)_{if} $ & $0$ & $0$ & $\frac{3}{2}$ & $0$ & $0$ & $0$ \\
        \hline
        $ Q_{q x}^{(8)} $ & $ (Y_x Y^\dagger_x)_{kj} $ & \multirow{3}{*}{$0$} & \multirow{3}{*}{$0$} & \multirow{3}{*}{$\frac{3}{2}$} & \multirow{3}{*}{$0$} & \multirow{3}{*}{$0$} & \multirow{3}{*}{$0$} \\
        \multirow{2}{*}{$ Q_{u d}^{(8)} $} & $ (Y_d Y^\dagger_d)_{kj} $, & & & & & & \\
         & $ (Y_u Y^\dagger_u)_{if} $ & & & & & & \\
        \hline
    \end{tabular}
    \caption{
    Coefficients of the set of counterterms needed in the finite renormalization of Green's functions carrying two scalar fields.
    We denote $x=u, d$, and $y=\nu, e$.
    The generation indices of the fields composing the operators are $fijk$.
    In the cases of the coefficients $\tilde{\gamma}^\mathcal{O}_1$, $\tilde{\gamma}^\mathcal{O}_2$ and $\tilde{\gamma}^\mathcal{O}_3$, the upper (lower) sign corresponds to $x=d$ or $y=e$ ($x=u$ or $y=\nu$, respectively).
    }
    \label{tab:CTs_two_scalars_1}
\end{table}

\begin{table}[]
    \centering\begin{tabular}{|cc||c|c|c|c|c|c|}
        \hline
        $\mathcal{O}$ & & $\gamma^\mathcal{O}_1$ & $\gamma^\mathcal{O}_2$ & $\gamma^\mathcal{O}_3$ & $\tilde{\gamma}^\mathcal{O}_1$ & $\tilde{\gamma}^\mathcal{O}_2$ & $\tilde{\gamma}^\mathcal{O}_3$ \\
        \hline
        \hline
        \multirow{2}{*}{$ Q_{\ell \ell} $} & $ (Y^\dagger_y Y_y)_{kf} $, & \multirow{2}{*}{$ \frac{1}{2} (4 y_L^\ell - y_R^y) $} & \multirow{2}{*}{$ 2 $} & \multirow{2}{*}{$ 0 $} & \multirow{2}{*}{$ \pm 2 (4 y_L^\ell - y_R^y) $} & \multirow{2}{*}{$ \pm 2 $} & \multirow{2}{*}{$ 0 $} \\
         & $ (Y^\dagger_y Y_y)_{ij} $ & & & & & & \\
        \hline
        \multirow{2}{*}{$ Q_{q q}^{(1)} $} & $ (Y^\dagger_x Y_x)_{kf} $, & \multirow{2}{*}{$ \frac{1}{2} (4 y_L^q - y_R^x) $} & \multirow{2}{*}{$ 2 $} & \multirow{2}{*}{$ \frac{3}{2} $} & \multirow{2}{*}{$ \pm 2 (4 y_L^q - y_R^x) $} & \multirow{2}{*}{$ \pm 2 $} & \multirow{2}{*}{$ \pm 6 $} \\
         & $ (Y^\dagger_x Y_x)_{ij} $ & & & & & & \\
        \hline
        \multirow{2}{*}{$ Q_{q q}^{(3)} $} & $ (Y^\dagger_x Y_x)_{kf} $, & \multirow{2}{*}{$ \frac{3}{2} (4 y_L^q - y_R^x) $} & \multirow{2}{*}{$- 2$} & \multirow{2}{*}{$\frac{9}{2}$} & \multirow{2}{*}{$ \mp 2 (4 y_L^q - y_R^x) $} & \multirow{2}{*}{$\pm 6$} & \multirow{2}{*}{$\mp 6$} \\
         & $ (Y^\dagger_x Y_x)_{ij} $ & & & & & & \\
        \hline
        \multirow{2}{*}{$ Q_{y y} $} & $ (Y_y Y^\dagger_y)_{kf} $, & \multirow{2}{*}{$4 y_R^y - y_L^\ell$} & \multirow{2}{*}{$0$} & \multirow{2}{*}{$0$} & \multirow{2}{*}{$0$} & \multirow{2}{*}{$\mp 1$} & \multirow{2}{*}{$0$} \\
         & $ (Y_y Y^\dagger_y)_{ij} $ & & & & & & \\
        \hline
        \multirow{2}{*}{$ Q_{x x} $} & $ (Y_x Y^\dagger_x)_{kf} $, & \multirow{2}{*}{$4 y_R^x - y_L^q$} & \multirow{2}{*}{$0$} & \multirow{2}{*}{$3$} & \multirow{2}{*}{$0$} & \multirow{2}{*}{$\mp 1$} & \multirow{2}{*}{$0$} \\
         & $ (Y_x Y^\dagger_x)_{ij} $ & & & & & & \\
        \hline
        $ Q_{\nu e} $ & $ (Y_\nu Y^\dagger_e)_{ij} $ & \multirow{2}{*}{$0$} & \multirow{2}{*}{$0$} & \multirow{2}{*}{$0$} & \multirow{2}{*}{$0$} & \multirow{2}{*}{$-1$} & \multirow{2}{*}{$0$} \\ %$2 y_{R}^\nu + 2 y_{R}^e - y_L^\ell$
        $ Q_{\nu e}^\dagger $ & $ (Y_\nu Y^\dagger_e)_{fk} $ & & & & & & \\
        \hline
        $ Q_{u d}^{(1)} $ & $ (Y_u Y^\dagger_d)_{ij} $ & \multirow{2}{*}{$0$} & \multirow{2}{*}{$0$} & \multirow{2}{*}{$0$} & \multirow{2}{*}{$0$} & \multirow{2}{*}{$- 1$} & \multirow{2}{*}{$0$} \\ %$2 y_{R}^u + 2 y_{R}^d - y_L^q$ $3$
        $ Q_{u d}^{(1) \dagger} $ & $ (Y_u Y^\dagger_d)_{fk} $ & & & & & & \\
        \hline
        $ Q_{u d}^{(8)} $ & $ (Y_u Y^\dagger_d)_{ij} $ & \multirow{2}{*}{$0$} & \multirow{2}{*}{$0$} & \multirow{2}{*}{$0$} & \multirow{2}{*}{$0$} & \multirow{2}{*}{$- C_A$} & \multirow{2}{*}{$0$} \\ %$C_A ( 2 y_{R}^u + 2 y_{R}^d - y_L^q )$ $- \frac{3}{2 N_c}$
        $ Q_{u d}^{(8) \dagger} $ & $ (Y_u Y^\dagger_d)_{fk} $ & & & & & & \\
        \hline
    \end{tabular}
    \caption{Continuation of Tab.~\ref{tab:CTs_two_scalars_1}.}
    \label{tab:CTs_two_scalars_2}
\end{table}

We observe that some operator counterterms appear in the following combinations

\begin{equation}
    T^{(2)}_{h^\dagger h} \pm (4 y_L^q - y_R^x) \tilde{T}^{(1)}_{h^\dagger h} \;\; \text{or} \;\; T^{(2)}_{h^\dagger h} \pm (4 y_L^\ell - y_R^y) \tilde{T}^{(1)}_{h^\dagger h} \,,
\end{equation}
where the upper (lower) sign corresponds to a right-handed down-type (up-type, respectively) quark or lepton running inside the internal fermion line.

\subsection{Discussion}\label{sec:final_discussion}

%%% The virtues of our calculation, and some results %%%
Having explicitly found finite counterterms that cure the obstructions to the Slavnov-Taylor identities at the one-loop order,
we have then shown explicitly that the theory is anomaly free.
This is as
expected, as the fermion content of the SM is such that no anomalies in its local symmetries are generated (of course, adding a sterile right-handed neutrino as we did does not change this conclusion).
The required operators are clearly indicated in Eqs.~\eqref{eq:T1}-\eqref{eq:T33f}, Eqs.~\eqref{eq:Ftilde21}-\eqref{eq:Ftilde2AG}, Eqs.~\eqref{eq:F21}-\eqref{eq:F2AG}, Eqs.~\eqref{eq:TS111}-\eqref{eq:T2eps21}, separated according to the structure of the fermion bilinear into Secs.~\ref{sec:gamma_structure_CTs}-\ref{sec:gamma_Pauli_GellMann_structure_CTs}, and Eqs.~\eqref{eq:CT_Th}-\eqref{eq:T23sigma}, Eqs.~\eqref{eq:Thh1}-\eqref{eq:Ttildehh3}.
It turns out that some of them appear in specific combinations, reducing the total number of needed gauge-variant structures to 28, when omitting the nature of the fermions (whether quarks or leptons), their chiralities, and generation indices, in linear combinations that may depend on the various hypercharges.
The obtained finite counterterms, which are
not evanescent, are not gauge-invariant by construction, in order to compensate for the obstructions to the Slavnov-Taylor identities found at one loop.
They consist of a single fermion bilinear contracted with a varying number of gauge boson fields, divided into three categories according to the number of scalar fields (namely, Sec.~\ref{sec:no_scalars_results}, when no scalars are present, Sec.~\ref{sec:one_scalar_results}, when one single scalar is present, and Sec.~\ref{sec:two_scalars_results}, when two scalars are present).
They enter in the finite counterterm Lagrangian with the coefficients found in Tabs.~\ref{tab:gamma_gamma}-\ref{tab:CTs_two_scalars_2}. The zeros in these tables
correspond to the absence of a particular structure, due to various reasons, such as some fields being singlets under $SU(2)_L$ and/or $SU(3)$, the tracelessness of the $SU(2)_L$ and $SU(3)$ generators, or the impossibility of writing a particular gauge-variant structure of dimension-6 that could possibly act as a counterterm (for instance, in the case of four-fermion operators, which must respect global symmetries in our approach where fictitious fields are introduced);
we describe topologies that do not lead to symmetry-breaking amplitudes in
App.~\ref{sec:properties_Greens_functions}.
The relations that can be identified across different rows in those tables follow from Fierz identities, Eqs.~\eqref{eq:Fierz_1}-\eqref{eq:Fierz_4}, and identities holding for the $SU(2)_L$ and $SU(3)$ generators, Eqs.~\eqref{eq:GellMann_identity_1}-\eqref{eq:GellMann_identity_2}; also, for instance, by multiplying where appropriate by the number of colors (whether $1$ or $N_c$) and/or components of the weak-isospin multiplets (whether $1$ or $2$), and flipping signs according to the chirality of the fermion running inside the loop in cases indicated with a Levi-Civita symbol. Similarly, some different columns are related by the interchange of the involved gauge interaction, e.g., there are analogous operators carrying only $SU(2)_L$ or alternatively only $SU(3)$ gauge bosons. For similar reasons, some results across Tabs.~\ref{tab:gamma_gamma}-\ref{tab:gammaTTL_gammaTTL} are related.
The basis of operators is only defined up to ambiguities stemming from gauge invariant operators.

%%% Comparison to a different calculation %%%
We follow the well-established algebraic renormalization approach, which consists of restoring the Slavnov-Taylor identities for the BRST transformation; such transformation enhances the ghost number by one unit. A novel method, circumventing the need in our approach to perform the inverse of the BRST transformation
(i.e., to identify operators of ghost number 0 from the obstructions of ghost number 1 obtained by computing Feynman diagrams with an external ghost),
is presented in Ref.~\cite{Fuentes-Martin:2025meq}. This reference discusses the complete set of operators up to dimension-6 compatible with the SM local symmetries, while we focus on the sub-set consisting of four-fermion operators (extended to include right-handed neutrinos).

Both sets of results depend on diagrams calculated with single-insertions of the four-fermion operators, and insertions of the the fermion-gauge boson and Yukawa vertices, and the evanescent symmetry-breaking vertex, that in our case consists of a fermion-ghost vertex, while in the case of Ref.~\cite{Fuentes-Martin:2025meq} introduces a spurion field $\Omega$.
The background field gauge of Ref.~\cite{Fuentes-Martin:2025meq} leads to Feynman rules that are very similar to the ones shown in Eq.~\eqref{eq:SM_nuR_Lagrangian}, with gauge bosons and the Higgs replaced by background fields for the purpose of this calculation; the propagators in both setups are the same.
One would expect then to obtain the same set of finite symmetry-restoring counterterms at one loop as in Ref.~\cite{Fuentes-Martin:2025meq} up to BRST-invariant operators.
The reason for this is that
their regularized classical Lagrangian is equivalent to ours in the limit $\Omega \to \mathbb{I}$,
and their resulting finite amplitudes
are polynomial in $\Omega - \mathbb{I}$, such that interchanging the limit on the spurion field and the integration over loop momentum should commute.

However, we have identified a number of differences with respect to the first public version of Ref.~\cite{Fuentes-Martin:2025meq}. For instance, they obtain a relative minus sign between the two contributions proportional to $\mathcal{Q}_2$, defined in our Eq.~\eqref{eq:T1CS} for any fermion $\xi$, while we obtain two contributions that must come with a relative positive sign as can be appreciated from the use of the Fierz identity in our Eq.~\eqref{eq:Fierz_1}, see our Tab.~\ref{tab:gamma_gamma}.\footnote{Following private communication with the authors of Ref.~\cite{Fuentes-Martin:2025meq}, it seems that this sign issue is due to an inconsistent convention employed by them when performing simplifications based on the Chisholm identity.}
The differences cannot all be explained by the use of two different but equivalent bases of counterterms, i.e., a change of renormalization scheme; namely,
in the case of the $Q_{ee}$ operator the difference between our results and the ones of Ref.~\cite{Fuentes-Martin:2025meq} (Eq.~(2.37) in their supplemental material) in the finite counterterm Lagrangian is ($ y^e_R = -1 $):
\begin{eqnarray}
    \frac{C^{ee}_{fijj}}{16 \pi^2} \left[ \left( - \frac{4}{3} y^e_R -\frac{8}{3} \right) g_1 (\bar{e}_f \bar{\gamma}_\mu e_i) \bar{\partial}^2 B^\mu + \frac{16}{3} (y^e_R)^2 g_1^2 (\bar{e}_f \bar{\gamma}_\mu e_i) \epsilon^{\mu \nu \rho \sigma} B_{\sigma} \bar{\partial}_\rho B_{\nu} \right] + {\rm h.c.} \,,
\end{eqnarray}
which is not invariant under a BRST transformation.\footnote{We do not introduce in this text the linearized Slavnov-Taylor operator, see e.g. Ref.~\cite{Belusca-Maito:2020ala}, which acts non-trivially only on external fields with respect to the BRST transformation. From our analysis of four-fermion operators, there are no finite counter-terms depending on these external sources to the BRST transformations at one loop.} In this latter expression, the two displayed operators are the $\mathcal{Q}_1$ and $\mathcal{Q}_2$ operators, respectively, with right-handed charged leptons, and $ f, i, j $ are generation indices.

\section{Conclusions}\label{sec:conclusions}

%%% Presentation of the problem; two classes of anomalies %%%
A mathematically consistent treatment of the algebra of $\gamma$-matrices in dimensional regularization is provided by the Breitenlohner-Maison-'t Hooft-Veltman (BMHV) scheme.
In this scheme, Slavnov-Taylor identities, which follow from the existence of symmetries, are not ``automatically'' satisfied at the quantum level.
Instead, due to the presence of spurious symmetry-breaking obstructions, which do not fall into the category of the physical anomalies that invalidate a symmetry at the quantum level, one needs to add a set of finite counterterms to the renormalized Lagrangian.

%%% Higher orders %%%
Such finite counterterms do not impact the running of SMEFT operators at the one-loop level \cite{Jenkins:2013zja,Jenkins:2013wua,Alonso:2013hga,Alonso:2014zka}.
However, for consistent applications of the BMHV scheme at one loop, i.e., including non-logarithmic contributions,
and when moving beyond a one-loop analysis (as, like for any counterterm, they have to be inserted in one-loop, etc., diagrams when moving to higher orders),
they become a must.

%%% Technical details about the computation %%%
Due to the large number of operators, diagrams and counterterms, it is challenging to compute radiative corrections in SMEFT systematically (to any given order, for any dimensionality).
To easy
the identification of the needed finite counterterms at one-loop order in single insertions of four-fermion operators, we employ a technique rarely discussed in sufficient details in the literature of non-power counting renormalizable new physics operators, namely, the method of Bonneau \cite{Bonneau:1979jx,Bonneau:1980zp}:
since
the obstructions originate from $\varepsilon/\varepsilon$ terms, where $\varepsilon$ is the regulator of dimensional regularization, this technique identifies the associated $1/\varepsilon$ poles, thus avoiding the direct determination of the obstructions to the Slavnov-Taylor identities from the whole quantum effective action. Being poles in the dimensional regulator, one sees that the obstructions are local functionals of the fields.
To some extent, it means that we face a calculation analogous to the one needed to renormalize divergences in SMEFT at one loop.
One important additional difficulty in our case, however, arises in the need to identify those functionals of the fields of ghost number $0$ that, when BRST-transformed, compensate for the obstructions, which are calculated from amplitudes of ghost number $1$.
Since the obstructions reflect the spurious breaking of the Slavnov-Taylor identities in the BMHV scheme, the BRST-transformation of such finite counterterms of ghost number $0$ does not vanish.
We stress that,
although the original symmetry breaking comes from an evanescent structure in the fermion kinetic term of a chiral theory in the regularized Lagrangian already at the classical level, the set of finite counterterms is not evanescent, and built solely from physical fields.

%%% Concluding remarks %%%
%%% Phrase targeting phenomenologists %%%
In this work, we focused on four-fermion operators, and our results for the finite renormalization are summarized in Sec.~\ref{sec:final_discussion}.
In the coming future, we plan to discuss SMEFT operators other than four-fermion ones.
The computation of some two-loop anomalous dimension matrix elements is also on the horizon.
This programme is particularly pertinent given the widespread use of SMEFT
in diverse problems, such as low- (e.g., light and heavy flavour physics) and high-energy (e.g., physics at colliders) observables. Following current and foreseen precision in data, sub-leading effects are increasingly necessary.

\section*{Acknowledgements}

We thank Herm\`{e}s B\'{e}lusca-Ma\"{i}to for many clarifying discussions.
We are indebted to Antonio Pich for many inspiring discussions, and support during the entire realization of this project.
We thank the authors of Ref.~\cite{Fuentes-Martin:2025meq}
for exchanges that helped us clarifying some aspects of their work.
We also thank
Marco Ardu, Shankha Banerjee, Joachim Brod, Leandro Cieri, Mart\'{i}n Gonz\'{a}lez-Alonso, Kirsten Leslie, Zachary Polonsky, Jorge Portol\'{e}s, Marcos N. Rabelo, Marc Riembau, Germ\'{a}n Rodrigo for useful discussions.

This work is supported by the Spanish Government (Agencia Estatal de Investigaci\'{o}n MCIN/AEI/10.13039/501100011033) Grants No. PID2020–114473GB-I00 and No. PID2023-146220NB-I00, CEX2023-001292-S (Agencia Estatal de Investigaci\'{o}n MCIU/AEI (Spain) under grant IFIC Centro de Excelencia Severo Ochoa), and grant FPU20/04279.

The work of SJ is supported in part by the UK Science and Technology Facilities Council grant ST/X000796/1.

\appendix

\section{Proof of the theorem for meromorphic functions} \label{sec:Proof_Theorem}

Here we prove that for every meromorphic function \(f\left(\varepsilon\right)\) with a pole structure at \(\varepsilon=0\) we have that Eq.~\eqref{Meromorphic_Functions} is satisfied. In order to do so, we use the fact that we can write every function of this kind in terms of its Laurent expansion

\begin{equation}
    \label{Laurent_Expansion}
    f\left(\varepsilon\right)=\sum_{n=-p}^{\infty}f_n\varepsilon^n=\sum_{n=1}^p\frac{f_{-n}}{\varepsilon^n}+\sum_{n=0}^{\infty}f_n\varepsilon^n,
\end{equation}

\noindent where \(p\) is the order of the pole. Then, by definition of \(\textrm{p.p.}\), it is clear that

\begin{equation}
    \label{singular_part_1}
    -\textrm{p.p.}\left(\varepsilon f\left(\varepsilon\right)\right)=-\textrm{p.p.}\left(\sum_{n=1}^p\frac{f_{-n}}{\varepsilon^{n-1}}\right)=-\sum_{n=1}^{p-1}\frac{f_{-n-1}}{\varepsilon^n}.
\end{equation}

\noindent On the other hand, we also have 

\begin{equation}
    \label{singular_part_2}
    -\varepsilon\,\textrm{p.p.}\left(f\left(\varepsilon\right)\right)=-\varepsilon\sum_{n=1}^p\frac{f_{-n}}{\varepsilon^{n}}=-f_{-1}-\sum_{n=1}^{p-1}\frac{f_{-n-1}}{\varepsilon^n}.
\end{equation}

\noindent Thus, we find 

\begin{equation}
    \label{singular_part_3}
    -\textrm{p.p.}\left(\varepsilon f\left(\varepsilon\right)\right)=-\varepsilon\,\textrm{p.p.}\left(f\left(\varepsilon\right)\right)+f_{-1}=-\varepsilon\,\textrm{p.p.}\left(f\left(\varepsilon\right)\right)+\textrm{r.s.p.}\left(f\left(\varepsilon\right)\right).
\end{equation}

\section{Properties of the Feynman diagrams having one insertion of the symmetry-breaking operator}\label{sec:properties_Greens_functions}

One considers all possible Green's functions with a single insertion of the dimension-6 operator, together with a single insertion of the symmetry-breaking operator. We discuss in turn cases having different superficial degrees of divergence (SDDs).

The Feynman rule associated with the symmetry breaking operator in Eq.~\eqref{Kinetic_BRST_Breaking_1} has several properties that simplify the calculation of the divergent part of the one-loop diagrams. For instance, let us consider that the ghost with incoming momentum \(k_g\) is attached to two internal fermion propagators: one of them with incoming momentum \(k\) and the other with outgoing momentum \(k+k_g\). The symmetry breaking vertex transforms physical fermions into fictitious ones and vice versa, but, given the prescription for the \(D\)-dimensional Lagrangian that we have chosen, only physical particles can appear in any other interaction vertex of a Feynman diagram. Consequently, we need to consider a chiral flip (i.e., a transformation between physical and fictitious fields) in one and only one of the fermion propagators, see the last term in Eq.~\eqref{Kinetic_Term}. Therefore, the combination of the chiral flip with the interaction vertex will produce the Dirac structure 

\begin{equation}
    \label{Dirac_Structure_SB}
    (\bar{\slashed{k}}+\bar{\slashed{k}}_g)\left[\hat{\slashed{k}}P_{R,L}-(\hat{\slashed{k}}+\hat{\slashed{k}}_g)P_{L,R}\right]\hat{\slashed{k}}+(\hat{\slashed{k}}+\hat{\slashed{k}}_g)\left[\hat{\slashed{k}}P_{R,L}-(\hat{\slashed{k}}+\hat{\slashed{k}}_g)P_{L,R}\right]\bar{\slashed{k}}
\end{equation}

\noindent in the numerator of the integrands, where the first term represents the case with the outgoing fermion being always physical and the incoming one having experienced a chiral flip, and the second term represents the opposite situation.\footnote{
Eq.~\eqref{Dirac_Structure_SB} means that we have contributions linear in the symmetry breaking source as explained in Sec.~\ref{sec:Restoring_Symmetry}.
}

\begin{figure}
    \centering
    \includegraphics[scale=0.2]{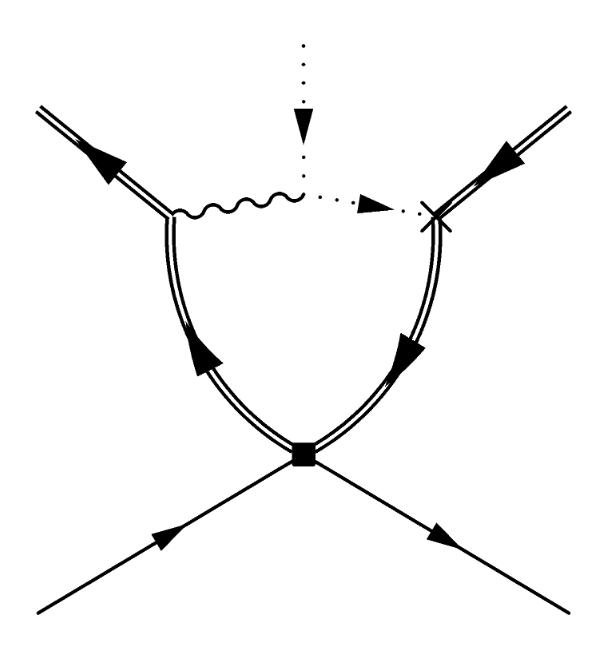} \hspace{15mm} \includegraphics[scale=0.182]{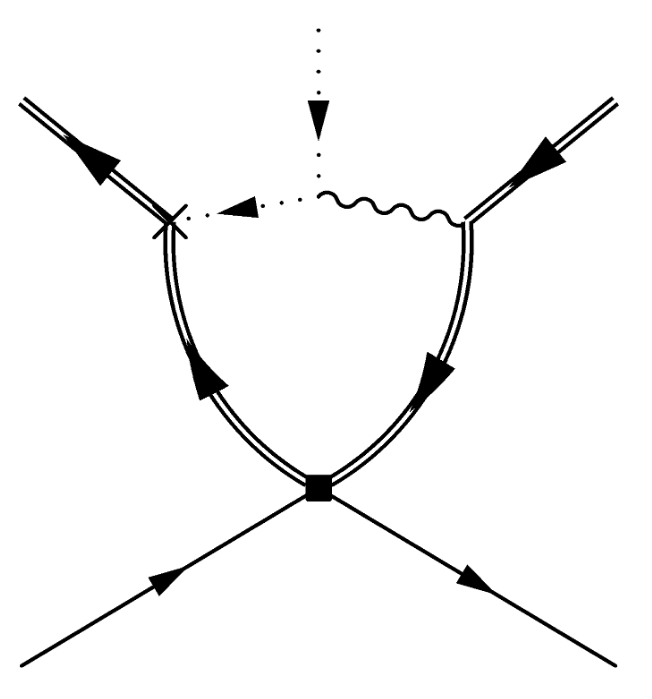}
    \caption{\textit{One-loop ``wine glass'' Feynman diagrams, where insertions of the symmetry breaking operator are indicated by a cross (thus alluding to the chiral flip discussed in the text). Their divergent part is canceled between each other, so they give no contribution to the symmetry breaking. The single (double) lines represent physical right-handed (left-handed) fermions; analogous diagrams with opposite chiralities are also possible, as well as diagrams expressing violation of baryon and lepton numbers. The arrows also indicate the flux of the four-momenta introduced in the text.}}
    \label{fig:Wine_Diagrams}
\end{figure}

Now, let us imagine that the previous Dirac structure appears in the calculation of the amplitude of a one-loop Feynman diagram with SDD=0. If \(k\) represents the integration variable, then the only contribution to the divergence comes from the term with only that momentum:

\begin{equation}
    \bar{\slashed{k}}\left[\hat{\slashed{k}}P_{R,L}-\hat{\slashed{k}}P_{L,R}\right]\hat{\slashed{k}}+\hat{\slashed{k}}\left[\hat{\slashed{k}}P_{R,L}-\hat{\slashed{k}}P_{L,R}\right]\bar{\slashed{k}}=\bar{\slashed{k}}\hat{\slashed{k}}\gamma_5\hat{\slashed{k}}+\hat{\slashed{k}}\hat{\slashed{k}}\gamma_5\bar{\slashed{k}} \, .
\end{equation}

\noindent Using the rules of the BMHV scheme it is easy to see that this is equal to \(\hat{k}^2\{\bar{\slashed{k}},\gamma_5\}=0\). As a consequence, there is no symmetry breaking contribution coming from diagrams with superficial degree of divergence zero, if the symmetry breaking operator is connected to an internal fermion.
Therefore, there are no contributions coming from diagrams with one insertion of a four-fermion operator and one insertion of the symmetry breaking operator, in topologies consisting of a fermion loop with the following sets of external legs attached: one Higgs, one ghost, and two gauge bosons; one ghost, and three gauge bosons; one ghost, and three Higgs fields.
The only diagrams with SDD=0 where the symmetry breaking operator appears attached to an external fermion leg are the ``wine glass'' diagrams, see Fig.~\ref{fig:Wine_Diagrams}. Nevertheless, they always appear in pairs and a direct calculation
taking the gauge-fixing functional
$\partial_\mu A^\mu_\alpha$ (i.e., when the coupling of ghosts to gauge bosons involves only the anti-symmetric structure constants \cite{Weinberg:1996kr}) for any $\xi$-gauge shows that the divergent part of each of them is the same except for the structure constant \(C^{abc}\) that comes from the interaction vertex of two ghosts and one gauge boson. Due to the fact that the structure constant is antisymmetric, the contributions cancel between each other.
Obviously, such diagrams are possible only in the non-Abelian cases. The $SU(3)$ case allows for diagrams where both left- and right-handed fermions interact with the internal ghost and gluon; with the same reasoning, in this case as well there is a cancellation among pairs of diagrams. The same holds for dimension-6 operators violating baryon and lepton numbers.
In conclusion, there is no symmetry breaking contribution coming from diagrams with SDD=0.

The previous reasoning can also be used to simplify the computation of the diagrams with SDD=1. We can use the well-known relation \cite{Chetyrkin:1997fm}

\begin{equation}
    \label{Propagator_Decomposition}
    \frac{1}{\left(k+p\right)^2}=\frac{1}{k^2-m_R^2}-\frac{p^2+2kp+m_R^2}{k^2-m_R^2}\frac{1}{\left(k+p\right)^2},
\end{equation}

\noindent where $m_R$ is an arbitrary mass (that can be used to regulate possible IR divergences), which allows the separation of the integrals in parts with different superficial degrees of divergence because the first term behaves like \(k^{-2}\), while the second one behaves like \(k^{-3}\) in the UV. Then, for the divergent part of the amplitudes with SDD=1, we have the contributions of the integral with all of the propagators substituted by \(1/(k^2-m^2_R)\) (SDD=1) and the case where one of them is substituted by the second term in the sum of Eq.~\eqref{Propagator_Decomposition} (SDD=0). Nonetheless, due to the fact that all the Feynman diagrams in our calculation with SDD=1 have the symmetry breaking vertex attached to an internal fermion, the previous reasoning shows that the second part of the decomposition (SDD=0) vanishes and only the case with all of the propagators written as \(1/(k^2-m^2_R)\) remains. Furthermore, the remaining term that only contains the momentum \(k\) in the numerator (SDD=1) also vanishes by the same argument used in the previous paragraph. Consequently, the only contribution comes from the case where all of the momenta of the amplitude are substituted by \(k\) except for the term in Eq.~\eqref{Dirac_Structure_SB} where we collect the terms with one \(k_g\):

\begin{equation}
    \label{Dirac_Structure_SB_1kg}
    \bar{\slashed{k}}_g\hat{\slashed{k}}\gamma_5\hat{\slashed{k}}-\bar{\slashed{k}}\hat{\slashed{k}}_gP_{L,R}\hat{\slashed{k}}+\hat{\slashed{k}}_g\hat{\slashed{k}}\gamma_5\bar{\slashed{k}}-\hat{\slashed{k}}\hat{\slashed{k}}_gP_{L,R}\bar{\slashed{k}}.
\end{equation}

We find agreement of these selection rules and the results displayed in Ref.~\cite{Belusca-Maito:2020ala}. 1PI diagrams with external BRST sources are convergent in our case, where our diagrams carry single insertions of dimension-6 four-fermion SMEFT operators.
(In other cases not discussed in this work, in general results depend on the specific choice of $\xi$-gauge, see e.g. Ref.~\cite{Belusca-Maito:2020ala}.)

\section{Candidate counterterms}\label{sec:list_CTs}

To compensate for the Green's functions of ghost number $1$ discussed in Sec.~\ref{sec:results_CTs}, we look for BRST-exact polynomials of the fields, i.e., that are written as the BRST transformation of a monomial of matter and gauge fields.
In the following, we employ that our theory in $D$ dimensions
is invariant under the global symmetry transformations that are part of any gauge symmetries, and thus consider candidates for counterterms that respect such global transformations. This is only possible due to the use of fictitious fields.
In practice, the global symmetry implies that only the terms for which at least one derivative acts on the ghost fields survive, which arise for instance when the BRST transformation acts on a field on which space-time derivatives act. This result can be understood from the fact that
the set of terms proportional to the ghost fields but not their derivatives matches the effect of a global symmetry transformation (e.g., dropping the term $\partial^{\mu}c_{\alpha}\left(x\right)$ in the transformation of $A^{\mu}_{\alpha}\left(x\right)$ in Eq.~\eqref{BRST_Transformation} one obtains an expression analogous to the one obtained from the transformation of $A^{\mu}_{\alpha}\left(x\right)$ under a global symmetry).
Implicitly, we will also exploit ghost number conservation.

Another global symmetry respected by the operators listed below is the set of Lorentz transformations.
The Lorentz indices are not indicated in this section.
Moreover, only the fermion $SU(2)_L$ index is indicated, i.e., its possible $SU(3)$ index is omitted. In the main text, Latin (Greek) characters are used for $SU(2)_L$ [$SU(3)$] indices, and also for summing over the group elements of the complete SM gauge group; capital Latin letters at the beginning (middle) of the alphabet are also used in the $SU(3)$ [$SU(2)_L$] case. Also in the main text, the ghosts of $U(1)_Y$, $SU(2)_L$, and $SU(3)$ are denoted $\lambda$, $\omega$, and $g$, respectively. Gauge couplings are always omitted in this section.
The letter $h$ denotes the Higgs doublet $H$ or $\widetilde{H}_j = \epsilon_{j k} H^\dagger {}^k$.

Cases proportional to powers of the EW scale (from the scalar potential) are not displayed, since no dimension-4 (in the dynamical fields) counterterm is being required in the finite renormalization of single insertions of dimension-6 operators.

Because the BRST transformation of a gauge invariant operator vanishes, when writing down the most complete set of monomials some of them are related, i.e., a set of linear combinations of these monomials is equal to gauge invariant operators;
for instance, we do not display in the following most cases for which only ordinary derivatives, and no gauge fields, appear (some are included for convenience, to give a more compact form to the final results of the main text).
(The counterterms of ghost number $0$ are also determined up to BRST-exact operators, i.e., operators resulting from the transformation acting on a monomial of ghost number $-1$; in the latter case the BRST transformation leads to structures having an anti-ghost with a ghost fields, or a Nakanishi-Lautrup field. We are dispensing everywhere with the introduction of Batalin-Vilkovisky antifields.)
One can exploit conservation of four-momentum to limit the form of the casted Green's functions; in other words, one looks for counterterms to be added to the Lagrangian density modulo operators that can be expressed as a total derivative.
In the following then, we do not keep structures resulting from ordinary derivatives acting on the fermion fields of the bilinear structure.
Furthermore, the diagrams of ghost number $1$ result in Green's functions proportional to the ghost four-momentum.
The potential counterterms displayed next are sufficient to perform the finite renormalization, as shown more explicitly in Sec.~\ref{sec:results_CTs}, and in details in the examples of App.~\ref{app:example}.

\subsection{No scalars}

\subsubsection{Two derivatives}

\begin{equation}
    ( \bar{\psi}_k \Gamma \xi_m )_b (\partial^2 A^b)
\end{equation}
(When not indicated otherwise, indices $k$ and $m$ are contracted together, i.e., the Kronecker symbol $ \delta_{k m} $ is omitted.)

\subsubsection{One derivative}

\begin{equation}
    ( \bar{\psi}_k \Gamma \xi_m )_{b c} (\partial A^b) A^c
\end{equation}

\subsubsection{No derivatives}

\begin{equation}
    ( \bar{\psi}_k \Gamma \xi_m )_{a b c} A^a A^b A^c
\end{equation}

\subsection{One scalar}

\subsubsection{Two derivatives}

\begin{equation}
    (\bar{\psi}_k \xi) (\partial^2 h_m)
\end{equation}

\subsubsection{One derivative}

\begin{equation}
    (\bar{\psi}_k \xi \partial h_m)_b A^b
\end{equation}
\begin{equation}
    (\bar{\psi}_k \xi h_m)_b (\partial A^b)
\end{equation}

\subsubsection{No derivatives}

\begin{equation}
    (\bar{\psi}_k \xi h_m)_{a b} A^a A^b
\end{equation}

\subsection{Two scalars}

\subsubsection{No derivatives}

\begin{equation}
    ( \bar{\psi}_{k'} \Gamma \xi_{m'} h_{l}^\dagger h_{n} )_{a} A^a
\end{equation}
(indices $k', m', l, n$ are contracted in the appropriate manner).

%%%%%%%%%%%%%%%%%%%%%%%%%%%%%%%

\section{Concrete example: four-fermion operators in a color octet structure}\label{app:example}

The explicit determination of finite counterterms when
inserting the four-fermion operators $ Q_{q x}^{(8)} $, $x=u, d$, and $ Q_{u d}^{(8)} $
was particularly complicated when compared to all other cases.
For the sake of illustration,
we now present the steps leading to the determination of symmetry-restoring finite counterterms in these more challenging cases, which should help in clarifying the discussion of the main text, while
simpler cases are found by adapting the following results.
In both subsections, $\xi$ denotes a fermion of left- or right-handed chirality, while the internal fermion carries left-handed chirality; cases proportional to the totally anti-symmetric symbol flip sign for an internal fermion carrying right-handed chirality. Consistently with this prescription, we have not
in this section
applied any identity involving the contraction of the Levi-Civita symbol with the external fermion bilinear structure.
Gauge couplings are omitted everywhere in this section.

The constant $c$ displayed below is

\begin{equation}
    c = 2 i \frac{C_{\rm NP}}{3 \cdot 16 \pi^2} \,,
\end{equation}
where $C_{\rm NP}$ is the Wilson coefficient of the dimension-6 SMEFT operator. The overall factor of $2$ is dropped when the internal degree of freedom in the loop is right-handed.

\subsection{$ U(1)_Y $ -- $SU(3)$ ghost and gauge fields}

Although this case is relatively simple, it serves as a warm up for the situation in the following subsection.
We first present the results for the distinct amplitudes, see Fig.~\ref{fig:diagrams_no_scalars}.

\vspace{2mm}
\noindent
$ \bullet $ Amplitude for $ U(1)_Y $ ghost, $ SU(3) $ gauge boson:

\begin{eqnarray}\label{eq:ghostU1_GBSU3}
    && 4 \, c \, y^q_L \, {\rm Tr} \{ T^A T^B \} \, [ \bar{\xi} \bar{\gamma}_\mu T^A \xi ] \, \bar{\partial}_\rho G^B_\nu \, \bar{\partial}_\sigma \lambda \, \epsilon^{\mu \nu \rho \sigma} \\
    && = 2 \, c \, y^q_L \, [ \bar{\xi} \bar{\gamma}_\mu T^A \xi ] \, \bar{\partial}_\rho G^A_\nu \, \bar{\partial}_\sigma \lambda \, \epsilon^{\mu \nu \rho \sigma} \,. \nonumber
\end{eqnarray}
This result is obtained after summing over the two possible diagrams; the trace over Gell-Mann matrices is the same in both cases.

\vspace{1mm}
\noindent
$ \bullet $ Amplitude for $ SU(3) $ ghost, $ U(1)_Y $ gauge boson:

\begin{eqnarray}\label{eq:ghostSU3_GBU1}
    && 4 \, c \, y^q_L \, {\rm Tr} \{ T^A T^B \} \, [ \bar{\xi} \bar{\gamma}_\mu T^A \xi ] \, \bar{\partial}_\rho B_\nu \, \bar{\partial}_\sigma g^B \, \epsilon^{\mu \nu \rho \sigma} \\
    && = 2 \, c \, y^q_L \, [ \bar{\xi} \bar{\gamma}_\mu T^A \xi ] \, \bar{\partial}_\rho B_\nu \, \bar{\partial}_\sigma g^A \, \epsilon^{\mu \nu \rho \sigma} \,. \nonumber
\end{eqnarray}

\vspace{1mm}
\noindent
$ \bullet $ ``Anti-symmetric'' part of the amplitude for $ U(1)_Y $ ghost, $ SU(3) $ -- $ SU(3) $ gauge bosons:

\begin{eqnarray}\label{eq:ghostU1_GBSU3_GBSU3_asym}
    && i \frac{3}{2} \, c \, y^q_L \, {\rm Tr} \{ T^A (T^B T^C - T^C T^B) \} \, [ \bar{\xi} \bar{\gamma}_\mu T^A \xi ] \, G^B_\nu \, G^C_\rho \, \bar{\partial}_\sigma \lambda \, \epsilon^{\mu \nu \rho \sigma} \\
    && = - \frac{3}{4} \, c \, y^q_L \, C_{ A B C } \, [ \bar{\xi} \bar{\gamma}_\mu T^A \xi ] \, G^B_\nu \, G^C_\rho \, \bar{\partial}_\sigma \lambda \, \epsilon^{\mu \nu \rho \sigma} \,. \nonumber
\end{eqnarray}
This result is obtained after summing over the three possible diagrams, together with
exchanging $B, C$ and $\nu, \rho$ indices.
The symmetry factor $1/2$ is due to having two identical particles.

\vspace{1mm}
\noindent
$ \bullet $ ``Symmetric'' part of the amplitude for $ U(1)_Y $ ghost, $ SU(3) $ -- $ SU(3) $ gauge bosons:

\begin{equation}\label{eq:ghostU1_GBSU3_GBSU3_sym}
    - \frac{1}{4} c \, y^q_L \, {\rm Tr} \{ T^A (T^B T^C + T^C T^B) \} \, G^B_\nu \, G^C_\rho \, \bar{\partial}_\sigma \lambda [ \bar{\xi} S^{\sigma \nu \rho} T^A \xi ] \,,
\end{equation}
where $ S^{\sigma \nu \rho} = \bar{\gamma}^\sigma (\bar{\gamma}^\nu \bar{\gamma}^\rho + \bar{\gamma}^\rho \bar{\gamma}^\nu) + (\bar{\gamma}^\nu \bar{\gamma}^\sigma \bar{\gamma}^\rho + \bar{\gamma}^\rho \bar{\gamma}^\sigma \bar{\gamma}^\nu) + (\bar{\gamma}^\nu \bar{\gamma}^\rho + \bar{\gamma}^\rho \bar{\gamma}^\nu) \bar{\gamma}^\sigma $ is totally symmetric. Again, a symmetry factor $1/2$ applies.

\vspace{1mm}
\noindent
$ \bullet $ ``Symmetric'' part of the amplitude for $ SU(3) $ ghost, $ U(1)_Y $ -- $ SU(3) $ gauge bosons:

\begin{equation}\label{eq:ghostSU3_GBU1_GBSU3_sym}
    - \frac{1}{2} c \, y^q_L \, {\rm Tr} \{ T^A (T^B T^C + T^C T^B) \} \, G^B_\nu \, B_\rho \, \bar{\partial}_\sigma g^C [ \bar{\xi} S^{\sigma \nu \rho} T^A \xi ] \,.
\end{equation}

\vspace{1mm}
\noindent
$ \bullet $ ``Anti-symmetric'' part of the amplitude for $ SU(3) $ ghost, $ U(1)_Y $ -- $ SU(3) $ gauge bosons:

\begin{eqnarray}\label{eq:ghostSU3_GBU1_GBSU3_asym}
    && i c \, y^q_L \, {\rm Tr} \{ T^A (T^B T^C - T^C T^B) \} \, [ \bar{\xi} \bar{\gamma}_\mu T^A \xi ] \, B_\nu \, G^C_\rho \, \bar{\partial}_\sigma g^B \, \epsilon^{\mu \nu \rho \sigma} \\
    && = - \frac{1}{2} c \, y^q_L \, C_{ A B C } \, [ \bar{\xi} \bar{\gamma}_\mu T^A \xi ] \, B_\nu \, G^C_\rho \, \bar{\partial}_\sigma g^B \, \epsilon^{\mu \nu \rho \sigma} \,. \nonumber
\end{eqnarray}

Before looking for counterterms, Wess-Zumino consistent conditions offer sanity checks. For instance, by taking the BRST transformation of Eqs.~\eqref{eq:ghostU1_GBSU3}, \eqref{eq:ghostU1_GBSU3_GBSU3_asym} and \eqref{eq:ghostSU3_GBU1_GBSU3_asym}, which involve different numbers of gauge bosons, it is straightforward to check that the coefficient accompanying the structure

\begin{equation}
    C_{A B C} \epsilon^{\mu \nu \rho \sigma} \bar{\partial}_\rho g^C G^B_\nu \bar{\partial}_\sigma \lambda \, [ \bar{\xi} \bar{\gamma}_\mu T^A \xi ]
\end{equation}
vanishes in the sum of these amplitudes.

Concerning the counterterms, we first have:

\begin{eqnarray}
    && s [ g^{(0)} ( \bar{\xi} \bar{\gamma}_\mu T^A \xi ) \, ( B_\sigma \bar{\partial}_\rho G^A_\nu - f C_{A B C} B_\sigma G^B_\rho G^C_\nu ) \, \epsilon^{\mu \nu \rho \sigma} ] \\
    && = g^{(0)} ( \bar{\xi} \bar{\gamma}_\mu T^A \xi ) \, [ (\bar{\partial}_\rho G^A_\nu + f C_{A B C} G^B_\nu G^C_\rho) \bar{\partial}_\sigma \lambda - C_{A B C} (1 + 2 f) G^C_\rho B_\nu \bar{\partial}_\sigma g^B ] \, \epsilon^{\mu \nu \rho \sigma} \,. \nonumber
\end{eqnarray}
If $ f = - 3 / 8 $, then we have counterterms for Eqs.~\eqref{eq:ghostU1_GBSU3} and \eqref{eq:ghostU1_GBSU3_GBSU3_asym} after adjusting the value of the coefficient $g^{(0)}$. This automatically provides a counterterm for Eq.~\eqref{eq:ghostSU3_GBU1_GBSU3_asym}.

Eqs.~\eqref{eq:ghostSU3_GBU1_GBSU3_sym} and \eqref{eq:ghostU1_GBSU3_GBSU3_sym} are closely connected, as shown by the candidate counterterm:

\begin{eqnarray}
    && s [ g^{(2)} {\rm Tr} \{ T^A (T^B T^C + T^C T^B) \} \, G^B_\nu \, G^C_\rho \, B_\sigma ( \bar{\xi} S^{\sigma \nu \rho} T^A \xi ) ] \\
    && = g^{(2)} {\rm Tr} \{ T^A (T^B T^C + T^C T^B) \} \, ( 2 \bar{\partial}_\sigma g^C \, G^B_\nu \, B_\rho + G^B_\nu \, G^C_\rho \, \bar{\partial}_\sigma \lambda ) ( \bar{\xi} S^{\sigma \nu \rho} T^A \xi ) \,, \nonumber
\end{eqnarray}
whose coefficient $g^{(2)}$ has to be adjusted.

Finding a counterterm is simpler in the case of Eq.~\eqref{eq:ghostSU3_GBU1}: we just need to adjust the coefficient $g^{(1)}$ of:

\begin{equation}
    s [ g^{(1)} ( \bar{\xi} \bar{\gamma}_\mu T^A \xi ) \, \bar{\partial}_\rho B_\nu G^A_\sigma \, \epsilon^{\mu \nu \rho \sigma} ] = g^{(1)} ( \bar{\xi} \bar{\gamma}_\mu T^A \xi ) \, \bar{\partial}_\rho B_\nu \bar{\partial}_\sigma g^A \, \epsilon^{\mu \nu \rho \sigma} \,.
\end{equation}

To summarize, we find:

\begin{equation}
    g^{(0)} = - 2 \, c \, y^q_L \,, \;\; g^{(1)} = - 2 \, c \, y^q_L \,, \;\; g^{(2)} = \frac{1}{4} c \, y^q_L \,,
\end{equation}
with an overall $-1$ for counterterms. For an internal right-handed fermion, the hypercharge $y^q_L$ is obviously replaced by $y^u_R$ or $y^d_R$.

\subsection{$SU(3)$ ghost and gauge fields}

We now analyze a substantially more complicated case, presenting first the calculated amplitudes, see Fig.~\ref{fig:diagrams_no_scalars}, and then discussing how counterterms were determined.

\vspace{2mm}
\noindent
$ \bullet $ Amplitude with no gauge boson fields:

\begin{equation}\label{eq:ghostSU3}
    c \, {\rm Tr} \{ T^A T^B \} \, [ \bar{\xi} \bar{\gamma}_\mu T^A \xi ] \, \bar{\partial}^2 \bar{\partial}^\mu g^B = c \frac{1}{2} \, [ \bar{\xi} \bar{\gamma}^\mu T^A \xi ] \, \bar{\partial}^2 \bar{\partial}_\mu g^A \,.
\end{equation}

\vspace{1mm}
\noindent
$ \bullet $ ``Anti-symmetric'' part of the amplitude with $ SU(3) $ ghost, $ SU(3) $ gauge boson:

\begin{eqnarray}\label{eq:ghostSU3_GBSU3_asym}
    && -i c \, {\rm Tr} \{ T^A (T^B T^C - T^C T^B) \} \, [ \bar{\xi} \bar{\gamma}^\mu T^A \xi ] \nonumber \\
    && \times \left( \bar{\partial}^2 g^B G^C_\mu + 2 \bar{\partial}_\nu g^B \bar{\partial}^\nu G^C_\mu - 2 \bar{\partial}_\mu g^B \bar{\partial}^\nu G^C_\nu - 2 \bar{\partial}^\nu \bar{\partial}_\mu g^B G^C_\nu - 2 \bar{\partial}^\nu g^B \bar{\partial}_\mu G^C_\nu \right) \nonumber \\
    && = \frac{1}{2} c \, C_{A B C} \, [ \bar{\xi} \bar{\gamma}^\mu T^A \xi ] \\
    && \times \left( \bar{\partial}^2 g^B G^C_\mu + 2 \bar{\partial}_\nu g^B \bar{\partial}^\nu G^C_\mu - 2 \bar{\partial}_\mu g^B \bar{\partial}^\nu G^C_\nu - 2 \bar{\partial}_\mu \bar{\partial}^\nu g^B G^C_\nu - 2 \bar{\partial}^\nu g^B \bar{\partial}_\mu G^C_\nu \right) \,. \nonumber
\end{eqnarray}

\vspace{1mm}
\noindent
$ \bullet $ ``Symmetric'' part of the amplitude with $ SU(3) $ ghost, $ SU(3) $ gauge boson:

\begin{eqnarray}\label{eq:ghostSU3_GBSU3_sym}
    2 c \, {\rm Tr} \{ T^A ( T^B T^C + T^C T^B ) \} \, [\bar{\xi} \bar{\gamma}_\mu T^A \xi] \bar{\partial}_\rho G^B_\nu \bar{\partial}_\sigma g^C \epsilon^{\mu \nu \rho \sigma} \,.
\end{eqnarray}

\vspace{1mm}
\noindent
$ \bullet $ We have various contributions to the amplitude with one ghost and two gauge bosons. We have first the symmetric structure:

\begin{eqnarray}\label{eq:symmetric_structure}
    && - \frac{1}{2} c \left( {\rm Tr} \{ T^A T^B T^C T^D + T^A T^B T^D T^C \} \bar{\partial}_\alpha g^B G^C_\mu G^D_\beta \right. \\
    && \qquad\qquad + {\rm Tr} \{ T^A T^C T^B T^D + T^A T^D T^B T^C \} G^C_\alpha \bar{\partial}_\mu g^B G^D_\beta \nonumber \\
    && \qquad\qquad + \left. {\rm Tr} \{ T^A T^C T^D T^B + T^A T^D T^C T^B \} G^C_\alpha G^D_\mu \bar{\partial}_\beta g^B \right) [ \bar{\xi} \bar{\gamma}^\alpha \bar{\gamma}^\mu \bar{\gamma}^\beta T^A \xi ] \,. \nonumber
\end{eqnarray}

\vspace{1mm}
\noindent
$ \bullet $ We also have the following ``$\mathcal{A}$'' structure:

\begin{equation}\label{eq:A_totally_antisymmetric}
    i \frac{1}{2} c \, [\bar{\xi} \bar{\gamma}_\mu T^A \xi] G^B_\nu G^C_\rho \bar{\partial}_\sigma g^D \epsilon^{\mu \nu \rho \sigma} \mathcal{A}_{A}^{B C D} \,,
\end{equation}
where $\mathcal{A}_{A}^{B C D}$, which is totally antisymmetric
in $B, C, D$, is defined in Eq.~\eqref{eq:ASU3_definition}.

\vspace{1mm}
\noindent
$ \bullet $ The ``$D \cdot C$'' structure accompanying a Levi-Civita symbol:

\begin{eqnarray}\label{eq:d_f_structure}
    && i c \, [\bar{\xi} \bar{\gamma}_\mu T^A \xi] G^B_\nu G^C_\rho \bar{\partial}_\sigma g^D \epsilon^{\mu \nu \rho \sigma} \, {\rm Tr} \{ T^A ( T^B T^D T^C - T^C T^D T^B ) \} \\
    && = - c \, [\bar{\xi} \bar{\gamma}_\mu T^A \xi] G^B_\nu G^C_\rho \bar{\partial}_\sigma g^D \epsilon^{\mu \nu \rho \sigma} \, \frac{1}{4} \left( D_{A B E} C_{D C E} + D_{A D E} C_{B C E} + D_{A C E} C_{B D E} \right) \nonumber \\
    && = - c \, [\bar{\xi} \bar{\gamma}_\mu T^A \xi] G^B_\nu G^C_\rho \bar{\partial}_\sigma g^D \epsilon^{\mu \nu \rho \sigma} \, \frac{1}{4} \left( 2 D_{A B E} C_{D C E} + D_{A D E} C_{B C E} \right) \,, \nonumber
\end{eqnarray}
where we have used the property that (see Ref.~\cite{Haber:2019sgz} for a systematic presentation)

\begin{eqnarray}\label{eq:identity_Haber}
    && {\rm Tr} \{ T^A T^B T^C T^D \} = \frac{1}{4 N} \left( \delta_{A B} \delta_{C D} - \delta_{A C} \delta_{B D} + \delta_{A D} \delta_{B C} \right) \\
    && + \frac{1}{8} \left( D_{A B E} D_{C D E} - D_{A C E} D_{B D E} + D_{A D E} D_{B C E} \right) \nonumber \\
    && + \frac{1}{8} i \left( D_{A B E} C_{C D E} + D_{A C E} C_{B D E} + D_{A D E} C_{B C E} \right) \,, \nonumber
\end{eqnarray}
with $N=N_c=3$ (and $N=2$ when adapting the discussion of this section to the operators $ Q_{\ell q}^{(3)}, Q_{q q}^{(3)} $, etc.).

\vspace{1mm}
\noindent
$ \bullet $ The ``$C \cdot C$'' structure:

\begin{eqnarray}\label{eq:f_times_f}
    && c \, [\bar{\xi} \bar{\gamma}^\mu T^A \xi] G^B_\mu G^C_\nu \bar{\partial}^\nu g^D \, {\rm Tr} \{ T^A T^D ( T^B T^C - T^C T^B ) \} \\
    && + c \, [\bar{\xi} \bar{\gamma}^\mu T^A \xi] G^B_\nu G^C_\mu \bar{\partial}^\nu g^D \, {\rm Tr} \{ T^A ( T^B T^C - T^C T^B ) T^D \} \nonumber \\
    && = c \, [\bar{\xi} \bar{\gamma}^\mu T^A \xi] G^B_\mu G^C_\nu \bar{\partial}^\nu g^D \, {\rm Tr} \{ ( T^A T^D - T^D T^A ) ( T^B T^C - T^C T^B ) \} \nonumber \\
    && = - c \, [\bar{\xi} \bar{\gamma}^\mu T^A \xi] G^B_\mu G^C_\nu \bar{\partial}^\nu g^D \frac{1}{2} C_{A D E} C_{B C E} \,. \nonumber
\end{eqnarray}

Before looking for counterterms, we consider sanity checks based on the Wess-Zumino consistency conditions. For instance, it is straightforward to check explicitly that the coefficients that accompany the structures

\begin{eqnarray}
    && C_{A B C} C_{C E F} G^E_\mu \bar{\partial}^\nu g^F \bar{\partial}_\nu g^B [ \bar{\xi} \bar{\gamma}^\mu T^A \xi ] \,, \quad
    C_{A B C} C_{C E F} G^E_\nu \bar{\partial}_\mu g^F \bar{\partial}^\nu g^B [ \bar{\xi} \bar{\gamma}^\mu T^A \xi ] \,, \nonumber \\
    && C_{A B C} \bar{\partial}^\nu \bar{\partial}_\mu g^C \bar{\partial}_\nu g^B [ \bar{\xi} \bar{\gamma}^\mu T^A \xi ] \,, \quad
    C_{A B C} \bar{\partial}^2 g^C \bar{\partial}_\mu g^B [ \bar{\xi} \bar{\gamma}^\mu T^A \xi ] \,,
\end{eqnarray}
resulting from the combination of the amplitudes in Eqs.~\eqref{eq:ghostSU3} and \eqref{eq:ghostSU3_GBSU3_asym}, which involve different numbers of gauge bosons, vanish.

For the candidate counterterm, we first have:

\begin{eqnarray}\label{eq:CT_f1_1GB}
    && s [ f_1^{(1)} \bar{\partial}^2 G^A_\mu ( \bar{\xi} \bar{\gamma}^\mu T^A \xi ) ] = f_1^{(1)} \bar{\partial}^2 ( \bar{\partial}_\mu g^A + C_{A B C} G^B_\mu g^C ) ( \bar{\xi} \bar{\gamma}^\mu T^A \xi ) \\
    && = f_1^{(1)} ( \bar{\partial}^2 \bar{\partial}_\mu g^A + C_{A B C} G^B_\mu \bar{\partial}^2 g^C + 2 C_{A B C} \bar{\partial}^\nu G^B_\mu \bar{\partial}_\nu g^C ) ( \bar{\xi} \bar{\gamma}^\mu T^A \xi ) \,, \nonumber
\end{eqnarray}

\begin{eqnarray}\label{eq:CT_f2_1GB}
    && s [ f_2^{(1)} \bar{\partial}_\mu \bar{\partial}^\nu G^A_\nu ( \bar{\xi} \bar{\gamma}^\mu T^A \xi ) ] = f_2^{(1)} \bar{\partial}_\mu \bar{\partial}^\nu ( \bar{\partial}_\nu g^A + C_{A B C} G^B_\nu g^C ) ( \bar{\xi} \bar{\gamma}^\mu T^A \xi ) \\
    && = f_2^{(1)} ( \bar{\partial}^2 \bar{\partial}_\mu g^A + C_{A B C} G^B_\nu \bar{\partial}_\mu \bar{\partial}^\nu g^C + C_{A B C} \bar{\partial}_\mu G^B_\nu \bar{\partial}^\nu g^C + C_{A B C} \bar{\partial}^\nu G^B_\nu \bar{\partial}_\mu g^C ) ( \bar{\xi} \bar{\gamma}^\mu T^A \xi ) \,, \nonumber
\end{eqnarray}
so that the amplitude in Eq.~\eqref{eq:ghostSU3} fixes the value of $ f_1^{(1)} + f_2^{(1)} = - c / 2 $.

Apart from these operators of coefficients $ f_1^{(1)} $ and $ f_2^{(1)} $, we also have as candidate counterterms:

\begin{eqnarray}\label{eq:CT_f1_2GB}
    && s [ f_1^{(2)} \bar{\partial}_\mu G^B_\nu G^{C \nu} ( \bar{\xi} \bar{\gamma}^\mu C_{A B C} T^A \xi ) ] \\
    && = f_1^{(2)} \left( G^C_\nu \bar{\partial}_\mu \bar{\partial}^\nu g^B + C_{B E F} G^{C \nu} G^E_\nu \bar{\partial}_\mu g^F + \bar{\partial}_\mu G^B_\nu \bar{\partial}^\nu g^C \right) ( \bar{\xi} \bar{\gamma}^\mu C_{A B C} T^A \xi ) \,, \nonumber
\end{eqnarray}

\begin{eqnarray}\label{eq:CT_f2_2GB}
    && s [ f_2^{(2)} \bar{\partial}^\nu G^B_\mu G^C_\nu ( \bar{\xi} \bar{\gamma}^\mu C_{A B C} T^A \xi ) ] \\
    && = f_2^{(2)} \left( G^C_\nu \bar{\partial}_\mu \bar{\partial}^\nu g^B + C_{B E F} G^C_\nu G^E_\mu \bar{\partial}^\nu g^F + \bar{\partial}^\nu G^B_\mu \bar{\partial}_\nu g^C \right) ( \bar{\xi} \bar{\gamma}^\mu C_{A B C} T^A \xi ) \,, \nonumber
\end{eqnarray}

\begin{eqnarray}\label{eq:CT_f3_2GB}
    && s [ f_3^{(2)} \bar{\partial}^\nu G^B_\nu G^C_\mu ( \bar{\xi} \bar{\gamma}^\mu C_{A B C} T^A \xi ) ] \nonumber \\
    && = f_3^{(2)} \left( G^C_\mu \bar{\partial}^2 g^B + C_{B E F} G^E_\nu G^C_\mu \bar{\partial}^\nu g^F + \bar{\partial}^\nu G^B_\nu \bar{\partial}_\mu g^C \right) ( \bar{\xi} \bar{\gamma}^\mu C_{A B C} T^A \xi ) \nonumber \\
    && = f_3^{(2)} \left( G^C_\mu \bar{\partial}^2 g^B + \bar{\partial}^\nu G^B_\nu \bar{\partial}_\mu g^C \right) ( \bar{\xi} \bar{\gamma}^\mu C_{A B C} T^A \xi ) \\
    && + f_3^{(2)} \left( C_{B C E} C_{A D E} G^B_\mu G^C_\nu \bar{\partial}^\nu g^D + C_{B E F} C_{A B C} G^C_\nu G^E_\mu \bar{\partial}^\nu g^F \right) ( \bar{\xi} \bar{\gamma}^\mu T^A \xi ) \,, \nonumber
\end{eqnarray}
where in the last equation we have employed

\begin{equation}
    C_{B E F} C_{A B C} = C_{E C B} C_{A B F} + C_{F C B} C_{A E B} \,.
\end{equation}

\noindent Looking at the structure $ C_{A B C} G^C_\mu \bar{\partial}^2 g^B ( \bar{\xi} \bar{\gamma}^\mu T^A \xi ) $, we have that

\begin{equation}
    - f_1^{(1)} + f_3^{(2)} = - \frac{1}{2} c \,,
\end{equation}
is fixed by the amplitude shown in Eq.~\eqref{eq:ghostSU3_GBSU3_asym}. Similarly,

\begin{equation}
    - f_2^{(1)} + f_1^{(2)} + f_2^{(2)} = c \,,
\end{equation}
after looking at the structure $ C_{A B C} G^C_\nu \bar{\partial}_\mu \bar{\partial}^\nu g^B ( \bar{\xi} \bar{\gamma}^\mu T^A \xi ) $;

\begin{equation}
    2 f_1^{(1)} + f_2^{(2)} = c \,,
\end{equation}
after looking at the structure $ C_{A B C} \bar{\partial}^\nu G^B_\mu \bar{\partial}_\nu g^C ( \bar{\xi} \bar{\gamma}^\mu T^A \xi ) $;

\begin{equation}
    f_2^{(1)} + f_1^{(2)} = - c \,,
\end{equation}
after looking at the structure $ C_{A B C} \bar{\partial}_\mu G^B_\nu \bar{\partial}^\nu g^C ( \bar{\xi} \bar{\gamma}^\mu T^A \xi ) $;

\begin{equation}
    f_2^{(1)} + f_3^{(2)} = - c \,,
\end{equation}
after looking at the structure $ C_{A B C} \bar{\partial}^\nu G^B_\nu \bar{\partial}_\mu g^C ( \bar{\xi} \bar{\gamma}^\mu T^A \xi ) $. These relations among the coefficients are consistent, and imply

\begin{eqnarray}\label{eq:sols_0_1_2_GBs}
    && f_2^{(1)} = - \frac{c}{2} - f_1^{(1)} \,, \;\; f_1^{(2)} = f_1^{(1)} - \frac{c}{2} \,, \\
    && f_2^{(2)} = - 2 f_1^{(1)} + c \,, \;\; f_3^{(2)} = f_1^{(1)} - \frac{c}{2} = f_1^{(2)} \,. \nonumber
\end{eqnarray}
One particular case is $ f_1^{(1)} = c/2 $, but this is not the most general solution. This point is further elaborated below.

The symmetric structure of Eq.~\eqref{eq:symmetric_structure} requires the introduction of the following counterterm:

\begin{eqnarray}\label{eq:CT_f1_3GB}
    f^{(3)}_1 {\rm Tr} \{ T^A T^B T^C T^D \} G^B_\alpha G^C_\mu G^D_\beta ( \bar{\xi} \bar{\gamma}^\alpha \bar{\gamma}^\mu \bar{\gamma}^\beta T^A \xi ) \,,
\end{eqnarray}
with $ f^{(3)}_1 = c $, in which case it acts as a counterterm for the amplitude in Eq.~\eqref{eq:symmetric_structure} and also the amplitude in Eq.~\eqref{eq:f_times_f}. However, it also introduces new ``$ D \cdot C $'' terms without a Levi-Civita symbol

\begin{eqnarray}
    && s [ f^{(3)}_1 {\rm Tr} \{ T^A T^B T^C T^D \} G^B_\alpha G^C_\mu G^D_\beta ( \bar{\xi} \bar{\gamma}^\alpha \bar{\gamma}^\mu \bar{\gamma}^\beta T^A \xi ) ] + \text{Eq.}~\eqref{eq:symmetric_structure} + \text{Eq.}~\eqref{eq:f_times_f} \nonumber \\
    && = i \frac{f_1^{(3)}}{8} ( \bar{\xi} \bar{\gamma}^\alpha \bar{\gamma}^\mu \bar{\gamma}^\beta T^A \xi ) \left[ C_{C D E} D_{A B E} \left( \bar{\partial}_\alpha g^B G^C_\mu G^D_\beta + G^C_\alpha G^D_\mu \bar{\partial}_\beta g^B \right) \right. \nonumber \\
    && \left. + \left( C_{C B E} D_{A D E} + C_{C D E} D_{A B E} + C_{B D E} D_{A C E} \right) G^C_\alpha \bar{\partial}_\mu g^B G^D_\beta \right] \,.
\end{eqnarray}
These contributions are canceled by the addition of

\begin{eqnarray}\label{eq:CT_f1_3GB_fd}
    -i \frac{f_1^{(3)}}{8} ( \bar{\xi} \bar{\gamma}^\alpha \bar{\gamma}^\mu \bar{\gamma}^\beta T^A \xi ) \left( G^B_\alpha G^C_\mu G^D_\beta + G^C_\alpha G^B_\mu G^D_\beta + G^C_\alpha G^D_\mu G^B_\beta \right) C_{C D E} D_{A B E} \,.
\end{eqnarray}

Focusing on structures proportional to ``$ C \cdot C $'', introduced by the counterterms of coefficients $ f^{(2)}_{1,2,3} $, we also have

\begin{eqnarray}\label{eq:CT_f4_3GB}
    && s [ f^{(3)}_4 ( \bar{\xi} \bar{\gamma}^\mu T^A \xi ) G^B_\mu G^C_\nu G^{D \nu} C_{A D E} C_{B C E} ] \\
    && = f^{(3)}_4 ( \bar{\xi} \bar{\gamma}^\mu T^A \xi ) ( \bar{\partial}_\mu g^B G^C_\nu G^{D \nu} + G^B_\mu \bar{\partial}_\nu g^C G^{D \nu} + G^B_\mu G^C_\nu \bar{\partial}^\nu g^D ) C_{A D E} C_{B C E} \nonumber \\
    && = f^{(3)}_4 ( \bar{\xi} \bar{\gamma}^\mu T^A \xi ) G^{C \nu} G^E_\nu \bar{\partial}_\mu g^F C_{A B C} C_{B E F} - f^{(3)}_4 ( \bar{\xi} \bar{\gamma}^\mu T^A \xi ) G^C_\nu G^E_\mu \bar{\partial}^\nu g^F C_{A B C} C_{B E F} \nonumber \\
    && + f^{(3)}_4 ( \bar{\xi} \bar{\gamma}^\mu T^A \xi ) G^B_\mu G^C_\nu \bar{\partial}^\nu g^D C_{A D E} C_{B C E} \,. \nonumber
\end{eqnarray}
Note that a similar structure identically vanishes

\begin{equation}
    ( \bar{\xi} \bar{\gamma}^\mu T^A \xi ) G^{E \nu} G^F_\nu G^C_\mu C_{A B C} C_{B E F} = 0 \,.
\end{equation}
We then have that $ f^{(3)}_4 + f^{(2)}_3 = 0 $,
$ f^{(3)}_4 + f^{(2)}_1 = 0 $, $ - f^{(3)}_4 + f^{(2)}_2 + f^{(2)}_3 = 0 $. Together with Eq.~\eqref{eq:sols_0_1_2_GBs}, this linear system of equations admits the following solution:

\begin{eqnarray}
    f^{(3)}_4 = f^{(1)}_2 + c \,.
\end{eqnarray}
In the following, we consider the choice $ f_1^{(1)} = c/2 $. Shifting this choice of $ f_1^{(1)} $ by a constant amounts to introducing a counterterm in the finite renormalization proportional to the gauge invariant structure:

\begin{equation}
    i \, \bar{\xi} \bar{\gamma}_\mu [\bar{D}_\nu, F^{\mu \nu}_A] T^A \xi = i \, \bar{\xi} \bar{\gamma}_\mu T^A \xi \, [\bar{D}_\nu, F^{\mu \nu}_A] \,,
\end{equation}
where $ \bar{D}_\mu $ is the covariant derivative in four dimensions.

Turning to the remaining simplest structure,
Eq.~\eqref{eq:ghostSU3_GBSU3_sym} fixes the coefficient of:

\begin{eqnarray}\label{eq:CT_f4_2GB}
    && s [ f^{(2)}_4 {\rm Tr} \{ T^A ( T^B T^C + T^C T^B ) \} \, [\bar{\xi} \bar{\gamma}_\mu T^A \xi] \bar{\partial}_\rho G^B_\nu G^C_\sigma \epsilon^{\mu \nu \rho \sigma} ] \\
    && = f^{(2)}_4 {\rm Tr} \{ T^A ( T^B T^C + T^C T^B ) \} \, [\bar{\xi} \bar{\gamma}_\mu T^A \xi] \left( C_{B E F} G^C_\sigma G^E_\nu \bar{\partial}_\rho g^F + \bar{\partial}_\rho G^B_\nu \bar{\partial}_\sigma g^C \right) \epsilon^{\mu \nu \rho \sigma} \,, \nonumber
\end{eqnarray}
i.e., $ f^{(2)}_4 = - 2 c $.
We introduce the following counterterm to address the ``$ D \cdot C $'' structure of Eq.~\eqref{eq:d_f_structure}:

\begin{eqnarray}\label{eq:CT_f3_3GB}
    && s [ f^{(3)}_3 ( \bar{\xi} \bar{\gamma}_\mu T^A \xi ) G^B_\nu G^C_\rho G^D_\sigma \epsilon^{\mu \nu \rho \sigma} D_{A D E} C_{B C E} ] \\
    && = f^{(3)}_3 ( \bar{\xi} \bar{\gamma}_\mu T^A \xi ) \left( 2 \bar{\partial}_\nu g^B G^C_\rho G^D_\sigma + G^B_\nu G^C_\rho \bar{\partial}_\sigma g^D \right) \epsilon^{\mu \nu \rho \sigma} D_{A D E} C_{B C E} \nonumber \\
    && = f^{(3)}_3 ( \bar{\xi} \bar{\gamma}_\mu T^A \xi ) G^B_\nu G^C_\rho \bar{\partial}_\sigma g^D \epsilon^{\mu \nu \rho \sigma} D_{A D E} C_{B C E} \nonumber \\
    && - 2 f^{(3)}_3 ( \bar{\xi} \bar{\gamma}_\mu T^A \xi ) G^B_\nu G^C_\rho \bar{\partial}_\sigma g^D \epsilon^{\mu \nu \rho \sigma} D_{A B E} C_{D C E} \,, \nonumber
\end{eqnarray}
i.e., $ f^{(3)}_3 = c / 4 $, by direct comparison with Eq.~\eqref{eq:d_f_structure}. For the other piece of this amplitude,
since

\begin{equation}
    {\rm Tr} \{ T^A ( T^B T^C + T^C T^B ) \} = \frac{1}{2} D_{A B C} \,,
\end{equation}
and

\begin{eqnarray}
    && f^{(2)}_4 {\rm Tr} \{ T^A ( T^B T^C + T^C T^B ) \} \, [\bar{\xi} \bar{\gamma}_\mu T^A \xi] C_{B E F} G^C_\sigma G^E_\nu \bar{\partial}_\rho g^F \epsilon^{\mu \nu \rho \sigma} \\
    && = - f^{(2)}_4 \frac{1}{2} D_{A B E} \, [\bar{\xi} \bar{\gamma}_\mu T^A \xi] C_{D C E} G^B_\nu G^C_\rho \bar{\partial}_\sigma g^D \epsilon^{\mu \nu \rho \sigma} \,, \nonumber
\end{eqnarray}
we have that $ - f^{(2)}_4 / 2 - 2 f^{(3)}_3 = c / 2 $, which holds true.

Finally, for the $ \mathcal{A} $-structure in Eq.~\eqref{eq:A_totally_antisymmetric} we have:

\begin{equation}\label{eq:CT_f2_3GB}
    s [ f^{(3)}_2 ( \bar{\xi} \bar{\gamma}_\mu T^A \xi ) G^B_\nu G^C_\rho G^D_\sigma \epsilon^{\mu \nu \rho \sigma} \mathcal{A}_{A}^{B C D} ] = 3 f^{(3)}_2 ( \bar{\xi} \bar{\gamma}_\mu T^A \xi ) G^B_\nu G^C_\rho \bar{\partial}_\sigma g^D \epsilon^{\mu \nu \rho \sigma} \mathcal{A}_{A}^{B C D} \,,
\end{equation}
where $ 3 f^{(3)}_2 = -i c / 2 $.

To summarize, counterterms are found in
Eqs.~\eqref{eq:CT_f1_1GB},
\eqref{eq:CT_f2_1GB},
\eqref{eq:CT_f1_3GB},
\eqref{eq:CT_f1_3GB_fd},
\eqref{eq:CT_f4_2GB},
\eqref{eq:CT_f3_3GB},
\eqref{eq:CT_f2_3GB}.

%%%%%%%%%%%%%%%%%%%%%%%%%%%%%%%

\bibliography{mybib}{}
\bibliographystyle{unsrturl}

\end{document}